\documentclass[twoside,12pt,a4paper]{report}

\usepackage[english]{babel}
\usepackage{epsfig,graphicx}
\graphicspath{{Figures/}} 
\DeclareGraphicsExtensions{.eps,.eps.gz}
\usepackage{makeidx}
\makeindex     
\usepackage{amsmath,amssymb,amsfonts,amsxtra}
\usepackage{euscript}
\usepackage{multicol}
\usepackage{layout}
\usepackage{moreverb,ulem}
\usepackage{setspace}
\usepackage{fancybox,boxedminipage,psboxit}
\usepackage{fancyheadings}
\usepackage{pifont}

\makeatletter  
\renewcommand{\section}{
\@startsection{section}{1}{0pt}{\baselineskip}{2ex}%
   {\Large\bfseries\uline}%
}
\renewcommand{\subsection}{
\@startsection{subsection}{2}{0pt}{\baselineskip}{1ex}%
   {\Large\bfseries\itshape\uline}%
}
\renewcommand{\subsubsection}{
\@startsection{subsubsection}{3}{0pt}{\baselineskip}{1ex}%
   {\large\itshape\uline}%
}
\renewcommand{\paragraph}{
\@startsection{paragraph}{4}{0pt}{\parindent}{-1em}%
   {\large\ttfamily\bfseries\uline}%
}
\renewcommand{\subparagraph}{
\@startsection{subparagraph}{5}{0pt}{\parindent}{-1em}%
   {\large\ttfamily\bfseries}%
}
\makeatother  

\renewcommand{\thesection}{\Roman{section}}

\renewcommand{\theequation}{\thesection.\arabic{equation}}

\newcommand{\mysecnp}[1]{\ref{#1}}

\newcommand{\encadre}[1]{
\begin{boxitpara}{box 0.7}\begin{center}#1\end{center}\end{boxitpara}
}

\newcommand{\Main}[0]{{\large{\ding{253} }}}

\newcommand{\somme}[2]{\underset{#1}{\overset{#2}{\sum}}}

\newcommand{\eg}[0]{\;=\;}
\newcommand{\x}[0]{\times}
\newcommand{\V}[1]{\overrightarrow{#1}}

\newcommand{\abs}[1]{\mid{#1}\mid }

\newcommand{\EnsR}[0]{\mathbb{R}}
\newcommand{\EnsC}[0]{\mathbb{C}}

\newcommand{\dsd}[1]{\frac{d}{d{#1}}}

\newcommand{\integ}[2]{\int\limits_{#1}^{#2}}

\newcommand{\conj}[1]{\overline{#1}}

\newcommand{\pl}[1]{P_{#1}}

\newcommand{\plm}[2]{P_{#1}^{#2}}

\newcommand{\hplm}[2]{\lambda_{#1}^{#2}}
\newcommand{\ylm}[2]{Y_{#1}^{#2}}

\newcommand{\alm}[2]{a_{{#1}{#2}}}
\newcommand{\cl}[1]{C_{#1}}
\newcommand{\clest}[1]{\widehat{C}_{#1}}
\newcommand{\blm}[2]{b_{{#1}{#2}}}
\newcommand{\bl}[1]{\mathcal{B}_{#1}}
\newcommand{\aplm}[2]{\widetilde{a}_{{#1}{#2}}}
\newcommand{\cpl}[1]{\widetilde{C}_{#1}}
\newcommand{\cplr}[1]{\widetilde{C}_{#1}^{R}}
\newcommand{\fcor}[0]{\xi}
\newcommand{\fcorp}[0]{\widetilde{\xi}}
\newcommand{\fcorest}[0]{\hat{\xi}}

\newcommand{\clg}[2]{<#1\mid #2>}
\newcommand{\wigj}[6]{\begin{pmatrix} #1&#2&#3\\ #4&#5&#6 \end{pmatrix}}
\newcommand{\drot}[2]{\mathfrak{D}^{(#1)}_{#2}}

\newcommand{\teafi}[0]{(\theta,\phi)}

\newcommand{\undemi}[0]{\frac{1}{2}}

\begin{document}

\begin{titlepage}

\begin{center}

{\Huge{\bf
\hrule
\vskip 1.5cm
Notes on bias and covariance matrix \\
\vskip 0.25cm
of the angular power spectrum \\
\vskip 0.25cm
on small sky maps
\vskip 1.5cm
\hrule
}}

{\Large
\vskip 2cm
C.~Magneville and J.P.~Pansart
\vskip 1cm
DSM/DAPNIA , CEA/SACLAY , F-91191 Gif-sur-Yvette.
}

\end{center}

\vskip 2cm
{\noindent \it Abstract:} \\
We compute the effects induced by the use of 
small CMB maps on the measurement of the
$\cl{l}$ coefficients of the angular power spectrum
and show that small systematic effects have
to be taken into account.
We also compute numerically the cosmic variance and covariance
of the $\cl{l}$ spectrum for various spherical cap like maps.
Comparisons with simulations are presented.
The calculations are done using the standard method based
on the spherical harmonic transform
or using the temperature angular correlation spectrum.

\begin{center}
\vskip 1.5cm
\end{center}

\end{titlepage}

\cleardoublepage
\pagenumbering{Roman}
\tableofcontents

\cleardoublepage
\pagenumbering{arabic}
\section{Introduction}

The field of cosmic microwave background (CMB) anisotropies
has dramatically advanced over the last decade especially
on its observational front.
Satellite experiments (COBE,WMAP) observed the whole sky.
But to study the high multipoles of the power spectrum,
balloon-borne experiments have emerged. \\
The BOOMERANG (\cite{Bern00}),
the MAXIMA (\cite{Han00})
and the ARCHEOPS (\cite{Ben03})
balloon experiments made measurements
of the first doppler peak of the CMB spectrum.
Due to a limited observing time and technical constraints,
these experiments only observed a small fraction of the sky.
BOOMERANG observed a region of $1800\;deg^2$ and measured
the spectrum up to multipoles $l_{max}\sim 1000$
and MAXIMA observed $124\;deg^2$ and measured
the spectrum up to $l_{max}\sim 785$.
In the future, the OLIMPO experiment (\cite{Mass06})
will observe $300\;deg^2$ and measure the spectrum
up to $l_{max}\sim 2500$.

In this paper we compute the effects induced by the use of 
small CMB maps on the measurement of the
$\cl{l}$ spectrum at high multipole moments.
Two methods are used to compute the $\cl{l}$ coefficients.
The first one uses the Fast Fourier Transform (FFT)  technics
and the second one the angular correlation spectrum.
These two technics are complementary and lead to different biases.
We show how the variance behaves with respect to the map size
and how the apodization works in the case of the
angular correlation spectrum.
These calculations will be compared with simulated
CMB maps of various sizes and shapes.
Numerical calculations provide a fast tool to assess
most of the biases.

\clearpage
\section{Estimating the power spectrum: formalism \\ for the full sphere.}\label{basic}

\subsection{Spectrum and angular correlation fonction \\ definition.}\label{basicdef}

The observed temperature $T$ on the $2$-sphere is a random
field that can be expanded on the spherical harmonic basis:
\begin{equation}
T(\theta,\phi) \eg 
\somme{l=0}{\infty}\;\somme{m=-l}{+l}\;\alm{l}{m}\;\ylm{l}{m}(\theta,\phi)
\label{tfralm}
\end{equation}
where the $\alm{l}{m}$ coefficients are random variables.
With $\V{\Omega}\equiv (\theta,\phi)$
one has\footnote{The overbar means complex conjugation}
\begin{equation}
\alm{l}{m}
\eg
\integ{S^2}{} T(\V{\Omega}) \;\conj{\ylm{l}{m}}(\V{\Omega)}\;d\V{\Omega}
\label{almfrt}
\end{equation}
$T(\V{\Omega})$ is a real field so $\;\conj{\alm{l}{-m}}=(-1)^m\alm{l}{m}\;$
which is a consequence of the relation
$\;\conj{\ylm{l}{-m}}=(-1)^m\ylm{l}{m}\;$. \\
The observed sky is a particular realisation of that random field.
If one assumes uncorrelated $\alm{l}{m}$ coefficients and isotropy,
one has:
\begin{equation}
<\conj{a_{l\,m}}a_{l'm'}>_{ens}
\eg
\delta _{l\,l'} \,\delta _{m\,m'} \;\cl{l}
\label{malmalm}
\end{equation}
where the $\cl{l}$ are specified by the cosmological theory
and the symbol $<>_{ens}$ means averaging over many sky realisations.

\noindent The temperature angular correlation fonction is defined as:
\begin{equation}
\fcor(\V{\Omega},\V{\Omega'})
\eg
<T (\V{\Omega}) T(\V{\Omega'}) >_{ens}
\label{ksio1o2}
\end{equation}
where $\V{\Omega},\V{\Omega'}$ are the directions of measurement.
The temperature angular correlation function is related to the spectrum of
the primordial fluctuations.
With the isotropy hypothesis, the angular correlation function is only a
function of the angular separation $\gamma$ with
$\;\cos(\gamma)=\V{\Omega}.\V{\Omega'}\;$. We have:
$$ \begin{array}{rcl}
\fcor(\gamma) &=& <\conj{T}(\Omega)\;T(\Omega')>_{ens} \\
 &=&
 \somme{l,m}{}\;\somme{l',m'}{}\;
 <\conj{\alm{l}{m}}\alm{l'}{m'}>_{ens}
 \;\conj{\ylm{l}{m}}(\Omega)\ylm{l'}{m'}(\Omega') \\
 &=&
 \somme{l,m}{}\;\somme{l',m'}{}\; \cl{l}\;\delta_{ll'}\delta_{mm'}
 \;\conj{\ylm{l}{m}}(\Omega)\ylm{l'}{m'}(\Omega') \\
 &=&
 \somme{l,m}{} \cl{l}\;\conj{\ylm{l}{m}}(\Omega)\ylm{l}{m}(\Omega') \\
 &=&
 \frac{1}{4\pi}\;\somme{l}{} (2l+1)\;\cl{l}\;\pl{l}(\cos(\gamma)) \\
\end{array} $$
This relation can be inverted and finally:
\encadre{
\begin{eqnarray}
\fcor(\gamma)
  &=&
  \frac{1}{4\pi}\;\somme{l=0}{\infty}\;(2l+1)\;\cl{l}\;\pl{l}(\cos(\gamma)) 
  \notag  \\
  && \label{cltoksi}  \\
\cl{l}
  &=&
  2\pi\;\integ{\gamma=0}{\pi}\;\fcor(\gamma)\;\pl{l}(\cos(\gamma))
           \;d\cos(\gamma)
  \notag
\end{eqnarray}
}
In the rest of these notes, the isotropy hypothesis (\ref{malmalm})
will always be assumed.

\subsection{Power spectrum estimator.}\label{basicpow}

In practice, one observes a unique sky realisation
and estimators of the power spectrum and the angular correlation function
have to be constructed: 
instead of averaging over many sky realisations,
we can use the sky isotropy hypothesis.
The bias and variance of such estimators have to be computed.

\ \newline
\noindent We can have an estimation of
the $\alm{l}{m}$ by using~(\ref{almfrt}).
Then we get an estimator $\clest{l}$ of the $\cl{l}$ spectrum
by averaging over $m$:
\begin{equation}
\clest{l}\eg \frac{1}{2l+1}\;\somme{m=-l}{l}\;\abs{\alm{l}{m}}^2
\label{clest4p}
\end{equation}
For the whole sky, this estimator is unbiased:
$$
<\clest{l}>_{ens}
  \eg 
\frac{1}{2l+1}\;\somme{m=-l}{+l} <\conj{\alm{l}{m}}\;\alm{l}{m}>_{ens}
  \eg
\frac{1}{2l+1}\;\somme{m=-l}{+l} \cl{l}
  \eg
\cl{l}
$$
We can compute the variance of the estimator:
$$ \begin{array}{rcl}
V(\clest{l},\clest{l'})
 &=&
<\clest{l}\clest{l'}>_{ens}\;-\;<\clest{l}><\clest{l'}>_{ens} \\
 &=&
\frac{1}{(2l+1)(2l'+1)}\;\somme{mm'}{}
 \;<\alm{l}{m}\conj{\alm{l}{m}}\alm{l'}{m'}\conj{\alm{l'}{m'}}>_{ens}
 \;-\; \cl{l}\cl{l'}
\end{array} $$
Provided that the $\alm{l}{m}$ are gaussian random variables,
their fourth order moments can be expressed with their second order moments
and we get:
$$ \begin{array}{rcl}
V(\clest{l},\clest{l'})
 &=&
\frac{1}{(2l+1)(2l'+1)}\;\somme{mm'}{}
\left(
\cl{l}\cl{l'}(1+\delta_{ll'}\delta_{mm'})
  \;+\;
(-1)^{m+m'}\cl{l}\cl{l'}\delta_{ll'}\delta_{m-m'}
\right)
 \;-\; \cl{l}\cl{l'} \\
 &=&
\frac{1}{(2l+1)(2l'+1)}\;\somme{mm'}{}
\left(
\cl{l}\cl{l'} \;+\; \cl{l}^2\delta_{ll'}\delta_{mm'})
  \;+\;
\cl{l}^2\delta_{ll'}\delta_{m-m'}
\right)
 \;-\; \cl{l}\cl{l'} \\
 &=&
\frac{1}{(2l+1)(2l'+1)}\;\somme{mm'}{}\;2\cl{l}^2\delta_{ll'}\delta_{mm'} \\
 &=&
\frac{2\cl{l}^2}{2l+1}\;\delta_{ll'}
\end{array} $$
\encadre{
Thus the $\clest{l}$ are unbiased and independants,
they follow a $\chi^2$ distribution with $2l+1$ degrees of freedom
and their variances are:
\begin{equation}
\sigma_{\clest{l}}^2 \eg \frac{2\cl{l}^2}{2l+1}
\label{varclsph}
\end{equation}
}
Figure~\ref{basicclcor} shows an example of a power spectrum $\cl{l}$
and the variance of its estimator $\clest{l}$
obtain with the CMBFAST code~(\cite{Sel96} and~\cite{Zal00})
with the standard set of cosmological parameters.

\subsection{Angular correlation function estimator.}\label{basicksi}

\begin{figure}[hhh]
\begin{center}
\includegraphics[width=150mm]{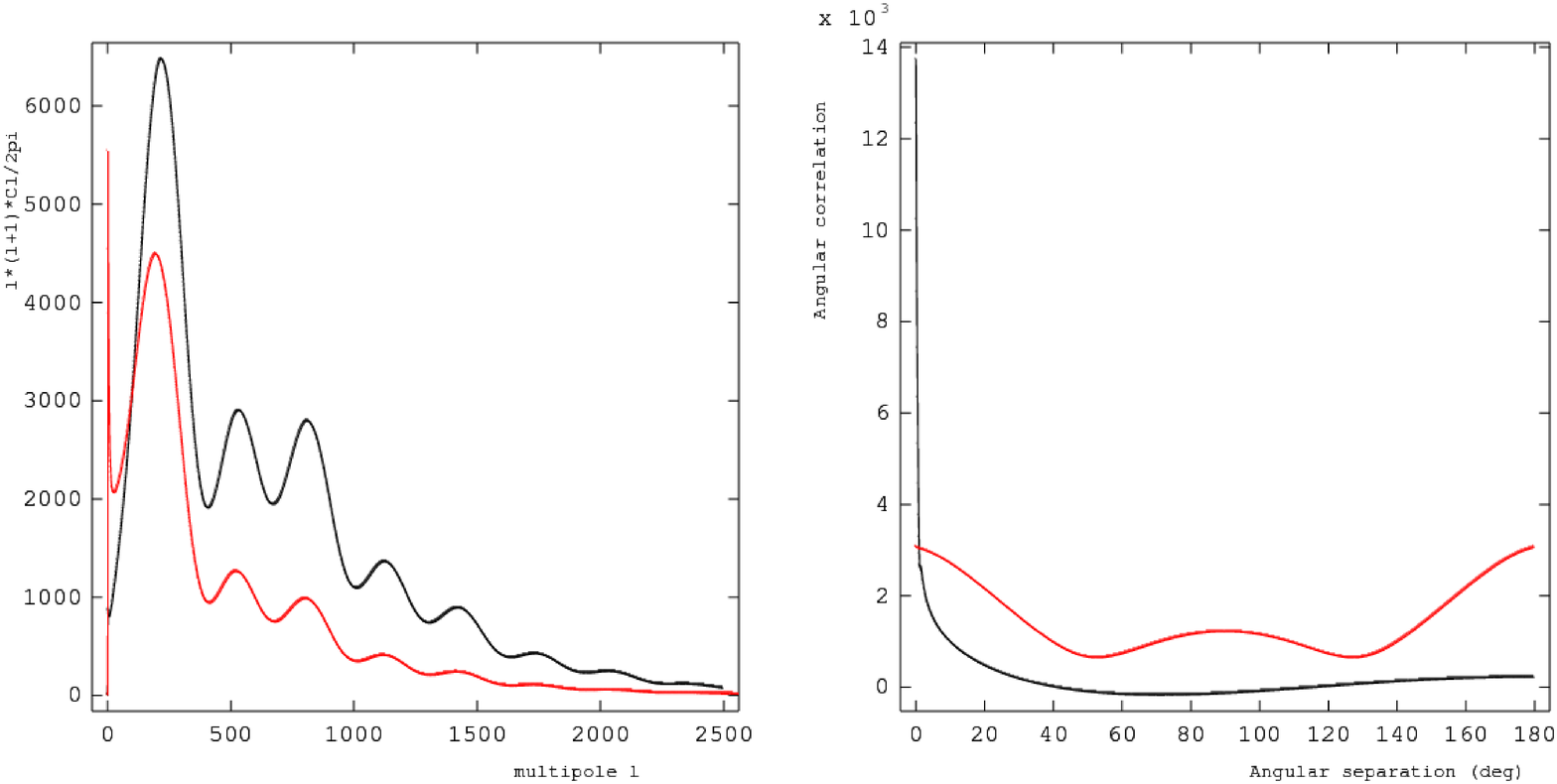}
\includegraphics[width=80mm]{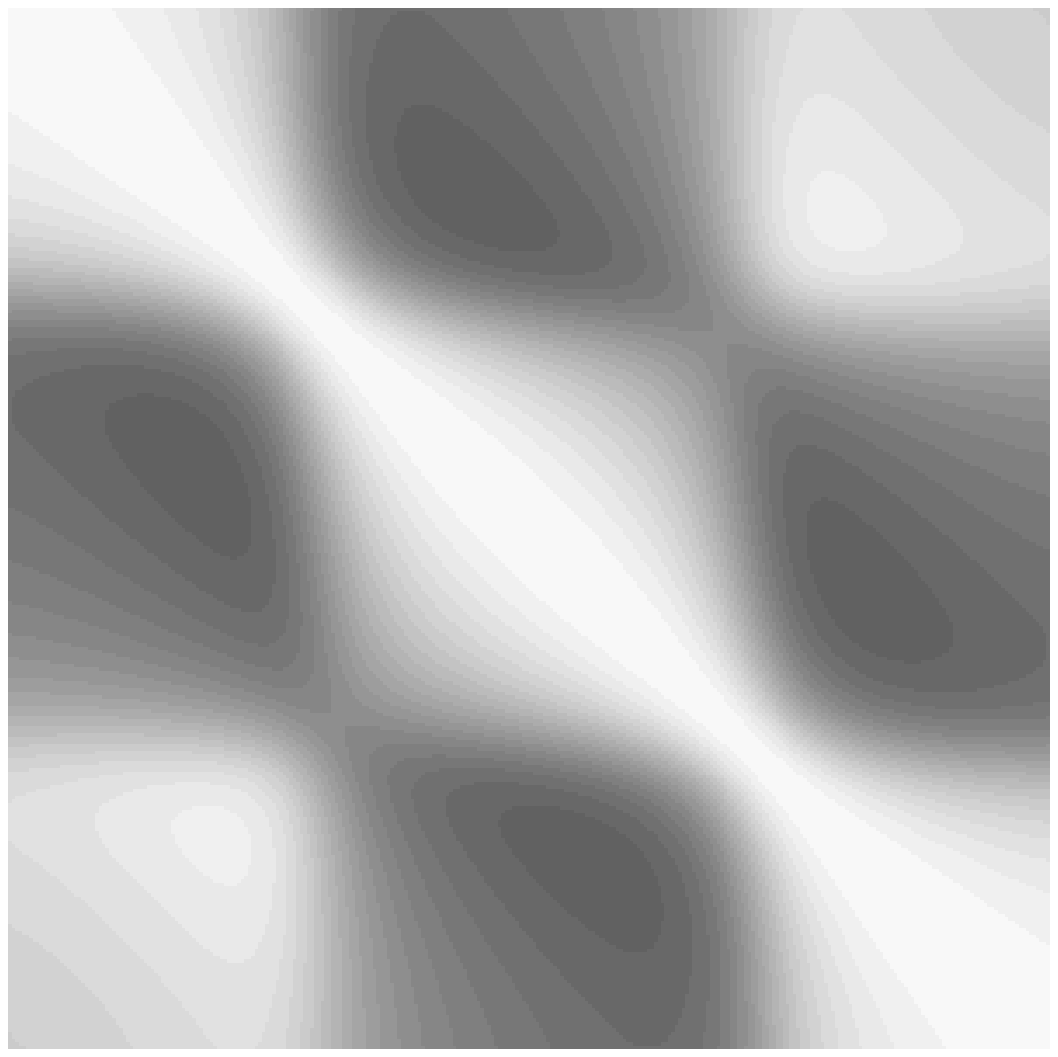}
\end{center}
\caption[Power spectra and angular correlation function]
  {
\ \newline
\Main {\it Top left:} power spectra versus the multipole moment \\
\Main {\it Top right:} angular correlation function versus the separation angle (deg) \\
The red curves shows the dispersion of the estimator due to cosmic variance
(values have been mutiply by $10$ for readability)
\ \newline
\Main {\it Bottom:} the correlation matrix of the angular correlation function estimator. \\
The upper left corner is $(0,0)$ degree and lower right corner is $(180,180)$ degres. \\
The LUT ranges from $-1$ (dark) to $1$ (white).
  }
\label{basicclcor}
\end{figure}

In order to obtain an estimator $\fcorest$ of the angular correlation function,
we can average over all pairs of directions $\;\V{\Omega},\V{\Omega'}\;$
of the observed sky (keeping the angle $\gamma$ between them fixed):
$$
\fcorest(\gamma) \eg
\frac{1}{\mathcal{N}(\gamma)}
\;\integ{S^2\x S^2}{}
\;\conj{T}(\Omega_1)T(\Omega_2)\;d\Omega_1 d\Omega_2
\;\delta(\V{\Omega_1}.\V{\Omega_2}-\cos(\gamma))
$$
with
$$
\mathcal{N}(\gamma) \eg
\integ{S^2\x S^2}{}\;d\Omega_1 d\Omega_2
\;\delta(\V{\Omega_1}.\V{\Omega_2}-\cos(\gamma))
$$
As $\;T(\Omega)\eg \somme{l,m}{}\;\alm{l}{m}\;\ylm{l}{m}(\Omega)\;$, we get:
$$
\fcorest(\gamma)
\eg
\frac{1}{\mathcal{N}(\gamma)}
\somme{l_1,m_1}{}\somme{l_2,m_2}{}
\;\conj{\alm{l_1}{m_1}}\;\alm{l_2}{m_2}
\;\integ{S^2\x S^2}{}
\;\conj{\ylm{l_1}{m_1}}(\Omega_1)\ylm{l_2}{m_2}(\Omega_2)\;d\Omega_1 d\Omega_2
\;\delta(\V{\Omega_1}.\V{\Omega_2}-\cos(\gamma))
$$
The internal integral $\mathcal{I}$ is computed in appendix~\ref{appyyd}
and leads to the estimator value:
$$ \begin{array}{rcl}
\fcorest(\gamma)
 &=&
\frac{1}{\mathcal{N}(\gamma)}
\somme{l_1,m_1}{}\somme{l_2,m_2}{}
\;\conj{\alm{l_1}{m_1}}\;\alm{l_2}{m_2}
\;2\pi\;\delta_{l_1l_2}\;\delta_{m_1m_2}\;\pl{l_2}(\cos(\gamma)) \\
 &=&
\frac{1}{\mathcal{N}(\gamma)}
\somme{l,m}{}
\;\conj{\alm{l}{m}}\;\alm{l}{m}
\;2\pi\;\pl{l}(\cos(\gamma))
\end{array} $$
Averaging over realisations gives us:
$$ \begin{array}{rcl}
< \fcorest(\gamma) >_{ens}
 &=&
\frac{1}{\mathcal{N}(\gamma)}
\somme{l,m}{}
\;< \conj{\alm{l}{m}}\;\alm{l}{m} >_{ens}
\;2\pi\;\pl{l}(\cos(\gamma)) \\
 &=&
\frac{1}{\mathcal{N}(\gamma)}
\;2\pi\;\somme{l}{}\;(2l+1)\cl{l}\;\pl{l}(\cos(\gamma))
\end{array} $$
Now $\mathcal{N}(\gamma)$ has to be computed.
$\mathcal{N}(\gamma)$ is equal to the previously computed
integral with $T(\Omega)=1$.
So, replacing 
$T(\Omega)$ by $\sqrt{4\pi}\ylm{0}{0}(\Omega)$,
or $\alm{l}{m}=\sqrt{4\pi}\delta_{l0}\delta_{m0}$
and $\cl{l}=4\pi\delta_{l0}$.
As $\pl{0}(\cos(\gamma))=1$,
one finally obtains $\mathcal{N}(\gamma)\eg \mathcal{N}\eg 8\pi^2$
and:
\encadre{
\begin{equation}
< \fcorest(\gamma) >_{ens}
\eg
\fcor(\gamma)
\eg
\frac{1}{4\pi}\;\somme{l}{}\;(2l+1)\cl{l}\;\pl{l}(\cos(\gamma))
\label{ksisph}
\end{equation}
\begin{center}
The estimator $\fcorest(\gamma)$ is unbiased.
\end{center}
}
Figure~\ref{basicclcor} shows the angular correlation function $\fcor(\gamma)$
and the variance of its estimator $\fcorest(\gamma)$.

\ \newline
Defining $K=\left(\frac{2\pi}{\mathcal{N}}\right)^2$,
the covariance of the estimator is:
$$ \begin{array}{l}
\frac{1}{K}\;\left(
<\fcorest(\gamma)\fcorest(\gamma')>_{ens}
  \;-\;
<\fcorest(\gamma)>_{ens}<\fcorest(\gamma')>_{ens}
\right) \\
\qquad\qquad \eg
\somme{lm}{}\somme{l'm'}{}\;\pl{l}(\gamma)\;\pl{l'}(\gamma')
<\alm{l}{m}\conj{\alm{l}{m}}\alm{l'}{m'}\conj{\alm{l'}{m'}}>_{ens} \\
\qquad\qquad\qquad\qquad \;-\;
\somme{lm}{}\somme{l'm'}{}\;\pl{l}(\gamma)\;\pl{l'}(\gamma')
<\alm{l}{m}\conj{\alm{l}{m}}>_{ens}<\alm{l'}{m'}\conj{\alm{l'}{m'}}>_{ens}
\end{array} $$
As the temperature field is real, and provided that the $\alm{l}{m}$
are gaussian random variables, one has:
$$ \begin{array}{l}
\frac{1}{K}\;\left(
<\fcorest(\gamma)\fcorest(\gamma')>_{ens}
  \;-\;
<\fcorest(\gamma)>_{ens}<\fcorest(\gamma')>_{ens}
\right) \\
\quad \eg
\somme{lm}{}\somme{l'm'}{}\;\pl{l}(\gamma)\;\pl{l'}(\gamma')
       \cl{l}\cl{l'}(1+\delta_{ll'}\delta_{mm'})
\;+\;
\somme{lm}{}\somme{l'm'}{}\;\pl{l}(\gamma)\;\pl{l'}(\gamma')
       \cl{l}^2\delta_{ll'}\delta_{m-m'} \\
\qquad \qquad \qquad \qquad \;-\;
\somme{lm}{}\somme{l'm'}{}\;\pl{l}(\gamma)\;\pl{l'}(\gamma')\cl{l}\cl{l'} \\
\quad \eg
2\;\somme{l}{}\;\cl{l}^2\;\pl{l}(\gamma)\;\pl{l}(\gamma')
\end{array} $$
Finally, by replacing $K$ by its value, we obtain:
\encadre{
\begin{equation}
<\fcorest(\gamma)\fcorest(\gamma')>_{ens}
  -
<\fcorest(\gamma)>_{ens}<\fcorest(\gamma')>_{ens}
 \eg
\frac{2}{(4\pi)^2}\somme{l}{}(2l+1)\cl{l}^2
\;\pl{l}(\gamma)\pl{l}(\gamma')
\label{covksisph}
\end{equation}
\begin{center}
The estimators of the angular correlation function values \\
in different directions are correlated.
\end{center}
}
Figure~\ref{basicclcor} shows the correlation matrix of the
angular correlation estimator $\fcorest(\gamma)$.

\subsection{What about computing the $\cl{l}$ on a portion \\ of sphere?}\label{basicpart}

There are two ways to compute the $\cl{l}$ coefficients.
Given a temperature field on the sphere, one can compute the $\alm{l}{m}$
with~(\ref{almfrt}) and then the $\cl{l}$ using the unbiased
estimator~(\ref{clest4p}).
Equation ~(\ref{almfrt}) is:
$
\alm{l}{m}\sim \integ{}{}\;\plm{l}{m}(\cos(\theta))d\cos(\theta)
\;\integ{}{}\;T\teafi e^{-im\phi}d\phi
$. \\
The inner integral is a Fourier transform which can be fastly 
computed with the Fast Fourier Transform (FFT) technics~(see \cite{Muc97}).
Note that this is possible only if the pixels lie on 
isolatitude lines.
The other way is to compute the angular correlation function $\fcorest(\gamma)$,
which is unbiased, and then use the second formula
in~(\ref{cltoksi}).

If the observed map is not the full sphere we can adapt these methods.
For the first one we can still compute pseudo $\alm{l}{m}$ by setting the
temperature to zero outside the observation zone.
This leads to a systematic bias that will be studied
in section~\ref{sec_clpartial}.
This also introduces correlations among the measured $\cl{l}$ and
increases their variances. Numerical calculations can be performed in the
case where the map is a spherical cap.
This is explored in section~\ref{sec_clpartial}
and its last section presents detailed comparisons with simulations.
This will show that the variance does not depend very much on the
map shape. The evolution of the variance with the map size is 
interpreted in this section.

In section~\ref{sec_ksipartial} we shall discuss the use of the angular
correlation function $\fcorest(\gamma)$ measured on the observed map
(so the angular separation $\gamma$ is limited to $\gamma_{lim}$).
We will show that this quantity is still
unbiased, as one intuitively guesses. One can still use
the formula~(\ref{cltoksi}) 
to extract the $\cl{l}$ but, since the correlation function
is defined only on some interval $[0,\gamma_{lim}]$,
the Legendre polynomials are no longer orthogonal. This introduces
wild fluctuations in the $\cl{l}$.
This is cured by "apodizing" the angular correlation function
and will be described in section~\ref{sec_clfrksi}.

Four appendices explain technical details to make these notes 
self contained.

\clearpage

\clearpage
\section{Spectrum estimation for a portion of sphere.}\label{sec_clpartial}

\subsection{Estimator of the $\cl{l}$ spectrum.}\label{clpartdef}

The temperature field is measured on a piece $A$ of the sphere
of solid angle $\Omega_{obs}$.
We will study the following estimator for the $\cl{l}$:
\begin{equation}
\cpl{l} \eg \frac{1}{2l+1}\;\somme{m=-l}{l} \abs{\aplm{l}{m}}^2
\label{tclfrtalm}
\end{equation}
where the $\aplm{l}{m}$ are computed on $\Omega_{obs}$:
\begin{equation}
\aplm{l}{m}
\eg
\integ{\Omega_{obs}}{} T(\Omega)\x \conj{\ylm{l}{m}}(\Omega)\;d\Omega
\label{talmfrt}
\end{equation}
The $\cpl{l}$ will therefore represent an estimation of the spectrum
of a temperature map which is set to zero outside the observed region.
We define the function $W(\V{\Omega})$ on the sphere
such that $W=1$ on $A$ and $0$ anywhere else. \\
The temperature field under study is:
$T'(\Omega)\eg (T\x W)(\Omega)$.

\subsection{Computation of the bias of the estimator.}\label{clpartbias}

The function $W$ can be  expanded on spherical harmonics basis as:
\begin{eqnarray}
W(\Omega) &=& \somme{l=0}{\infty}\;\somme{m=-l}{l} \blm{l}{m}\ylm{l}{m}(\Omega) \\
 && \notag \\
\blm{l}{m} &=& \integ{S^2}{}\;W(\Omega)\;\conj{\ylm{l}{m}}(\Omega)\;d\Omega
          \eg \integ{\Omega_{obs}}{}\;\conj{\ylm{l}{m}}(\Omega)\;d\Omega
\end{eqnarray}
Then the $\aplm{l}{m}$ are:
$$
\aplm{l}{m} 
\eg
  \integ{S^2}{} T(\Omega) W(\Omega)\;\conj{\ylm{l}{m}(\Omega)}\;d\Omega
\eg
  \somme{l',m',l'',m''}{ }\;\alm{l'}{m'}\blm{l''}{m''}
  \integ{S^2}{} \ylm{l'}{m'} \ylm{l''}{m''} \conj{\ylm{l}{m}}\;d\Omega
$$
Using the property $\;\conj{\ylm{l}{m}}=(-1)^m \ylm{l}{-m}$:
$$
\aplm{l}{m} \eg
  \somme{l',m',l'',m''}{ }\;\alm{l'}{m'}\blm{l''}{m''} (-1)^m
  \integ{S^2}{} \ylm{l'}{m'} \ylm{l}{-m} \ylm{l''}{m''} d\Omega
$$
The integral over the sphere of the product of three spherical harmonics is
(\cite{Brink62} and \cite{Mess64}):
$$
\integ{S^2}{} \ylm{a}{\alpha} \ylm{b}{\beta} \ylm{c}{\gamma}
       \; \sin(\theta) d\theta d\phi
  \eg
\sqrt{\frac{(2a+1)(2b+1)(2c+1)}{4\pi}}
  \wigj{a}{b}{c}{\alpha}{\beta}{\gamma}\wigj{a}{b}{c}{0}{0}{0}
$$
where $\;\wigj{a}{b}{c}{\alpha}{\beta}{\gamma}\;$ are the Wigner $3j$ symbols
related to the Clebsch-Gordan coefficients by:
$$
\clg{ab\alpha\beta}{c-\gamma}
\eg
(-1)^{a-b-\gamma}\sqrt{2c+1} \wigj{a}{b}{c}{\alpha}{\beta}{\gamma}
$$
Recall that these coefficients are real.
$$ \begin{array}{lcl}
\aplm{l}{m} &=& 
  \somme{l',m',l'',m''}{ }\;\alm{l'}{m'}\blm{l''}{m''} (-1)^m
  \sqrt{\frac{(2l+1)(2l'+1)(2l''+1)}{4\pi}}
    \wigj{l'}{l}{l''}{m'}{-m}{m''}\wigj{l'}{l}{l''}{0}{0}{0} \\
\conj{\aplm{l}{m}} &=& 
  \somme{L',M',L'',M''}{ }\;\conj{\alm{L'}{M'}}\conj{\blm{L''}{M''}} (-1)^m
  \sqrt{\frac{(2l+1)(2L'+1)(2L''+1)}{4\pi}}
    \wigj{L'}{l}{L''}{M'}{-m}{M''}\wigj{L'}{l}{L''}{0}{0}{0} \\
\end{array} $$
\begin{eqnarray*}
\abs{\aplm{l}{m}}^2
  &=& \aplm{l}{m} \conj{\aplm{l}{m}} \\
  &=&
  \somme{}{}\;\;
  \left( \alm{l'}{m'}\conj{\alm{L'}{M'}} \right)
  \left( \blm{l''}{m''}\conj{\blm{L''}{M''}} \right)
  \x \frac{2l+1}{4\pi} \sqrt{(2l'+1)(2l''+1)(2L'+1)(2L''+1)} \\
  & &
  \qquad\qquad\qquad
  \x \wigj{l'}{l}{l''}{m'}{-m}{m''}\wigj{L'}{l}{L''}{M'}{-m}{M''}
  \wigj{l'}{l}{l''}{0}{0}{0}\wigj{L'}{l}{L''}{0}{0}{0}
\end{eqnarray*}
where the sum runs on $l',m', l'',m'', L',M', L'',M''$.

The $\alm{l}{m}$ are random variables. We want to compute the average effect of
finite size maps, therefore we shall perform the average over sky realisations
(ensemble average\footnote{
In the following, we will drop the subscript for ensemble average
notations: $\;<\cdots>_{ens}\;\rightarrow\;<\cdots>$
}).
Recalling equation~(\ref{malmalm}), one obtains:
\begin{eqnarray*}
<\cpl{l}>
  &=&
  \frac{1}{4\pi}
  \somme{m=-l}{l}\;\somme{ l',m', l'',m'', L'',M'' }{ }
  \cl{l'} \blm{l''}{m''}\conj{\blm{L''}{M''}}
  (2l'+1)\sqrt{(2l''+1)(2L''+1)} \\
  & &
  \qquad\qquad\qquad
  \x \wigj{l'}{l}{l''}{m'}{-m}{m''}\wigj{l'}{l}{L''}{m'}{-m}{M''}
  \wigj{l'}{l}{l''}{0}{0}{0}\wigj{l'}{l}{L''}{0}{0}{0}
\end{eqnarray*}
Now, remark that the coefficients of the Wigner $3j$ 
symbol product do not depend on $m$ nor $m'$, therefore one can use the 
orthogonality relations
$$
\somme{\alpha\beta}{ }
\sqrt{(2c+1)(2c'+1)}
\wigj{a}{b}{c}{\alpha}{\beta}{\gamma}
\wigj{a}{b}{c'}{\alpha}{\beta}{\gamma'}
\eg \delta_{cc'}\delta_{\gamma\gamma'}
$$
Defining:
\begin{equation}
\bl{l''}\eg \frac{1}{2l''+1}\somme{m''=-l''}{l''}\blm{l''}{m''}\conj{\blm{l''}{m''}}
  \label{blmask}
\end{equation}
we obtain:
\encadre{
\begin{eqnarray}
<\cpl{l}>
  &=&
  \frac{1}{4\pi}\somme{l',l''}{ }
  \cl{l'} \bl{l''}
  (2l'+1) (2l''+1)
  \wigj{l'}{l}{l''}{0}{0}{0}^2
\label{cltilde}
\end{eqnarray}
}
The expression of the $3j$ symbols can be found for instance
in~\cite{Brink62}.
We found recently that this calculation has already been published in~\cite{Hiv02}. \\
The above equation provides the relation between the true 
$\cl{l}$ and the average $<\cpl{l}>$ over many sky realisations
of the  $\cpl{l}$  coefficients measured over a patch $A$
of any shape of the sphere.
This can be written in matrix form:
\begin{eqnarray}
\left[ <\cpl{l}> \right]
\eg
\left[ M \right]_{ll'}
\x
\left[ \cl{l'} \right]
\; \mbox{with} \;
M_{ll'}
\eg
\frac{2l'+1}{4\pi}
\;\somme{l''=0}{\infty}
(2l''+1)\bl{l''}
\wigj{l'}{l}{l''}{0}{0}{0}^2
\label{matmll}
\end{eqnarray}
This result is valid for any weighting function $W$ on the sphere.
The relation between the $\cpl{l}$ and the $\cl{l}$ is linear.
The matrix $M_{ll'}$ depends only on the map shape and not
on the initial $\cl{l'}$ spectrum.
Permutations of $2$ columns of the $3j$ symbols change their values
by a phase (\cite{Mess64}) so $\;(2l+1) M_{ll'}= (2l'+1) M_{l'l}$.
The $M_{ll'}$ elements are positive or null.

When the full sky is observed, $\;\bl{l''}=\bl{0}\delta_{l''0}\;$
and the $3j$ symbols imply $l=l'$ whence $M_{ll'}$ is the identity matrix. \\
On the other hand, when observing a portion of the sky,
each measured $\cpl{l}$ is a mixture of the $\cl{l}$ whose
weights are given by the matrix $M_{ll'}$.

The properties of the matrix are illustrated
in figure~\ref{matrice_Mllp}.
The bottom graphs show line $l=1500$ of matrix $M_{ll'}$
for a $10\x 10 \;deg^2$ map and lines $l=500,2400$
where the peaks have been shifted to $l'=1500$ for easy comparison:
the width of the peak is nearly constant with $l$,
that is to say that the matrix is nearly band-diagonal. \\
The top graphs show the $M_{ll'}$ element values of line $l=1500$
for maps of various size and shapes.
As the map size decreases, the peak width gets larger
and a given $\cpl{l}$ measurement involves a larger band in $\cl{l'}$.
The width of the peak is inversely proportionnal to the caracteristic
width of the map and it does not depend very much on the shape
provided it is not excessively elongated.
This is somewhat equivalent to what we have in Fourier analysis
on small intervals. \\
The $\cl{l}$ are coefficients in the correlation function expansion~(\ref{cltoksi}). 
The $\pl{l}$ are orthogonal on $[0,\pi]$ and form a basis on this interval. 
On a partial map of characteristic size $\theta_c$ 
these functions are "nearly orthogonal" if they differ
by more than $\delta l \simeq \pi / \theta_c$:
loosely speaking if they have not the same number of roots in that interval.
Therefore, for a given correlation
function, the $\cpl{l}$ will be mixtures of the "true" $\cl{l}$ over a 
range $\simeq \delta l$. \\

If $\Omega_{obs}$ is the solid angle corresponding to the observed patch
of the sky, using the completude relation for the spherical harmonics,
we have for the $\bl{l}$ normalisation:
\begin{eqnarray*}
\somme{l=0}{\infty} (2l+1) \bl{l}
  \eg
  \somme{l=0}{\infty}\;\somme{m=-l}{l} \blm{l}{m}\conj{\blm{l}{m}}
  \eg
  \integ{S^2}{}\;\abs{W(\V{\Omega})}^2\;d\V{\Omega}
\end{eqnarray*}
Until the end of this section, we assume that $W=1$ on $A$
and zero elsewhere.
Thus we have:
\begin{eqnarray}
\somme{l=0}{\infty} (2l+1) \bl{l} \eg \Omega_{obs}
\label{blnorm}
\end{eqnarray}
The $\cpl{l}$ are biased relative to the $\cl{l}$
since one observes only a fraction of the sphere, the rest being $0$.
Assuming that the function $T$ is isotropic, to compare the $\cpl{l}$
with $\cl{l}$ one has to take into account the sky coverage normalisation:
\begin{equation}
<\cplr{l}>
\eg  \frac{4\pi}{\Omega_{obs}}\x <\cpl{l}>
\eg \somme{l'}{}\;M_{ll'}^{R} \x \cl{l'}
\quad\mbox{where}\quad
 M_{ll'}^{R} \eg \frac{4\pi}{\Omega_{obs}}M_{ll'}
\label{cltnorm}
\end{equation}
{\it In the following, when talking about comparison between $\cpl{l}$
and $\cl{l}$, we will assume that the $\cpl{l}$ have been renormalised
as describe above.} \\

Using the $3j$ orthogonality relation (\cite{Mess64}, appendix C relation $15b$)
which gives
$\quad
\somme{l'}{}(2l'+1)\wigj{l'}{l}{l''}{0}{0}{0}^2 = 1
\quad$
and equation~(\ref{blnorm}) we get
$\;\;\somme{l'}{}\;M_{ll'}^R=1$. \\
Therefore for a given $<\cplr{l}>$,
the sum of the weights of the $\cl{l'}$ is equal to one. \\

\begin{figure}
   \centering
\begin{center}
\includegraphics[width=100mm]{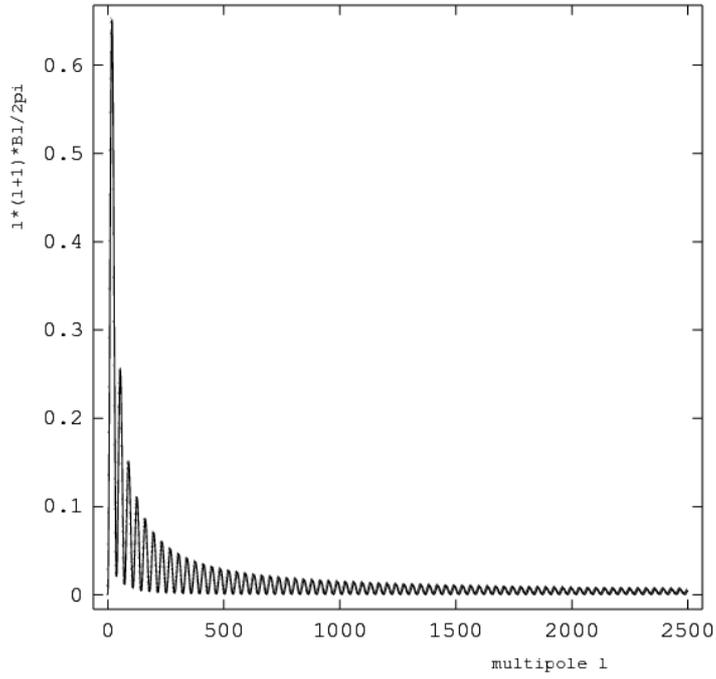}
\includegraphics[width=100mm]{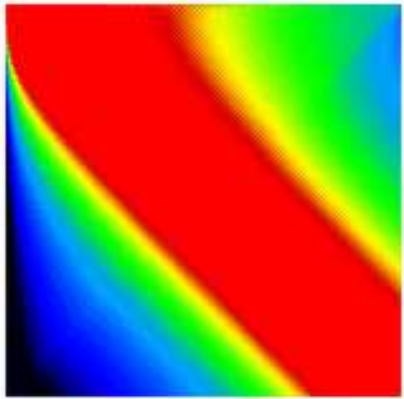}
\end{center}
\caption[Power spectra estimation for a $10\x10\;deg^2$ square map]
       {
The top picture shows the $\bl{l}$ spectrum for a $10\x 10\; deg^2$
square map. \\
The bottom picture shows the matrix $M_{ll'}$ for the same map
(The LUT as been optimised for readability purpose).
Axes range from upper corner left $(l=0,l'=0)$
to bottom right corner $(l=2500,l'=2500)$.
         }
   \label{blspec}
\end{figure}

The $\bl{l}$ can be obtain using usual computer technics (see \cite{Gors05}).
Then it is easy and fast to compute numerically the matrix $M_{ll'}$
and check the systematic effect of finite size map. \\
The figure~\ref{blspec} shows the $\bl{l}$ spectrum and the $M_{ll'}$ matrice
for a $10\x 10\;deg^2$ square map.

\subsection{Computation of the covariance of the $\cpl{l}$ \\ estimator for spherical cap maps.} \label{secclcov}

We have tried to compute the variance  of the $\cpl{l}$
estimator following the same line of calculation.
Although the result does not look like too complicated,
it is of little use, because it would cost
a lot of computer time to be numerically calculated in the general case.
However it can be calculated in the simple case of a spherical cap domain.

If, in equation~(\ref{talmfrt}),
we replace the temperature field by its expression~(\ref{tfralm}), we get:
$$
\aplm{l}{m}
\eg
\somme{l_1m_1}{}\;\alm{l_1}{m_1}\;B_{l_1m_1lm}
$$
where we have set:
$$
B_{l_1m_1lm}
\eg
\integ{S^2}{}\;
W(\V{\Omega})\;\ylm{l_1}{m_1}(\V{\Omega})\;\conj{\ylm{l}{m}}(\V{\Omega})\;d\V{\Omega}
$$
Averaging over sky realisations and using equation~(\ref{malmalm}), one gets
(see details in appendix~\ref{appclblmlm}):
\begin{equation*}
<\cpl{l}>
 \eg
\frac{1}{2l+1}\;\somme{ml'm'}{}\cl{l'}\abs{B_{lml'm'}}^2
\qquad\mbox{for}\quad
-l\le m \le +l \;,\; 0\le l' < \infty   \;,\; -l'\le m' \le +l'
\end{equation*}
and provided that the $\alm{l}{m}$ are {\it gaussian} random fields:
\begin{equation*}
<\cpl{l}\cpl{L}>-<\cpl{l}><\cpl{L}>
 \eg
\frac{2}{(2l+1)(2L+1)}
\somme{m=-l}{+l}\;\somme{M=-L}{+L}
\abs{
\somme{l'}{}\cl{l'}\somme{m'=-l'}{+l'}B_{l'm'lm}\conj{B_{l'm'LM}}
}^2
\end{equation*}
The last two results are valid for any weighting function $W$ on the sphere.
$W$ could include an additionnal weighting,
for instance to treat border effects.
The results of the former section can also be adapted to that case.
Until the end of this section, we will assume that $W=1$ on $A$ and zero elsewhere.

\begin{figure}
   \centering
   \includegraphics[width=150mm]{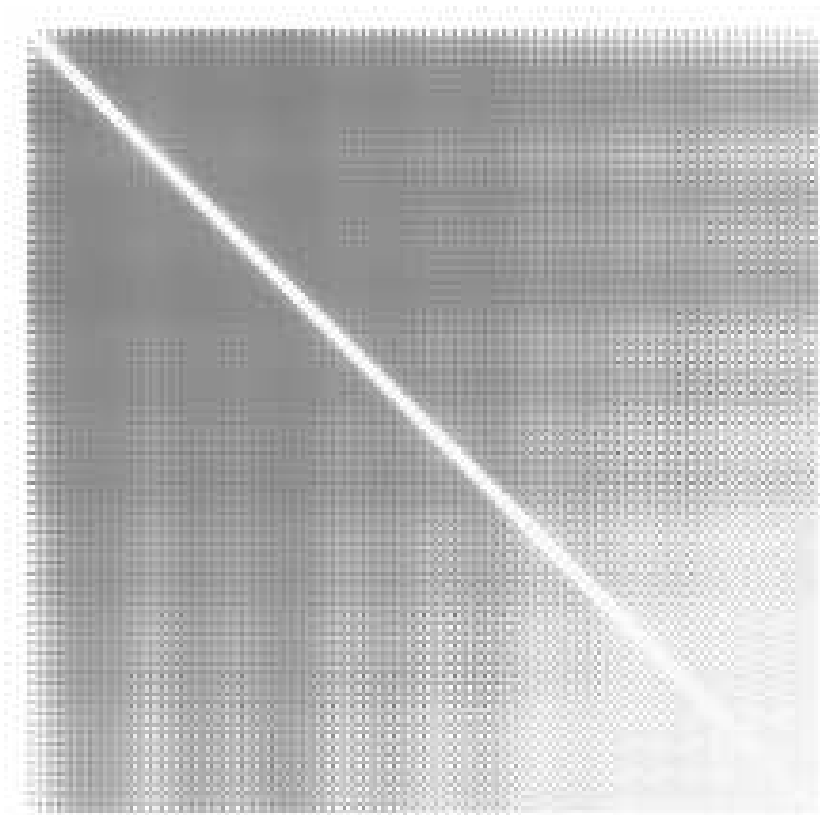}
   \caption[correlation matrix for the $\cpl{l}$ a $10\x10\;deg^2$ spherical cap]
         {
Correlation matrix for the $\cpl{l}$ a $10\x10\;deg^2$ spherical cap. \\
The top left corner is for $(l=0,l'=0)$
and the bottom left corner is for $(l=2500,l'=2500)$.
The LUT ranges from $-1$ (dark) to $+1$ (white).
         }
   \label{clcov}
\end{figure}

This equation simplifies a lot if one considers a 
spherical cap centered at the north pole with border at polar
angle $\theta_c$.
Using $\ylm{l}{m}\teafi\eg \hplm{l}{m}(\cos(\theta))\;e^{im\phi}\;$
(where $\hplm{l}{m}$ are the normalised associated Legendre polynomials),
we get in this case:
\begin{center}
$
B_{lmLM} \eg 2\pi\;\delta_{mM}\;B_{lLm}
\qquad$
where\footnote{
We have $\;B_{ll'm}\eg B_{l'lm}\eg B_{ll'-m}\in\EnsR\;$, 
$\;B_{000}\eg \frac{1}{4\pi}\;(1-\cos(\theta_c))\;$,
$B_{ll'm}=0$ for $\abs{m}> \min(l,l')$. \\
From the relation
$\;
\somme{m}{}\conj{\ylm{l}{m}}(\V{\Omega_1})\ylm{l}{m}(\V{\Omega_2})
=\frac{2l+1}{4\pi}\pl{l}(\V{\Omega_1}.\V{\Omega_2})
\;$ we get
$\;\somme{m}{}\;B_{llm} \eg \frac{2l+1}{8\pi^2}\;\Omega_{obs}$
}
$\qquad
B_{lLm}
\eg
\;\integ{0}{\theta_c}
\;\hplm{l}{m}(\cos(\theta))\;\hplm{L}{m}(\cos(\theta))\;d\cos(\theta)
$
\end{center}
Replacing in the general expression for the covariance, one obtains:
\begin{eqnarray}
<\cpl{l}>
  \eg
\frac{(2\pi)^2}{2l+1}
\;\somme{m=-l}{l}
\;\;\somme{l'\ge \abs{m}}{}\;\cl{l'}\;(B_{ll'm})^2
\label{cplbllm}
\end{eqnarray}
and for the covariance:
\begin{equation}
<\cpl{l}\cpl{L}>-<\cpl{l}><\cpl{L}>
 \eg
\frac{2\;(2\pi)^4}{(2l+1)(2L+1)}
\;\somme{m}{}
\;\left(
\somme{l'\ge \abs{m}}{}
\;\cl{l'}\;B_{ll'm}B_{Ll'm}
\;\right)^2
\label{eqcov}
\end{equation}
The $B_{ll'm}$ coefficients could be computed by numerical integration
but this would be very slow at high $l$.
The calculations were done using recurrence relations. \\
The figure~\ref{clcov} shows the correlation matrix computed from
the covariance formula above for a $10\x 10\;deg^2$ spherical cap. \\
\ \newline \noindent {\bf \underline{Recurrence relations}} \\
In order to derive the following recurrence relations for the $B_{ll'm}$
coefficients, we have used the usual recurrence relations
among the normalised associated Legendre functions
(\cite{Grad80})
which imply integrals of the form\footnote{
We set $\;c=\cos(\theta_c)\;$ and $\;x=\cos(\theta)$
}
$\;\integ{1}{c}x\hplm{l}{m}(x)\hplm{k}{m}(x)dx$.
The latter integration can be in turn related to integrals
of the type
$\;\integ{1}{c}\hplm{l}{m}(x)\hplm{k}{m}(x)dx\;$
by evaluating
$\;
\integ{1}{c}(1-x^2)(\hplm{l}{m}{'}\hplm{k}{m}{''}
+\hplm{l}{m}{''}\hplm{k}{m}{'})dx
\;$
and using the Legendre differential equation
(where $'$ and $''$ means first and second order derivations with respect to $x$).
One gets:
$$\begin{array}{rcl}
B_{l,l+1,m}
  &=&
  -c\hplm{l}{m}(c)\hplm{l+1}{m}(c)
  \;+\;
  \frac{\sqrt{(l+1-m)(l+1+m)}}{2(l+1)}
  \left(
  \sqrt{\frac{2l+3}{2l+1}}\;[\hplm{l}{m}(c)]^2
  \;+\;
  \sqrt{\frac{2l+1}{2l+3}}\;[\hplm{l+1}{m}(c)]^2
  \right)
\end{array}$$
And for $\;\abs{k-l}\ge 2$, setting $\;b=(l+1)(k-l)(k-m)$, one has:
\begin{eqnarray}
b\x B_{lkm}
 &=&
  k\sqrt{\frac{2k+1}{2k-1}}\sqrt{\frac{2l+1}{2l+3}}
  \sqrt{\frac{k-m}{k+m}}\sqrt{\frac{l+1-m}{l+1+m}}
  (k-l-2)(l+1+m)\;B_{l+1,k-1,m} \nonumber \\
 &&
  -\;
  (k-m)\left(
    (l+1)c\hplm{l}{m}(c)
    -\sqrt{\frac{2l+1}{2l+3}}
      \sqrt{\frac{l+1+m}{l+1-m}}(l+1-m)\hplm{l+1}{m}(c)
  \right)\hplm{k}{m}(c) \nonumber \\
 &&
  -\;
  \sqrt{\frac{2k+1}{2k-1}}\sqrt{\frac{k-m}{k+m}}
  \;\hplm{k-1}{m}(c) \label{eqblk} \\
 &&
  \qquad\x \left(
    k\sqrt{\frac{2l+1}{2l+3}}\sqrt{\frac{l+1-m}{l+1+m}}(l+1+m)c\hplm{l+1}{m}(c)
    -(k(l+1)-m^2)\hplm{l}{m}(c)
    \right) \nonumber
\end{eqnarray}
This heavy recurrence relation has been verified to be numerically stable
up to $l\simeq 2500$. \\
The only numerical integral to perform is $\;B_{llm}$.
As $\hplm{l}{m}(x) \;\sim\; (1-x^2)^{\undemi m}\;Q(l-m)$
(where $Q(n)$ stands for a polynome of degree $n$),
the integrand of $B_{llm}$ is a polynome of degree $2l$
and it may be exactly computed using
the Gauss-Legendre integration method with $l+1$ points.
All other $B_{lkm}$ can be computed using the recurrence relations~(\ref{eqblk}). \\
With similar technics \cite{Wan01} derived different equivalent recurrence relations. \\
\ \newline \noindent {\bf \underline{Approximation for small maps}} \label{clbessel} \\
For small $\theta_c$ ($\theta_c\lesssim 0.15$),
one can make the approximation (\cite{Grad80}):
$$
\hplm{l}{m}\left(\cos(\theta)\right)
   \;\rightarrow\;
\frac{(-1)^m}{l^m}\;\sqrt{\frac{(2l+1)(l+m)!}{4\pi(l-m)!}}\;J_m(l\theta)
$$
where the $J_m$ are the cylindrical Bessel functions.
Therefore:
$$
B_{l_1lm}
 \;\simeq\;
\sqrt{
\frac{2l+1}{4\pi}\;\frac{2l_1+1}{4\pi}
\;\frac{(l+m)!}{(l-m)!}\frac{(l_1+m)!}{(l_1-m)!}
}
\;\;\frac{1}{l_1^ml^m}
\;\;\integ{0}{\theta_c}\theta\;J_m(l_1\theta)\;J_m(l\theta)\;d\theta
$$
where the last integral is a Lommel integral~(\cite{Grad80}). Then:
\begin{eqnarray}
B_{l_1lm}
   \simeq
\sqrt{
\frac{2l+1}{4\pi}\;\frac{2l_1+1}{4\pi}
\;\frac{(l+m)!}{(l-m)!}\frac{(l_1+m)!}{(l_1-m)!}
}
\;\;\frac{1}{l_1^ml^m}\;\;\frac{\theta_c}{l_1^2-l^2} \notag \\
\qquad\qquad \x \left\{
lJ_m(l_1\theta_c)J'_m(l\theta_c)
-l_1J_m(l\theta_c)J'_m(l_1\theta_c)
\right\}
\ \label{eqblkb}
\end{eqnarray}
where $J'_m(z)$ stands for the derivative of $J_m(z)$ relative to $z$. \\
For numerical purposes it is easier to replace
$
\left\{
lJ_m(l_1\theta_c)J'_m(l\theta_c)
-l_1J_m(l\theta_c)J'_m(l_1\theta_c)
\right\}
$
by
$
\left\{
l_1J_m(l\theta_c)J_{m+1}(l_1\theta_c)
-lJ_m(l_1\theta_c)J_{m+1}(l\theta_c)
\right\}
$.
Figure~\ref{figure3} compares the variance of the $\cplr{l}$ computed
with relation~(\ref{eqblk}) with the calculation done with the
above approximation:
the agreement is excellent for
$f_{sky}\eg\frac{\Omega_{obs}}{4\pi}\lesssim 6\;10^{-3}\;\;$
($\theta_c\lesssim 0.15\;rd$). \\
\ \newline \noindent {\bf \underline{Behaviour of the variance as a function of $\theta_c$}} \\
If $l\theta_c<m$, $\;J_m(l\theta_c)$ behaves as $(l\theta_c)^m$
which gives a negligible contribution if $m$
is large. On the other hand, if $m<l\theta_c$ one can use
the following approximation (\cite{Grad80}):
$$
J_m(l\theta_c)\;\simeq\;\;
\sqrt{\frac{2}{\pi}}\;\frac{1}{\sqrt{l\theta_c}}\;\cos(l\theta_c+\phi_m)
$$
where $\phi_m$ is a phase depending on $m$ and slowly varying with
$l\theta_c$:
$\;\phi_m\rightarrow -m\frac{\pi}{2}-\frac{\pi}{4}\;\;$
for $\;\;l\theta_c\rightarrow \infty$. \\
In that approximation, from equation~(\ref{eqblkb}) one obtains\footnote{
This equation and~(\ref{cplbllm})
explain qualitatively the behaviour of the matrix $M_{ll'}$.
}:
\begin{equation}
B_{l_1lm} \;\sim\; \frac{\sin((l_1-l)\theta'_c)}{l_1-l}
\qquad\theta'_c=\theta_c \;+\;\mbox{corrections depending on}\;m
\label{bllsinc}
\end{equation}
One can use this approximation in equation~(\ref{eqcov}) in which
we set $l=L$ for the variance calculation.

If only the central peak of $B_{ll_1m}^2$ as a function of $l_1$
is considered
($\abs{l_1-l}\lesssim \pi/\theta_c$)
the partial sum $\somme{l'}{}$ in equation~(\ref{eqcov})
behaves as $\theta_c$ because
$B_{ll_1m}^2$ is a function with a peak height proportionnal
to $\theta_c^2$ and width $\pi/\theta_c$.
A more systematic calculation shows that,
due to the high values of $\cl{l}$ at low $l$,
one can not neglect the contribution
of $B_{ll_1m}^2$ at the left of the peak
($l_1 < l-\pi/\theta_c$). \\
The sum over $m$ in equation~(\ref{eqcov}) introduces a factor $\sim 2l\theta_c$. \\
If one considers only the central peak
(and using the normalisation in formula~(\ref{cltnorm})), one gets
$\;
\sigma^2_{\cplr{l}}
  \sim
\frac{\sigma^2_{\cl{l}}(4\pi)}{\theta_c}
\;$,
so $\;\sigma_{\cplr{l}}\;$ would scale as $\;f_{sky}^{-\frac{1}{4}}$. \\
Numerically we found that a good approximation for small $\theta_c$
is:
\begin{equation}
\sigma_{\cplr{l}}
  \sim
\frac{\sigma_{\cl{l}}(4\pi)}{\sqrt{f_{sky}}\sqrt{\frac{\pi}{\theta_c}}}
\label{sigclpi}
\end{equation}
For large $l$ values, the peak wing corrections lead to
corrections of higher order in $1/\theta_c$:
\begin{equation}
\sigma^2_{\cplr{l}}
  \sim
\frac{\sigma^2_{\cl{l}}(4\pi)}{\theta_c}
\x(1+\frac{b(l)}{\theta_c}+\cdots)
\label{sigclcor}
\end{equation}.

\subsection{Discussion and comparison with simulations}

In the following the $\cpl{l}$ coefficients are computed
from simulated CMB maps
using the well known Fast Fourier Transform (FFT) technics (\cite{Muc97}).

Simulations have been realised using the HEALPIX package (\cite{Gors05}).
Pure CMB maps of various sizes and shapes have been simulated using 
the program SYNFAST with $l_{max}=2500$ for the standard $\Lambda CDM$ spectrum.
The temperature field outside the various observed regions is set to zero.

The pixel size was chosen small enough ($NSIDE=2048$) in
order not to bias the $\cl{l}$ reconstruction up to $l_{max}=2500$.
The pixel smoothing windows has been chosen to be that of
a $NSIDE=8192$.
So the value in a pixel is very nearly the value of the temperature
field at the center of the pixel.
No detector noise nor telescope lobe effect was added.

The $\cl{l}$ were reconstructed using the program
ANAFAST up to $l_{max}=2500$.
That is to say that ANAFAST computes our $\cpl{l}$ as defined
in equation~(\ref{tclfrtalm}).

Using full sphere simulations we checked that the reconstructed
$\clest{l}$ were in perfect agreement, even at large $l$, with the generated ones.
Table~(\ref{tablegen}) shows the generations used in the following discussion.
In the following $f_{sky}=\frac{\Omega_{obs}}{4\pi}$
is the fraction of the observed sky.

\begin{table}
  \centering                          
  \begin{tabular}{c c c}
  \hline \hline
  map size  &  $f_{sky}$  &  number of generations  \\
  \hline
  $4\pi\;sr$ & $1$  & $579$ \\
  \hline
  $20\x 32$  & $1.54\;10^{-2}$  & $228$ \\
  \hline
  $17\x 17$  & $0.697\;10^{-2}$  & $579$ \\
  $R=9.585$  &       -           & $2279$ \\
  \hline
  $10\x 10$  & $0.242\;10^{-2}$  & $228$ \\
  \hline
  $6\x 6$    & $0.872\;10^{-3}$  & $228$ \\
  $4\x 9$    &       -           & $228$ \\
  $R=3.385$  &       -           & $228$ \\
  \hline
  \end{tabular}
  \caption
  {
  Description of simulations:
  $\;R=\theta_c$ means a spherical cap with polar angle $\theta_c\;deg$,
  other maps are labeled $\Delta \theta\x \Delta \phi\;deg^2$.
  }
  \label{tablegen}
\end{table}

\begin{figure}
   \centering
   \includegraphics[width=150mm]{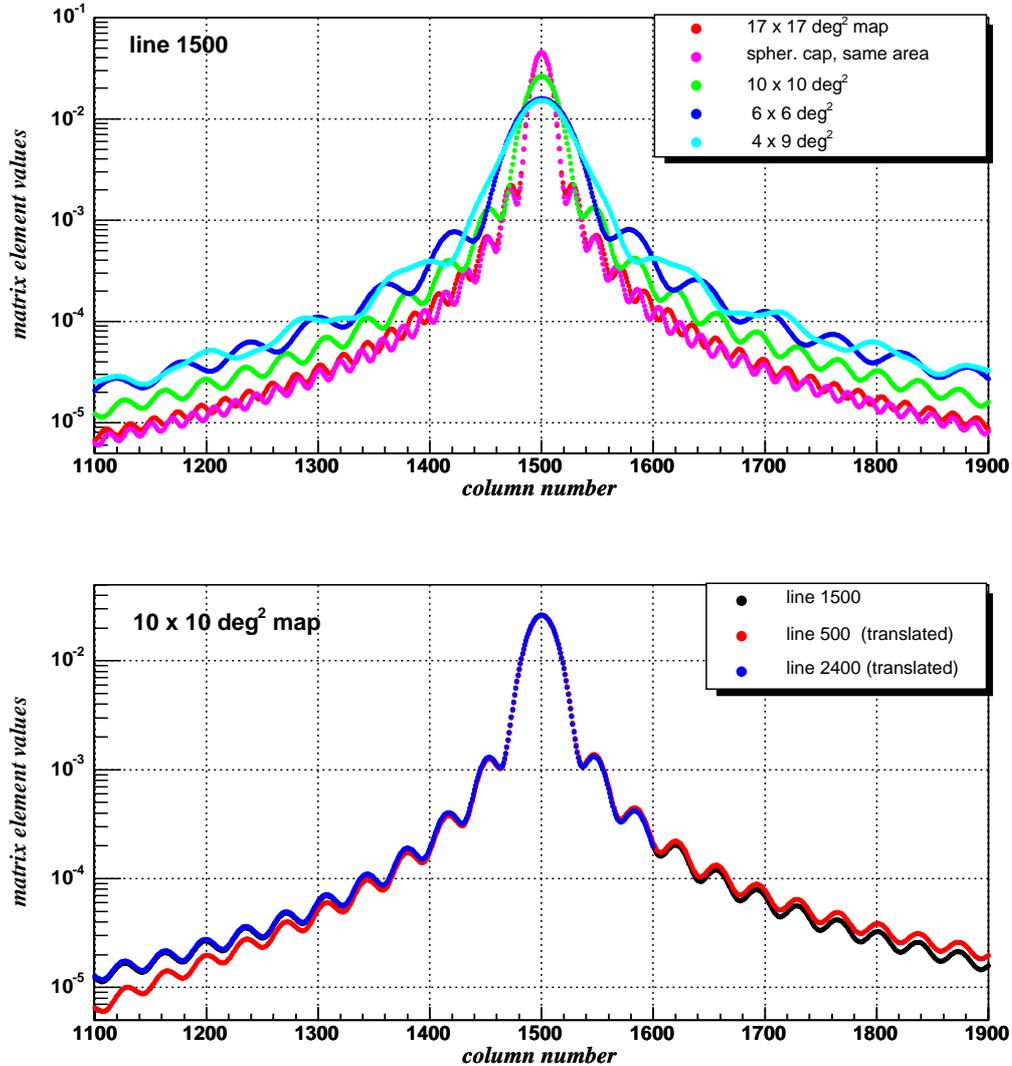}
   \caption[Power spectra estimation for a $10\x10\;deg^2$ square map]
         {
The top picture shows line $l=1500$ of matrix $M_{ll'}$
for $l'$ ranging from $1100$ to $1900$
and for maps of various sizes and shapes.
The bottom one shows line $l=1500$ of matrix $M_{ll'}$
for a $10\x 10 \;deg^2$ map
and lines $l=500,2400$ where the peaks have been shifted to $l'=1500$
for easy comparison.
         }
   \label{matrice_Mllp}
\end{figure}

\begin{figure}
   \centering
   \includegraphics[width=150mm]{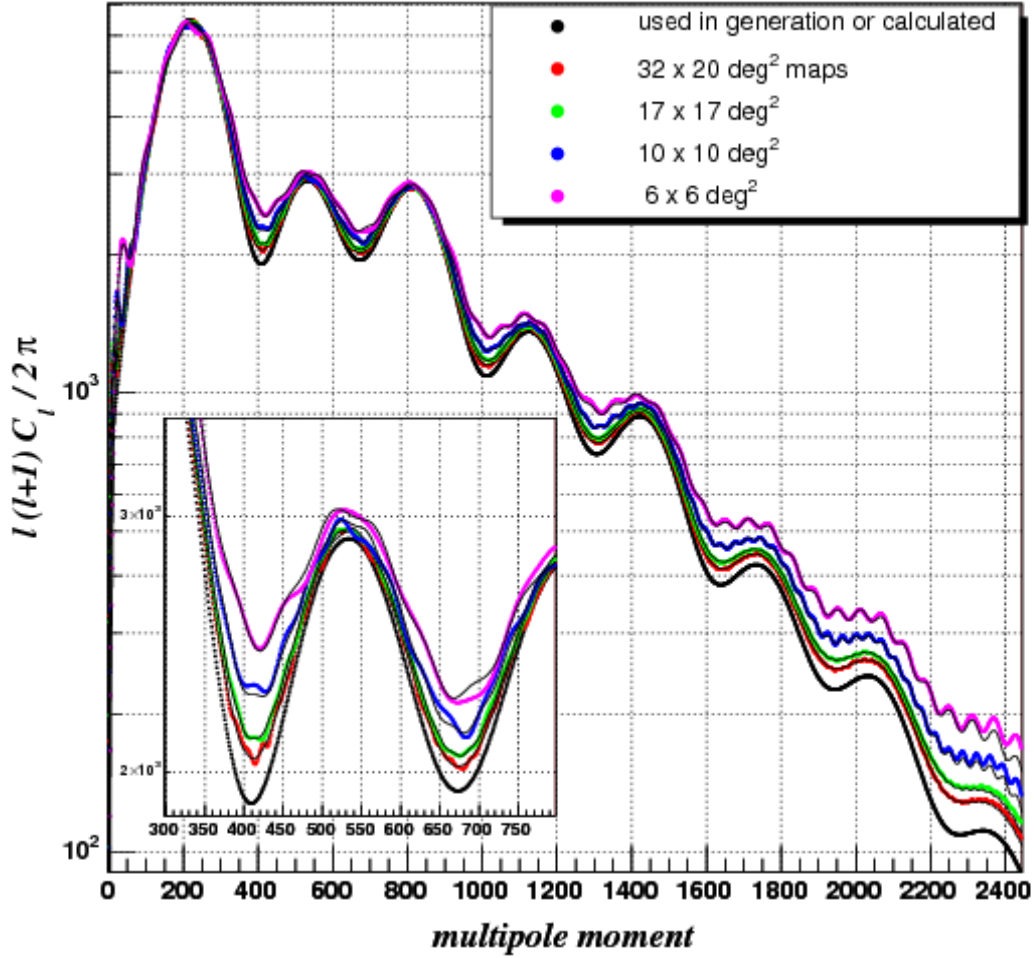}
   \caption[Systematic effect on $\cl{l}$ for various size maps]
         {Systematic effect on $\cl{l}$ for various size maps.
The thick black curve represents the $\cl{l}$ coefficients used in the simulations. 
The colored curves show the average $<\cplr{l}>$ reconstructed from simulations
for various map sizes.
The thin black curves superimposed on colored ones show the results
predicted by formula~(\ref{matmll}).
         }
   \label{figure1a}
\end{figure}

\begin{figure}
   \centering
   \includegraphics[width=150mm]{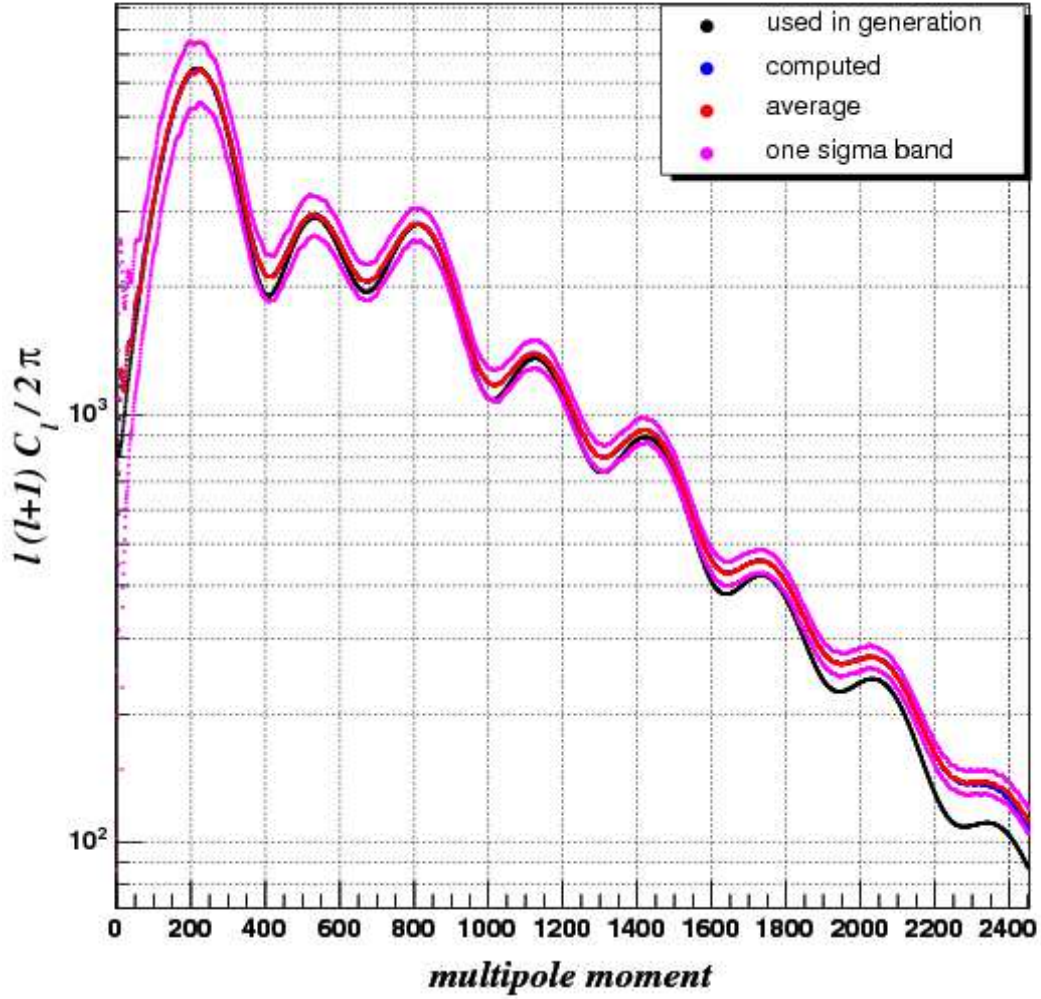}
   \caption[Finite size map systematic effect on $\cl{l}$ for a $17\x 17\;deg^2$ map]
         {Finite size map systematic effect on $\cl{l}$ for a $17\x 17\;deg^2$ map.
The thick black curve represents the $\cl{l}$ used in the simulations.
The red curve show the average $<\cplr{l}>$ reconstructed from simulations.
The two pink curves show the one sigma deviation reconstructed from simulations.
The blue curve shows the results predicted by formula~(\ref{matmll}).
         }
   \label{figure1b}
\end{figure}

Figure~\ref{figure1a} shows that the bias computed using
formulae~(\ref{matmll}) and~(\ref{cltnorm})
is in excellent agreement with the simulations.
Figure~\ref{figure1b} illustrates the fact,
with $17 \x 17 deg^2$ maps, that this systematic bias may not be small with 
respect to the variance. In the present figure it is of the order 
of $2 \sigma$ for $l \geq 2000$. We note also that the correction can be 
important at lower $l$ although dominated by the variance.
The $\cplr{l}$ values are systematically above the input $\cl{l}$
when $200\lesssim l$ and are oscillating with a period
$\delta l\simeq \pi/\theta_c$.
The matrix $M_{ll'}$ (\ref{matmll})
has long oscillating quasi-symetric tails (see the figure~\ref{matrice_Mllp}).
The input $\cl{l}$ spectrum used is a rapidly decreasing function.
Therefore the left tail contribution is much larger than the right tail one,
making the bias always positive at high $l$.
The observed oscillations are a consequence of the $M_{ll'}$ matrix
oscillating tails and of the global shape of the $\cl{l}$ spectrum,
but are not, at first order, the image of the acoustic peaks.
A smooth input $\cl{l}$ spectrum without peaks leads also to
an oscillating $\cplr{l}$ spectrum with only a sligthly
different pattern.
For a more rapidly decreasing input spectrum,
the amplitude of the oscillations increases.

\begin{figure}
   \centering
   \includegraphics[width=150mm]{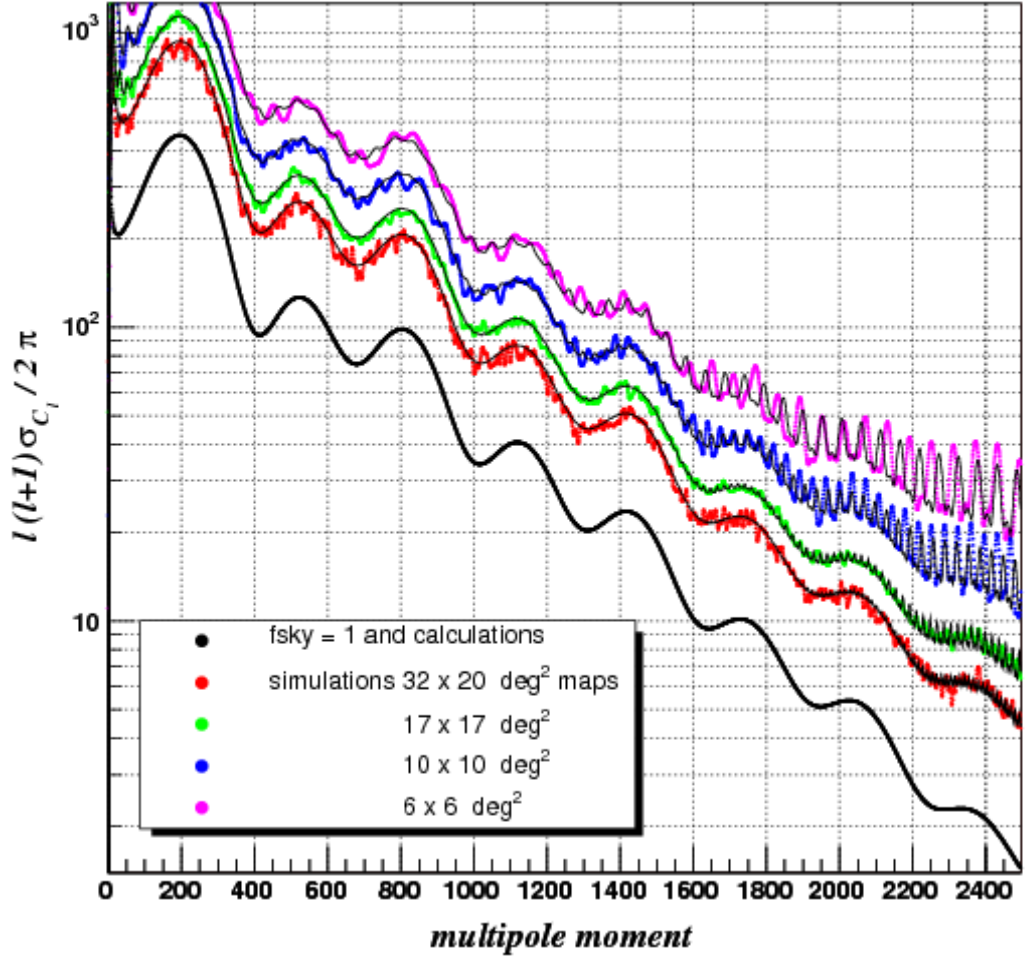}
   \caption[Finite size map effect on $\cl{l}$ variance for various maps]
         {Finite size map effect on $\cl{l}$ variance for various maps.
The thick black curve represents the variance of the $\clest{l}$ for the full sphere.
The colored curves show the variance of the $\cplr{l}$ reconstructed from simulations.
The thin black curves superimposed on colored ones show the results
predicted by formula~(\ref{eqblk}).
         }
   \label{figure2a}
\end{figure}

\begin{figure}
   \centering
   \includegraphics[width=150mm]{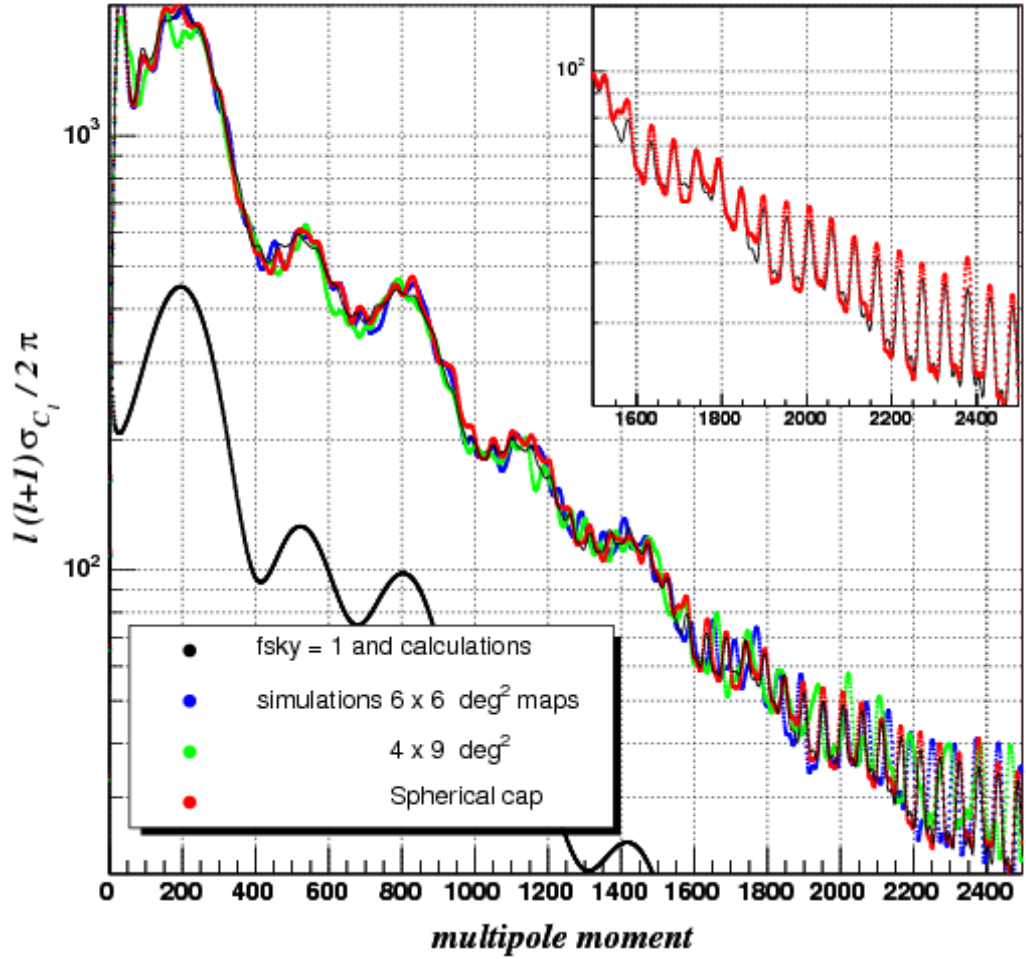}
   \caption[Finite size map effect on $\cl{l}$ variance for various $\sim 36\;deg^2$ maps
          with different shapes]
         {Finite size map effect on $\cl{l}$ variance for various $\sim 36\;deg^2$ maps
          with different shapes.
The zoom compares the result of simulations for spherical cap
with the formula~(\ref{eqblk}).
         }
   \label{figure2b}
\end{figure}

\begin{figure}
   \centering
   \includegraphics[width=150mm]{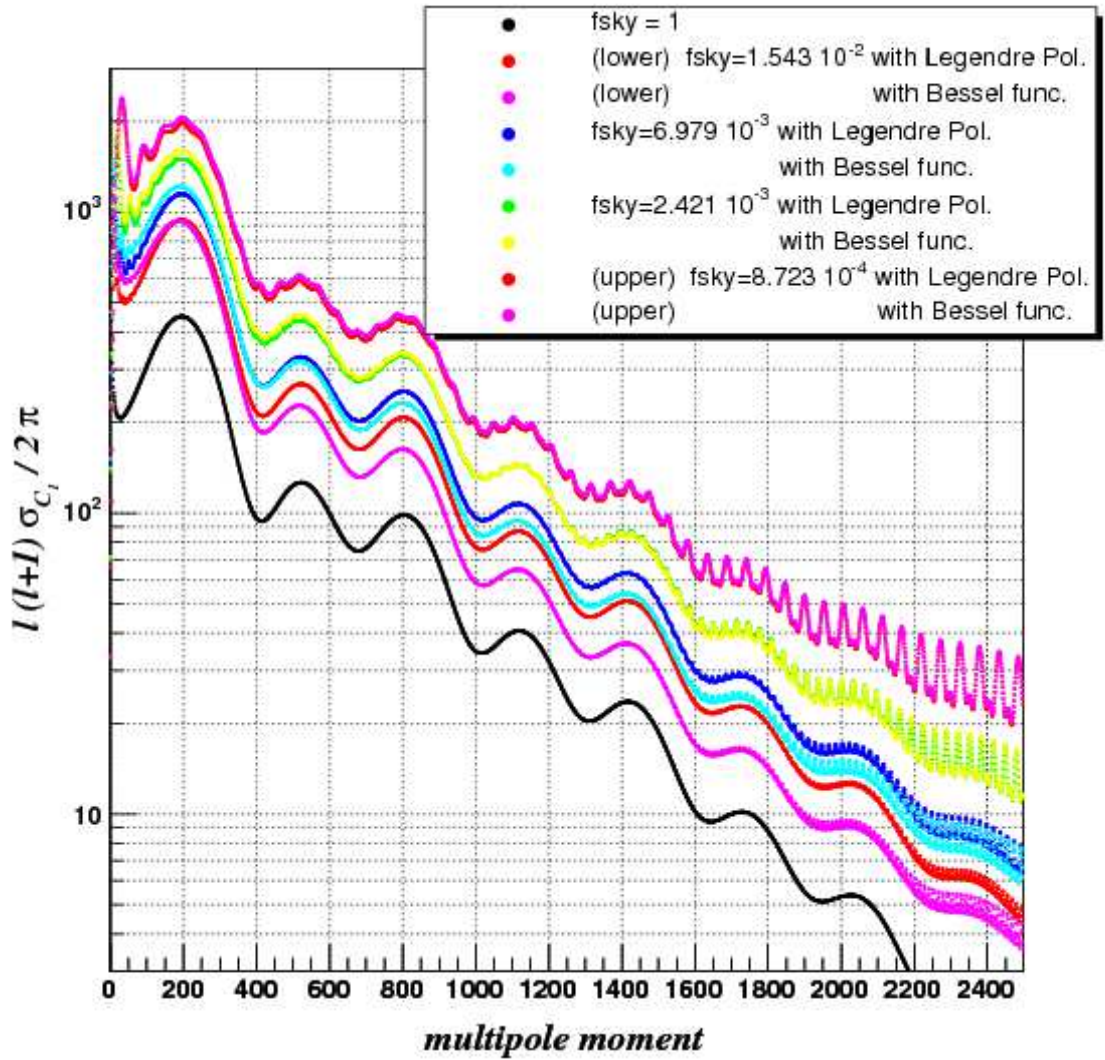}
   \caption[Validity of the small polar cap map approximation]
         {Validity of the small polar cap map approximation.
Comparison of the variance computed with Legendre polynomials
and Bessel function approximations (see section~(\ref{clbessel})).
         }
   \label{figure3}
\end{figure}

\begin{figure}
   \centering
   \includegraphics[width=150mm]{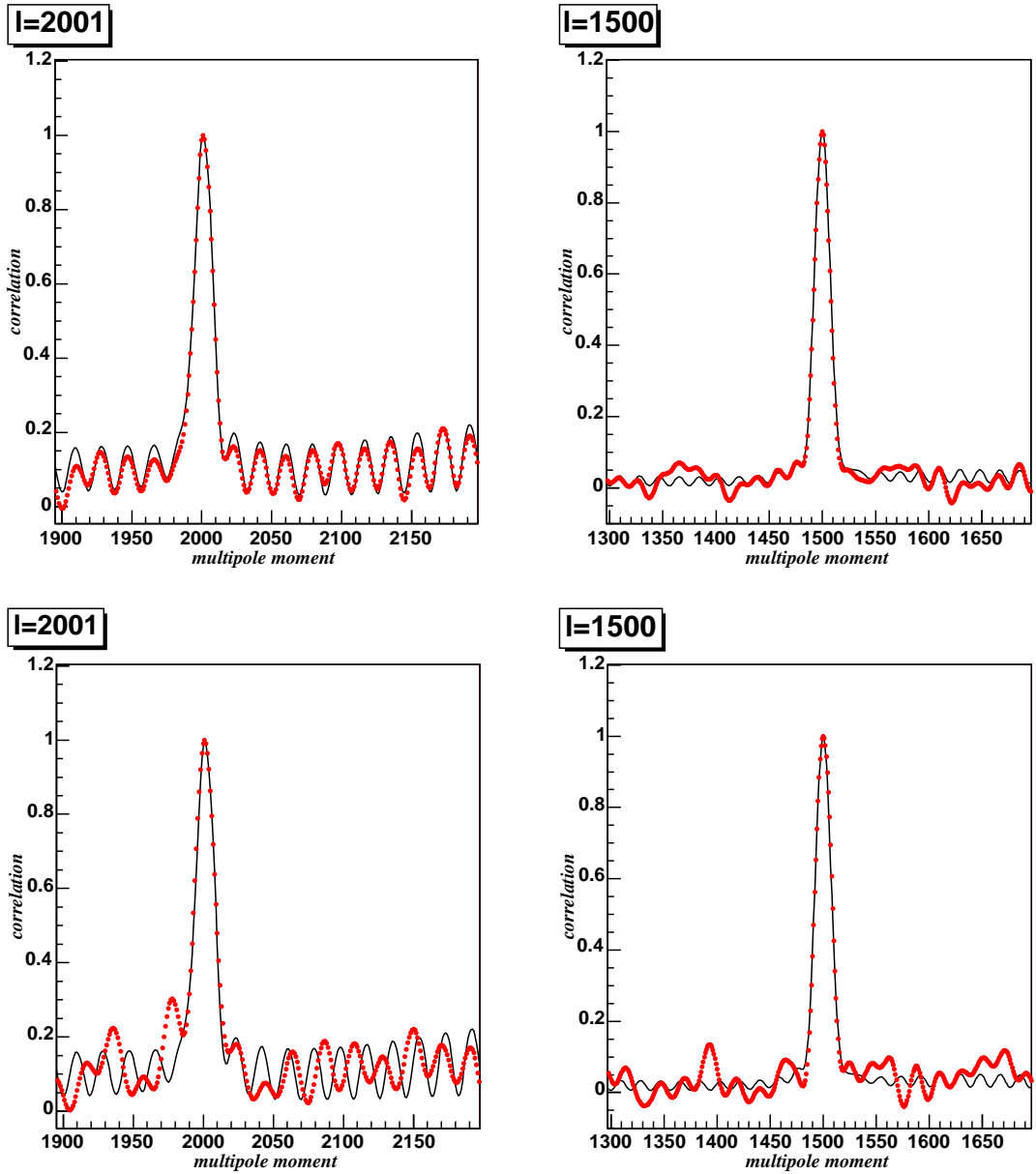}
   \caption[Comparison of the computed correlation matrix with simulations]
         {
  Comparison of the computed correlation matrix
  with simulations for $17\x 17\;deg^2$ maps.
  Black curves correspond to the computed correlation
  for a spherical cap.
  Red curves represent the correlation computed from simulations:
  the top figures correspond to a spherical cap
  and the bottom figures to a square map.
  The lines $l=2001$ and $l=1500$ of the correlation matrix
  $Cor(l,L)$ are drawn versus columns $L$
  around the correlation peak. 
         }
   \label{figure4}
\end{figure}

\begin{figure}
   \centering
   \includegraphics[width=150mm]{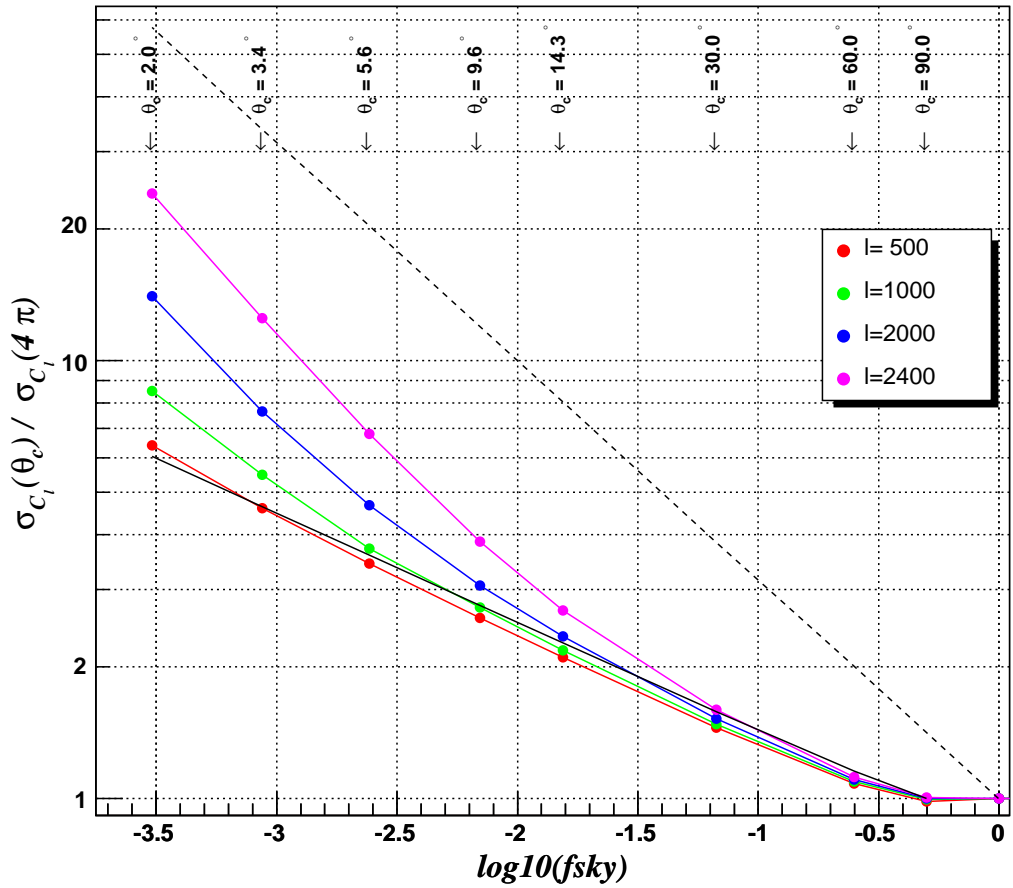}
   \caption[Variance of $\cl{l}$ relative to the full sphere variance]
         {
Square root of the variance of $\cl{l}$ relative to the full sphere variance
versus map aperture $\theta_c$ for various $l$.
The black dashed curve represents the simple
$f_{sky}^{-\undemi}$ scaling.
The black solid curve represents the first order scaling (equation~(\ref{sigclpi})).
         }
   \label{figure5}
\end{figure}

The variance has been computed numerically using~(\ref{eqcov})
and the recurrence relations~(\ref{eqblk}).
The results are compared with various simulations in 
figures~\ref{figure2a} and~\ref{figure2b}.
They show that the variance can be predicted accurately 
independently of the map shape, although the calculations were performed
for spherical caps. The figure~\ref{figure2b} shows in detail the agreement between 
spherical cap map simulations and calculations.
We can even go further and compare the correlation matrix obtained from 
simulations to the calculations for spherical caps:
\begin{eqnarray*}
Cor(l,L)\eg
\frac{<\cpl{l}\;\cpl{L}>-<\cpl{l}><\cpl{L}>}
{
\sqrt{<\cpl{l}^2>-<\cpl{l}>^2}
\sqrt{<\cpl{L}^2>-<\cpl{L}>^2}
}
\end{eqnarray*}
We will do the comparison for a few lines of the correlation matrix.
Figure~\ref{figure4} compares computation and simulation
for two lines ($l=2001$ and $l=1500$) of the correlation matrix
around the correlation peak for $f_{sky}\eg 0.697\;10^{-2}$
(see table~(\ref{tablegen})).
Top plots are for a spherical cap ($2279$ simulations)
and bottom ones are for a square map ($579$ simulations).
When $l,L$ are both larges, the dispersion is rather small.
The top left plot shows the very good agreement between
computation and simulations.
The bottom left plot shows that, even for a square map,
the correlation peak is well reproduced.
Outside the peak, the correlation level is fairly good but
out of phase due to the map shape.
The same conclusions can be drawn for the right plots
although dispersion in the simulations is larger
because of the large variance we have for small values of $l,L$.
In conclusion,the correlation peak shape is rather independent of the map shape.
We note also that the correlation matrix outside the diagonal region has a 
periodic structure of period $\simeq 2 \pi / \theta_c $ which is a consequence
of formula~(\ref{bllsinc}). \\
Figure~\ref{figure5} shows
$\sigma_{\cplr{l}}/\sigma_{\cl{l}}(4\pi)$
versus $f_{sky}$ for various $l$.
The variance ($\sigma_{\cplr{l}}^2$) does not scale as $f_{sky}^{-1}$
as often assumed (see the discussion in (\cite{Tegm97})).
The black dashed curve shows
$\sigma_{\cplr{l}}/\sigma_{\cl{l}}(4\pi)=\frac{1}{\sqrt{f_{sky}}}$. \\
For $l\sim 500$
the variance scale as $f_{sky}^{-\frac{1}{2}}$ as expected
from the lowest order calculation (formula~(\ref{sigclpi}))
represented by the solid black curve.
For larger $l$ values, we have to take into account the correction
terms of formula~(\ref{sigclcor}).
This indicates that the dependance of the variance on the map size
depends also on the $\cl{l}$ spectrum shape.

\clearpage
\section{Angular correlation function estimation for \\ a portion of sphere.}\label{sec_ksipartial}

\subsection{Estimator of the angular correlation function.}\label{ksipartdef}

The angular correlation function $\fcorp(\gamma)$ is estimated
as in section~\ref{basicksi} but with the integration
limited to the portion of sphere $A$. Let's define :
\begin{center}
$W^A(\Omega)=1\;$ if $\Omega\in A$ and $0$ elsewhere,
or $\; W(\Omega)>0\;$ in $A$ and $0$ elsewhere.
\end{center}

The angular correlation function can be computed without ponderation
on the temperature field. We define:
$$\begin{array}{rcl}
\fcorp(\gamma)
  &=&
\frac{1}{\mathcal{N}^A(\gamma)}
  \;\integ{A\x A}{}\;d\Omega_1 d\Omega_2
  \;\conj{T}(\Omega_1)T(\Omega_2)
  \;\delta(\V{\Omega_1}.\V{\Omega_2}-\cos(\gamma)) \\
  &=&
\frac{1}{\mathcal{N}^A(\gamma)}
  \;\integ{S^2\x S^2}{}\;d\Omega_1 d\Omega_2
  \;\conj{T}(\Omega_1)\conj{W^A(\Omega_1)}
    T(\Omega_2)W^A(\Omega_2)
  \;\delta(\V{\Omega_1}.\V{\Omega_2}-\cos(\gamma))
\end{array}$$
or more generally with a ponderation $W(\Omega)$:
$$\begin{array}{rcl}
\fcorp(\gamma)
  &=&
\frac{1}{\mathcal{N}^A(\gamma)}
  \;\integ{A\x A}{}\;d\Omega_1 d\Omega_2
  \;\conj{T}(\Omega_1)\conj{W(\Omega_1)}
  \;T(\Omega_2)W(\Omega_2)
  \;\delta(\V{\Omega_1}.\V{\Omega_2}-\cos(\gamma)) \\
  &=&
\frac{1}{\mathcal{N}^A(\gamma)}
  \;\integ{S^2\x S^2}{}\;d\Omega_1 d\Omega_2
  \;\conj{T}(\Omega_1)\conj{W(\Omega_1)}
  \;T(\Omega_2)W(\Omega_2)
  \;\delta(\V{\Omega_1}.\V{\Omega_2}-\cos(\gamma)) \\
\end{array}$$
with
$$\begin{array}{rcl}
\mathcal{N}^A(\gamma)
 &=&
\integ{A\x A}{}\;d\Omega_1 d\Omega_2
  \;\delta(\V{\Omega_1}.\V{\Omega_2}-\cos(\gamma)) \\
 &=&
\integ{S^2\x S^2}{}\;d\Omega_1 d\Omega_2
  \;\conj{W^A(\Omega_1)}\;W^A(\Omega_2)
  \;\delta(\V{\Omega_1}.\V{\Omega_2}-\cos(\gamma))
\end{array}$$
We have$\;T(\Omega) = \somme{LM}{}\;\alm{l}{m}\;\ylm{l}{m}(\Omega)$
and we define:
$$ \left\{ \begin{array}{rcl}
  W(\Omega) &=& \somme{LM}{}\;\blm{L}{M}\;\ylm{L}{M}(\Omega) \\
  W^A(\Omega) &=& \somme{LM}{}\;\blm{L}{M}^A\;\ylm{L}{M}(\Omega) \\
\end{array} \right. $$
The above integral becomes:
$$ \begin{array}{rcl}
\mathcal{N}^A(\gamma)\;\fcorp(\gamma)
 &=&
\somme{\cdots}{}
\;\conj{\alm{l_1}{m_1}}\conj{\blm{L_1}{M_1}}
\;\alm{l_2}{m_2}\blm{L_2}{M_2} \\
 &&
\quad\x
\integ{S^2\x S^2}{}\;d\Omega_1 d\Omega_2
\;\delta(\V{\Omega_1}.\V{\Omega_2}-\cos(\gamma))
\;\conj{\ylm{l_1}{m_1}}(\Omega_1)\conj{\ylm{L_1}{M_1}}(\Omega_1)
\;\ylm{l_2}{m_2}(\Omega_2)\ylm{L_2}{M_2}(\Omega_2)
\end{array} $$
where the sum $\somme{}{}$
goes over $\{l_1,m_1,L_1,M_1,l_2,m_2,L_2,M_2\}$. \\
The computation is described in details in appendix~\ref{appksipart}.
We get:
$$
<\fcorp(\gamma)>
  \eg
\left( \somme{l}{}\;\frac{2l+1}{4\pi}\;\cl{l}\;\pl{l}(\gamma) \right)
\frac{\somme{L}{}\;(2L+1)\;\bl{L}\;\pl{L}(\gamma)}
     {\somme{L}{}\;(2L+1)\;\bl{L}^A\;\pl{L}(\gamma)}
$$
where $\bl{l}$ and $\bl{l}^A$ are defined respectively with
$\blm{l}{m}$ and $\blm{l}{m}^A$ according to~(\ref{blmask}), and
$$
\mathcal{N}^{(A)}(\gamma)
  \eg
2\pi\;\somme{L}{}\;(2L+1)\;\bl{L}^{(A)}\;\pl{L}(\gamma)
$$
For $\;\gamma> \Theta_{max}$,
where $\Theta_{max}$ is the largest angular distance on the map $A$,
$\fcorp(\gamma)$ is not computable. \\
\ \newline
$\bullet$ If $W=W^A$, the temperature field is not ponderated
on $A$ and the estimator is unbiased:
$$
<\fcorp(\gamma)>
  \eg
\somme{l}{}\;\frac{2l+1}{4\pi}\;\cl{l}\;\pl{l}(\gamma)
  \eg
\fcor(\gamma)
$$
\ \newline
$\bullet$ If $W\ne W^A$, the temperature field is ponderated on $A$
by the positive function $W(\Omega)$
and the angular correlation function is biased. \\
$\bullet$ If the temperature field is ponderated on $A$
by the positive function $W(\Omega)$ and we replace
$\mathcal{N}^A(\gamma)$ by:
$$
\mathcal{N}^W(\gamma)\eg \integ{S^2\x S^2}{}\;d\Omega_1 d\Omega_2
\;\conj{W(\Omega_1)}\;W(\Omega_2)
\;\delta(\V{\Omega_1}.\V{\Omega_2}-\cos(\gamma))
 \eg
2\pi\;\somme{L}{}\;(2L+1)\;\bl{L}\;\pl{L}(\gamma)
$$
the estimator is unbiased. \\
Note that $\mathcal{N}^A(\gamma)$ is proportional to
the number of pairs of directions separated by an angle $\gamma$
that can be done on $A$.

\subsection{Comparison of the estimator with the $<\cpl{l}>$.}\label{ksipartcl}

$<\fcorp(\gamma)>$ can also be expressed relative to the ensemble average
$<\cpl{l}>$ of the bias $\cl{l}$ computed in section~\ref{clpartbias}.

\noindent We have ($c=\cos(\gamma)$):
$$ \begin{array}{rcl}
\mathcal{N}^A(c)\;\fcorp(c)
 &=&
\somme{lml'm'}{}\;\conj{\alm{l}{m}}\;\alm{l'}{m'}
\;\integ{A\x A}{}\;d\Omega d\Omega'
\;\delta(\V{\Omega}.\V{\Omega'}-c)
\;\conj{\ylm{l}{m}}(\Omega)\;\ylm{l'}{m'}(\Omega')
\end{array} $$
The integral is computed in appendix~\ref{appyydpart}.
Computing the ensemble average gives:
$$ \begin{array}{rcl}
\mathcal{N}^A(c)\;<\fcorp(c)>
 &=&
\somme{lml'm'LM}{}\;\cl{l}\delta_{ll'}\delta_{mm'}
\x\left(
2\pi\pl{L}(c)\;B_{LMlm}\;\conj{B_{LMl'm'}}
\right) \\
 &=&
2\pi\somme{lmLM}{}\;\cl{l}\pl{L}(c)\;\abs{B_{LMlm}}^2 \\
 &=&
2\pi\somme{L}{}\;\pl{L}(c)\;\somme{lmM}{}\;\cl{l}\abs{B_{LMlm}}^2 \\
 &=&
2\pi\somme{L}{}\;\pl{L}(c)\;(2L+1)<\cpl{L}> 
\end{array} $$
We finally obtain:
$$
<\fcorp(\gamma)>
  \eg
2\pi\somme{l}{}\;(2l+1)\;\frac{\pl{l}(\gamma)}{\mathcal{N}^A(\gamma)}\;<\cpl{l}> 
$$

\subsection{Covariance of the estimator of the angular \\ correlation function on partial map.}\label{ksipartcov}

\begin{figure}
   \centering
   \includegraphics[width=150mm]{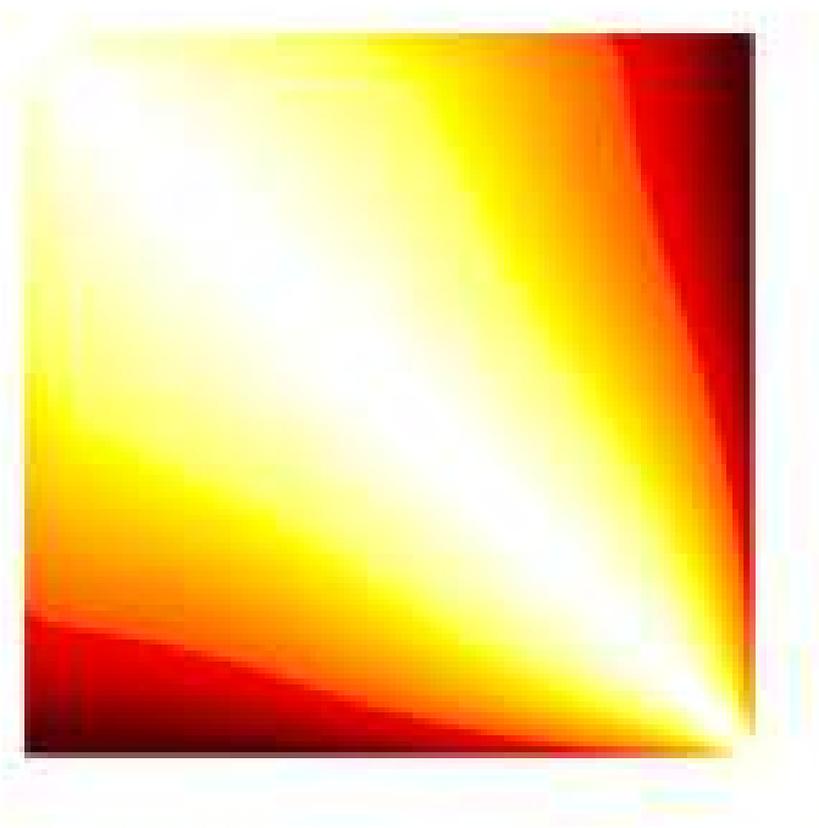}
   \caption[Correlation matrix of the $\fcorp(\gamma)$]
         {
Correlation matrix of the $\fcorp(\gamma)$ for a $17\x 17\;deg^2$ spherical cap.
The top left corner is for $(\gamma=0,\gamma'=0)$
and the bottom left corner is for $(\gamma=19.1,\gamma'=19.1\;deg)$.
The LUT ranges from $0.5$ (dark) to $+1$ (white).
         }
   \label{corksi}
\end{figure}

Let's define $c_1=\cos(\gamma_1)$ and $c_2=\cos(\gamma_2)$.
$$ \begin{array}{rcl}
\mathcal{N}^A(c_1)\;\fcorp(c_1)
 &=&
\somme{\cdots}{}\;\conj{\alm{l_1}{m_1}}\;\alm{l'_1}{m'_1}
\;\integ{A\x A}{}\;d\Omega_1 d\Omega'_1
\;\delta(\V{\Omega_1}.\V{\Omega'_1}-c_1)
\;\conj{\ylm{l_1}{m_1}}(\Omega_1)\;\ylm{l'_1}{m'_1}(\Omega'_1)
\end{array} $$
where the sum $\somme{}{}$ goes over $\{l_1,m_1,l'_1,m'_1\}$.
$$ \begin{array}{l}
\mathcal{N}^A(c_1)\mathcal{N}^A(c_2)\;<\fcorp(c_1)\fcorp(c_2)> \\
 \qquad\eg
 <\alm{l'_1}{m'_1}\conj{\alm{l_1}{m_1}}\alm{l'_2}{m'_2}\conj{\alm{l_2}{m_2}}>
 \integ{A\x A}{} d\Omega_1 d\Omega'_1
 \;\delta(\V{\Omega_1}.\V{\Omega'_1}-c_1) \\
 \qquad\qquad\x
 \integ{A\x A}{} d\Omega_2 d\Omega'_2
 \;\delta(\V{\Omega_2}.\V{\Omega'_2}-c_2)
 \;\conj{\ylm{l_1}{m_1}}(\Omega_1)\;\ylm{l'_1}{m'_1}(\Omega'_1)
 \;\conj{\ylm{l_2}{m_2}}(\Omega_2)\;\ylm{l'_2}{m'_2}(\Omega'_2)
\end{array} $$
If the temperature field is real and gaussian, we can compute
the fourth moment of the $\alm{l}{m}$.
That leads to $3$ terms:
$$ \begin{array}{rl}
1-/ & \cl{l'_1}\cl{l'_2}\cdots
      \ylm{l'_1}{m'_1}(\Omega'_1)\conj{\ylm{l'_1}{m'_1}}(\Omega_1)
      \ylm{l'_2}{m'_2}(\Omega'_2)\conj{\ylm{l'_2}{m'_2}}(\Omega_2) \\
2-/ & \cl{l'_1}\cl{l'_2}\cdots
      \ylm{l'_1}{m'_1}(\Omega'_1)\conj{\ylm{l'_2}{m'_2}(\Omega_1)}
      \ylm{l'_2}{m'_2}(\Omega'_2)\conj{\ylm{l'_1}{m'_1}(\Omega_2)} \\
3-/ & (-1)^{m'_1+m_1} \cl{l'_1}\cl{l_1}\cdots
      \ylm{l'_1}{m'_1}(\Omega'_1)\conj{\ylm{l_1}{m_1}(\Omega_1)}
      \ylm{l'_1}{-m'_1}(\Omega'_2)\conj{\ylm{l_1}{-m_1}(\Omega_2)}
\end{array} $$
The first term gives
$<\mathcal{N}^A(c_1)\fcorp(c_1)><\mathcal{N}^A(c_2)\fcorp(c_2)>$.
Using $\ylm{l}{m}=(-1)^m\conj{\ylm{l}{-m}}$ and playing with indices,
it is easy to demonstrate that the second and third terms are equal.
We have:
$$ \begin{array}{l}
\mathcal{N}^A(c_1)\mathcal{N}^A(c_2)\;\left(
<\fcorp(c_1)\fcorp(c_2)>-<\fcorp(c_1)><\fcorp(c_2)>
\right) \\
  \qquad\qquad\eg
\mathcal{N}^A(c_1)\mathcal{N}^A(c_2)\;V(\fcorp(c_1),\fcorp(c_2)) \\
  \qquad\qquad\eg
2\x\somme{l_1m_1l'_1m'_1}{}\;\cl{l'_1}\cl{l_1} \\
  \qquad\qquad\qquad\qquad\x
\;\integ{A\x A}{} d\Omega_1 d\Omega'_1\;\delta(\V{\Omega_1}.\V{\Omega'_1}-c_1)
\;\ylm{l'_1}{m'_1}(\Omega'_1)\conj{\ylm{l_1}{m_1}}(\Omega_1) \\
  \qquad\qquad\qquad\qquad\qquad\x
\;\integ{A\x A}{} d\Omega_2 d\Omega'_2\;\delta(\V{\Omega_2}.\V{\Omega'_2}-c_2)
\;\ylm{l_1}{m_1}(\Omega_2)\conj{\ylm{l'_1}{m'_1}}(\Omega'_2) \\
  \qquad\qquad\eg
2\x\somme{l_1m_1l'_1m'_1}{}\;\cl{l'_1}\cl{l_1}
I_{l'_1m'_1}^{l_1m_1}(c_1)\;I_{l_1m_1}^{l'_1m'_1}(c_2)
\end{array} $$
where:
$$
I_{l'm'}^{lm}(c)
  \eg
\integ{A\x A}{} d\Omega d\Omega'\;\delta(\V{\Omega}.\V{\Omega'}-c)
\;\ylm{l'}{m'}(\Omega')\conj{\ylm{l}{m}}(\Omega)
  \eg
\conj{I_{lm}^{l'm'}(c)}
$$
is computed in appendix~\ref{appyydpart}.
Going back to the computation of the $\fcorp(c)$ covariance:
$$ \begin{array}{l}
\mathcal{N}^A(c_1)\mathcal{N}^A(c_2)\;V(\fcorp(c_1),\fcorp(c_2)) \\
  \qquad\eg
2(2\pi)^2\somme{l_1m_1l'_1m'_1}{}\;\cl{l_1}\cl{l'_1}
\;I_{l'_1m'_1}^{l_1m_1}(c_1)\;I_{l_1m_1}^{l'_1m'_1}(c_2) \\
  \qquad\eg
2(2\pi)^2\somme{l_1m_1l'_1m'_1}{}\;\somme{LML'M'}{}
\;\cl{l_1}\cl{l'_1}
\;\pl{L}(c_1)\pl{L'}(c_2)
\;B_{LMl_1m_1}\conj{B_{LMl'_1m'_1}}
\;B_{L'M'l'_1m'_1}\conj{B_{L'M'l_1m_1}} \\
  \qquad\eg
2(2\pi)^2\somme{LML'M'}{}
\;\pl{L}(c_1)\pl{L'}(c_2)
\left(
  \somme{l_1m_1}{}
  \;\cl{l_1}
  \;B_{LMl_1m_1}\conj{B_{L'M'l_1m_1}}
\right)
\left(
  \somme{l'_1m'_1}{}
  \;\cl{l'_1}
  \;B_{L'M'l'_1m'_1}\conj{B_{LMl'_1m'_1}}
\right) \\
  \qquad\eg
2(2\pi)^2\somme{LML'M'}{}
\;\pl{L}(c_1)\pl{L'}(c_2)
\;\abs{\somme{lm}{}\;\cl{l}\;B_{L'M'lm}\conj{B_{LMlm}}}^2 \\
  \qquad\eg
2(2\pi)^2\somme{LL'}{}
\;\pl{L}(c_1)\pl{L'}(c_2)
\;\left(
\somme{MM'}{}
\abs{\somme{lm}{}\;\cl{l}\;B_{L'M'lm}\conj{B_{LMlm}}}^2
\right)
\end{array} $$
The term in parenthesis can be expressed relative to
the covariance $V_{\cpl{l}}(L,L')$ of the $\cpl{l}$
(see appendix~\ref{appclblmlm}):
$$ \begin{array}{rcl}
\mathcal{N}^A(c_1)\mathcal{N}^A(c_2)\;V(\fcorp(c_1),\fcorp(c_2))
  &=&
2(2\pi)^2\somme{LL'}{}
\;\pl{L}(c_1)\pl{L'}(c_2)
\;\left(
\frac{(2L+1)(2L'+1)}{2}\;V_{\cpl{l}}(L,L')
\right) \\
  &=&
(2\pi)^2\somme{LL'}{}\;(2L+1)(2L'+1)
\;V_{\cpl{l}}(L,L')\;\pl{L}(c_1)\pl{L'}(c_2)
\end{array} $$
Using the value of $\mathcal{N}^A(\gamma)$ computed
in section~\mysecnp{appksipart},
one finally obtain for a partial map of any shape:
\encadre{
$$\begin{array}{rcl}
V(\fcorp(c_1),\fcorp(c_2))
 &=&
<\fcorp(c_1)\fcorp(c_2)>-<\fcorp(c_1)><\fcorp(c_2)> \\
 && \\
 &=&
\frac{
(2\pi)^2
\;\somme{ll'}{}
\;(2l+1)(2l'+1)
\;V_{\cpl{l}}(l,l')
\;\pl{l}(\gamma_1)\pl{l'}(\gamma_2)
}
{
\mathcal{N}^A(\gamma_1)\;\mathcal{N}^A(\gamma_2)
} \\
 && \\
 &=&
\frac{
\somme{ll'}{}
\;(2l+1)(2l'+1)
\;V_{\cpl{l}}(l,l')
\;\pl{l}(\gamma_1)\pl{l'}(\gamma_2)
}
{
(\somme{L}{}\;(2L+1)\;\bl{L}^A\;\pl{L}(\gamma_1))
\;(\somme{L'}{}\;(2L'+1)\;\bl{L'}^A\;\pl{L'}(\gamma_2))
}
\end{array} $$
}
Recall that the $V_{\cpl{l}}(l,l')$ can only be computed easily if the
partial map is a spherical cap.
The figure~\ref{corksi} shows the correlation matrix of the $\fcorp(\gamma)$
for a spherical cap of area $17\x 17\;deg^2$:
one sees that the level of correlation is very high.

\clearpage
\section{$\cl{l}\;$ estimation using the angular correlation function \\ for a portion of sphere.}\label{sec_clfrksi}

\subsection{Using integration of the angular correlation \\ function.}\label{clfrksia}

Suppose that we measure the angular correlation function
$\fcorp(\gamma)$. \\
For the full sphere we have:
$$
\cl{l}
\eg
2\pi\;\integ{\gamma=0}{\pi}\;\fcor(\gamma)\;\pl{l}(\cos(\gamma))\;d\cos(\gamma)
$$
On a portion of sphere,
we may have an estimation of the $\cl{l}$ by performing the integration
\begin{equation}
\cpl{l}^{\xi}
\eg
2\pi\;\integ{\gamma=0}{\theta_{lim}}\;\fcorp(\gamma)\;\pl{l}(\cos(\gamma))\;d\cos(\gamma)
\label{clestksi}
\end{equation}
where $\theta_{lim}$ is the maximum separation angle obtainable
for that map. \\
We demonstrated that $\fcorp(\gamma)$ is unbiased, so:
$$
<\fcorp(\gamma)>
\eg
\fcor(\gamma)
\eg
\frac{1}{4\pi}\;\somme{l=0}{\infty}\;(2l+1)\;\cl{l}\;\pl{l}(\cos(\gamma))
$$
Thus we compute:
$$ \begin{array}{rcl}
<\cpl{l}^{\xi}>
&=&
  2\pi\;\integ{\gamma=0}{\theta_{lim}}\;<\fcorp(\gamma)>\;\pl{l}(\cos(\gamma))\;d\cos(\gamma) \\
&=&
  2\pi\;\integ{\gamma=0}{\theta_{lim}}\;\fcor(\gamma)\;\pl{l}(\cos(\gamma))\;d\cos(\gamma) \\
&=&
2\pi\;\frac{1}{4\pi}\;\somme{l'=0}{\infty}\;(2l'+1)\;\cl{l'}
  \;\integ{\gamma=0}{\theta_{lim}}\;\pl{l'}(\cos(\gamma))\;\pl{l}(\cos(\gamma))\;d\cos(\gamma)
\end{array} $$
Using the definition of the $B_{Llm}$, the $\pl{l}$ and the $\hplm{l}{0}$ we have:
$$ \begin{array}{rcl}
<\cpl{l}^{\xi}>
&=&
  2\pi\;\somme{l'=0}{\infty}\;\sqrt{\frac{2l'+1}{2l+1}}\;\cl{l'}\;B_{l'l0}
\end{array} $$
with $B_{l'l0} \equiv B_{l'l0}(\theta_{lim})$.
\ \newline
\Main As we have a sharp cut-off at $\theta_{lim}$, the obtained spectrum
oscillates strongly around the theorical value
(that is somewhat equivalent to the Gibbs phenomena we have with the Fourier transform),
and that method is not usable.
The figure~\ref{clfrksi1} shows the predicted $\cpl{l}^{\xi}$
for a spherical cap of angular aperture $\theta_{lim}=19\;deg$.
The level of oscillations is very high.
The ``wave length'' of the oscillations is about
$\Delta l \;\sim\;\frac{2\pi}{\theta_{lim}}$.
This problem is discussed in the next section.

\ \newline
One can compute the covariance of the $\cpl{l}^{\xi}$:
$$
<\cpl{l}^{\xi}\cpl{l'}^{\xi}>
 \eg
 (2\pi)^2
 \;\integ{0}{\theta_{lim}}
 \;<\fcorp(\gamma)\fcorp(\gamma')>
 \;\pl{l}(\cos(\gamma))\pl{l'}\fcorp(\cos(\gamma'))
 \;d\cos(\gamma) d\cos(\gamma')
$$
Using the variance of the $\fcorp$ computed in section~\ref{ksipartcov},
we obtain:
\begin{equation}
V(\cpl{l}^{\xi},\cpl{l'}^{\xi})
 \eg
 \somme{l_1l_1'}{}\;M_{ll_1}\;V_{\cpl{l}}(l_1,l_1')\;M_{l'l_1'}
\label{covclksi}
\end{equation}
with
\begin{equation}
M_{ll_1}
 \eg
 (2\pi)^2\;(2l_1+1)
 \;\integ{0}{\theta_{lim}}
 \;\frac{1}{\mathcal{N}^A(\gamma)}
 \;\pl{l}(\gamma) \pl{l_1}(\gamma)
 \;d\cos(\gamma)
\label{covclksimll}
\end{equation}

\subsection{Using integration of a smooth apodization of \\ the angular correlation function.}\label{clfrksiap}

\begin{figure}
   \centering
   \includegraphics[width=150mm]{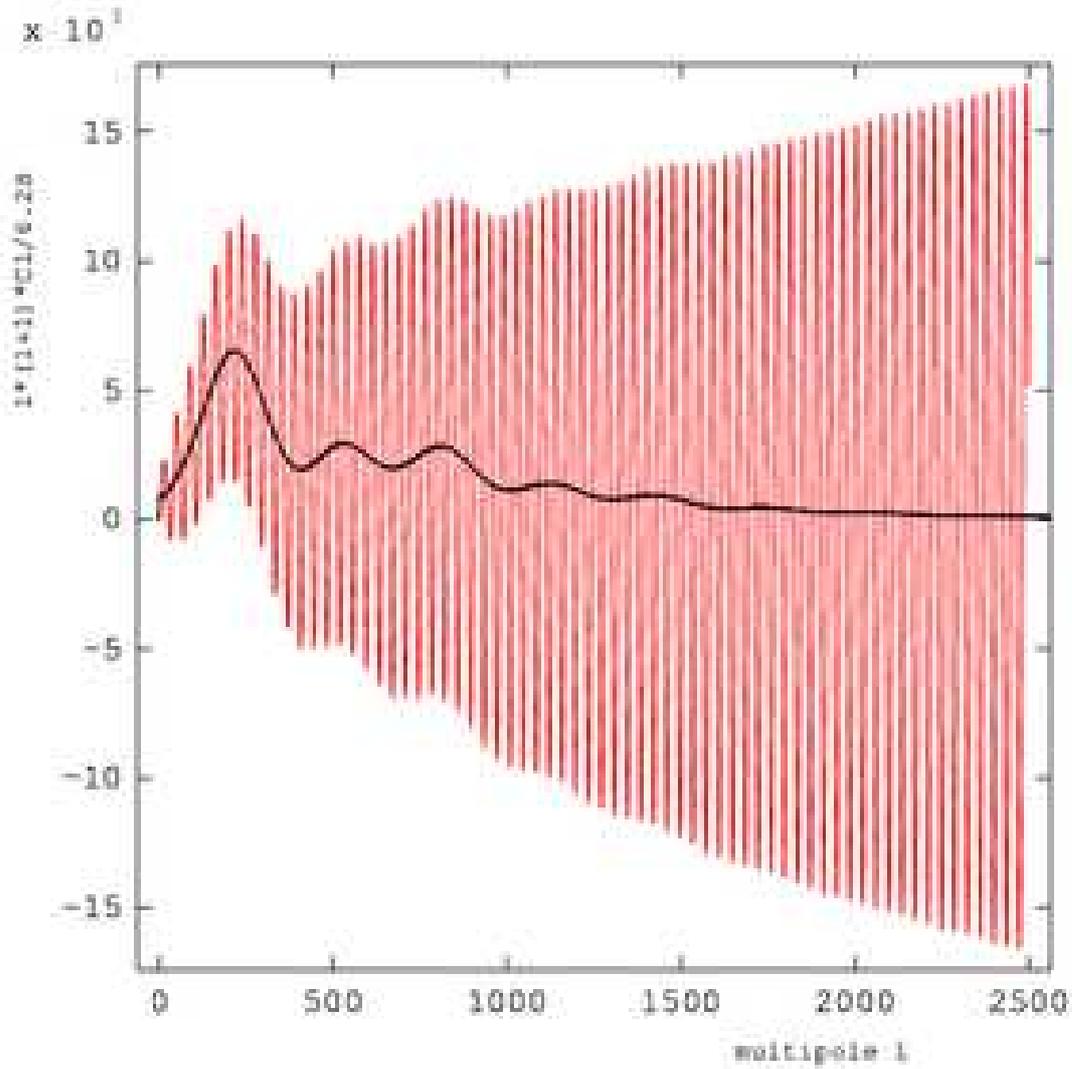}
   \caption[Predicted $<\cpl{l}^{\xi}>$ from $\fcorp(\gamma)$ integration]
         {
Predicted $<\cpl{l}^{\xi}>$ from $\fcorp(\gamma)$ integration
for a spherical cap of aperture $19\;deg$: \\
- the red curve shows the predicted $l(l+1)<\cpl{l}^{\xi}>/2\pi$ \\
- the black curve the input $l(l+1)\cl{l}/2\pi$ spectrum.
         }
   \label{clfrksi1}
\end{figure}

In order to avoid the problem shown in figure~\ref{clfrksi1}
due to the sharp cut-off at $\theta_{lim}$,
the correlation function $\fcorp(\gamma)$ is multiplied by a function
going smoothly to zero at $\theta_{lim}$. \\
\ \newline
A usefull function is:
$\;F_{tanh}(\theta) \eg \undemi (1-\tanh(\frac{\theta-\theta_0}{\Delta}))\;$
where $\theta_0$ is the cutting angle
and $\Delta$ caracterises the width of the smoothing. \\ 
One could also take
$\;F_{erf}(\theta) \eg \undemi (1-erf(\frac{\theta-\theta_0}{\Delta}))\;$. \\
The estimator in~(\ref{clestksi}) is changed to:
\begin{equation}
\cpl{l}^{\xi}
\eg
2\pi\;\integ{\gamma=0}{\theta_{lim}}\;\fcorp(\gamma)\;F(\gamma)\;\pl{l}(\cos(\gamma))\;d\cos(\gamma)
\label{clestksiap}
\end{equation}
In that case the average value of the estimator is:
$$\begin{array}{rcl}
<\cpl{l}^{\xi}>
 &=&
 2\pi
 \;\integ{\gamma=0}{\theta_{lim}}
 \;<\fcorp(\gamma)>\;F(\gamma)\;\pl{l}(\cos(\gamma))\;d\cos(\gamma) \\
 &=&
 \undemi\;\somme{l'=0}{\infty}\;(2l'+1)\;\cl{l'}
  \;\integ{\gamma=0}{\theta_{lim}}
  \;\pl{l'}(\cos(\gamma))\;\pl{l}(\cos(\gamma))
  \;F(\gamma)\;d\cos(\gamma)
\end{array}$$
It has to be computed numerically. \\
The covariance keeps the same form as~(\ref{covclksi})
if we replace $M_{ll_1}$ in the former section by:
\begin{equation}
M_{ll_1}
 \eg
 (2\pi)^2\;(2l_1+1)
 \;\integ{0}{\theta_{lim}}
 \;\frac{1}{\mathcal{N}^A(\gamma)}
 \;\pl{l}(\gamma) \pl{l_1}(\gamma) F(\gamma)
 \;d\cos(\gamma)
\label{covclksimlltap}
\end{equation}

{\bf \Main \underline{Apodization effect for small maps.}} \\

\begin{figure}
   \centering
   \includegraphics[width=150mm]{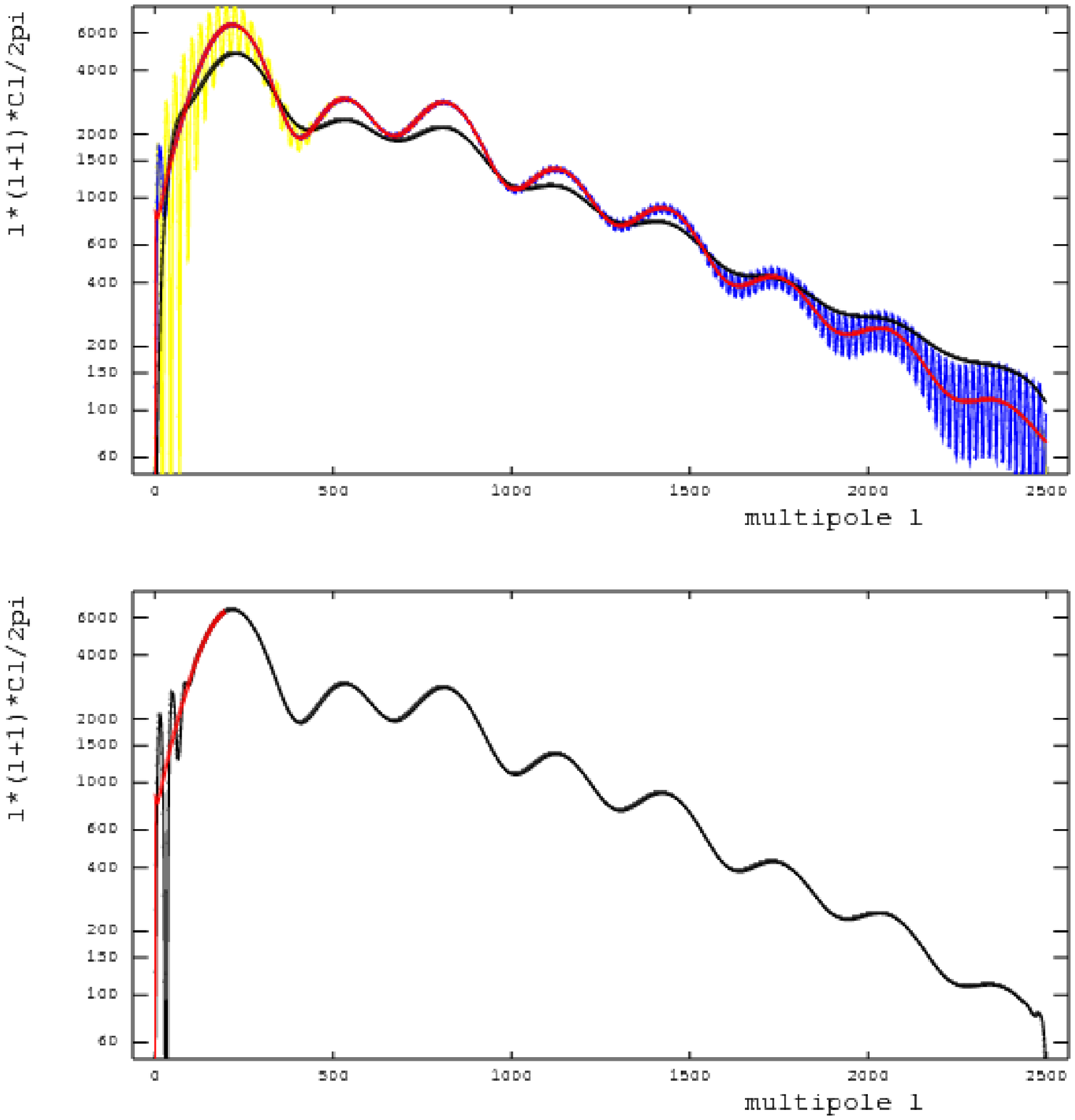}
   \caption[Predicted $<\cpl{l}^{\xi}>$ from $\fcorp(\gamma)$ integration with apodization]
         {
Predicted $l(l+1)<\cpl{l}^{\xi}>$ from $\fcorp(\gamma)$ integration with apodization
for a spherical cap of aperture $19\;deg$: \\
$\bullet$ {\it Top figure} \\
- the yellow curve is for apodization $(\theta_0= 15\;deg,\;\Delta = 0.5\;deg)$\\
- the blue curve is for apodization $(\theta_0= 10\;deg,\;\Delta = 5\;deg)$\\
- the black curve is for apodization $(\theta_0= 1\;deg,\;\Delta = 2\;deg)$\\
- the red curve is the $l(l+1)\cl{l}/2\pi$ input spectrum (the curve is undistinghishable
from the yellow one at large $l$).\\
$\bullet$ {\it Bottom figure} \\
- the black curve is for apodization $(\theta_0= 10\;deg,\;\Delta = 2\;deg)$\\
- the red curve is the $l(l+1)\cl{l}/2\pi$ input spectrum (the curve is not plotted above $l=200$
because it is undistinghishable from the black one).\\
}
   \label{clksith1}
\end{figure}

\begin{figure}
   \centering
   \includegraphics[width=150mm]{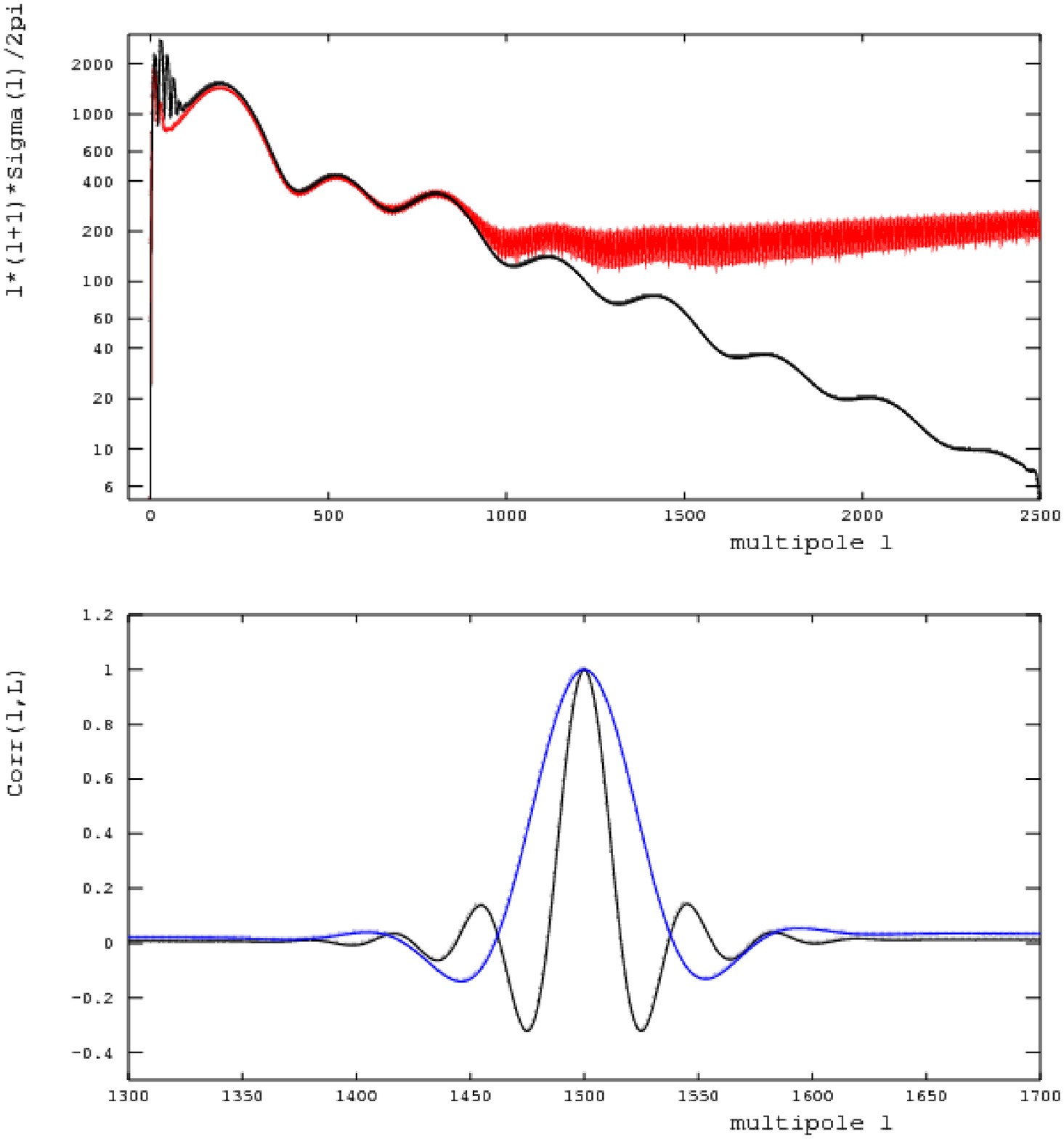}
   \caption[Predicted variance and correlation of $\cpl{l}^{\xi}$ with apodization]
         {
Predicted variance and correlation for $\cpl{l}^{\xi}$ with apodization
for a spherical cap of aperture $19\;deg$: \\
$\bullet$ {\it Top figure:} $\;l(l+1)\sigma_{\cpl{l}^{\xi}}/2\pi$ \\
- the black curve is for apodization $(\theta_0= 10\;deg,\;\Delta = 2\;deg)$\\
- the red curve is for apodization $(\theta_0= 10\;deg,\;\Delta = 5\;deg)$\\
$\bullet$ {\it Bottom figure:} correlation values for $l=1500$ \\
- the black curve is for apodization $(\theta_0= 10\;deg,\;\Delta = 2\;deg)$\\
- the blue curve is for apodization $(\theta_0= 5\;deg,\;\Delta = 2\;deg)$\\
         }
   \label{clksith2}
\end{figure}

We compute:
$\;
I \eg
\integ{\theta=0}{\theta_{lim}}
  \;\pl{l'}(\cos(\theta))\;\pl{l}(\cos(\theta))
  \;F(\theta)\;d\cos(\theta)
\;$.
For $\theta_{lim}$ small, we may approximate
$\;
\pl{l}(\cos(\theta)) \simeq J_0(l\theta)
\;$
and we have ($\sin(\theta)\simeq\theta $):
$$
I \eg
\integ{0}{\theta_{lim}}
  \;J_0(l'\theta)\;J_0(l\theta)
  \;F(\theta)\;\theta\;d\theta
$$
We define:
$\;
L(\theta)
 \eg
\integ{0}{\theta}
  \;J_0(l'\theta')\;J_0(l\theta')
  \;\theta'\;d\theta'
\;$
which is a Lommel integral.
Integrating $I$ by parts:
$$
I
 \eg
\left[ L(\theta) \;F(\theta) \right]_{0}^{\theta_{lim}}
  \;-\;
\integ{0}{\theta_{lim}}\;L(\theta')\;\dsd{\theta'}F(\theta')\;d\theta'
$$
But $L(0)=0$ and we set the smoothing function to be null
at $\theta_{lim}$ i.e. $F(\theta_{lim})=0$.
Thus $\;\left[\cdots\right]_{0}^{\theta_{lim}}=0$ and we have:
$$
I
 \eg
-\integ{0}{\theta_{lim}}\;L(\theta)\;\dsd{\theta}F(\theta)\;d\theta
$$
To perform analytical computations, we take:
$\;
F(\theta)
 =
F_{erf}(\theta)
 =
\undemi (1-erf(\frac{\theta-\theta_0}{\Delta}))
\;$
, we have:
$\;
\dsd{\theta}F_{erf}
 =
-\frac{1}{\sqrt{\pi}\Delta}
\;\exp\left(-\frac{(\theta-\theta_0)^2}{\Delta^2}\right)
\;$
and we obtain:
$$
I
 \eg
\frac{1}{\sqrt{\pi}\Delta}
\;\integ{0}{\theta_{lim}}
\;L(\theta)
\;\exp\left(-\frac{(\theta-\theta_0)^2}{\Delta^2}\right)
\;d\theta
$$
The Lommel integral $L(\theta)$ is~(\cite{Grad80}): \\
\begin{equation*}
L(\theta)
  \eg
\frac{\theta}{l'^2-l^2}
\;\left[
  lJ_0(l'\theta)J'_0(l\theta)
  \;-\;
  l'J_0(l\theta)J'_0(l'\theta)
\right]
\end{equation*}
where $J'_0(x)$ is the derivative with respect to $x$,
and we make the approximation (for $l\theta_{lim} \gg 1$):
$\quad
J_0(l\theta)
 \simeq
\sqrt{\frac{2}{\pi l \theta}}
\;\cos(l\theta + \phi)
\;$
with
$\;
\phi \rightarrow -(n+\undemi)\frac{\pi}{2}\;$
if $\;\theta \rightarrow \infty\;$.
For $J_0$, this remains a good approximation even
for values as low as $l\theta \simeq 4$. \\
We obtain:
$\;
L(\theta)
  \eg
\frac{1}{\pi}
\frac{1}{\sqrt{ll'}}
\left[
  \frac{\sin((l'-l)\theta)}{l'-l}
  \;+\;
  \frac{\sin((l'+l)\theta+2\phi)}{l'+l}
\right]
\;$ \\
The second term is of order $\frac{1}{l'+l}$
with respect to the first one and will be neglected.
We finally have:
$$\begin{array}{rcl}
I
 &\propto&
\frac{1}{\Delta\sqrt{ll'}}
\;\integ{0}{\theta_{lim}}
\;\frac{\sin((l'-l)\theta)}{l'-l}
\;\exp\left(-\frac{(\theta-\theta_0)^2}{\Delta^2}\right)
\;d\theta \\
 &\propto&
\frac{1}{\Delta\sqrt{ll'}}
\;\integ{-\infty}{+\infty}
\;\frac{\sin((l'-l)\theta)}{l'-l}
\;\exp\left(-\frac{(\theta-\theta_0)^2}{\Delta^2}\right)
\;d\theta
\end{array}$$
where we have replaced the integration limits
because $\Delta \ll \theta_0$ so that the integrand
is null outside the limits.
Setting $z=\theta-\theta_0$ we have:
$$
\sin((l'-l)\theta)
 =
\sin((l'-l)(\theta_0+z))
 =
\sin((l'-l)\theta_0)\cos((l'-l)z)
 +
\cos((l'-l)\theta_0)\sin((l'-l)z)
$$
Thus:
$$\begin{array}{rcl}
I
 & \propto &
  \frac{1}{\Delta\sqrt{ll'}}
  \left\{
  \;\frac{\sin((l'-l)\theta_0)}{l'-l}
  \;\integ{-\infty}{+\infty}
  \;\cos((l'-l)z)\exp(-\frac{z^2}{\Delta^2})\;dz
  \right. \\
 &&
  \left.
  \qquad +
  \cos((l'-l)\theta_0)
  \;\integ{-\infty}{+\infty}
  \;\frac{\sin((l'-l)z)}{l'-l}\exp(-\frac{z^2}{\Delta^2})\;dz
\right\}
\end{array}$$
The second integral is null by symetry:
$$
I
 \; \propto \;
  \frac{\theta_0}{\Delta\sqrt{ll'}}
  \;\frac{\sin((l'-l)\theta_0)}{(l'-l)\theta_0}
  \;\integ{-\infty}{+\infty}
  \;\cos((l'-l)z)\exp(-\frac{z^2}{\Delta^2})\;dz
$$
As
$\;
\integ{-\infty}{\infty}\;\cos((l'-l) x)\;\exp(-x^2/\Delta^2)\;dx
  \eg
\sqrt{\pi}\;\Delta\;\exp(-\frac{1}{4} (l'-l)^2 \Delta^2)
\;$,
we have:
$$
I
 \; \propto \;
  \frac{\theta_0}{\sqrt{ll'}}
  \;\frac{\sin((l'-l)\theta_0)}{(l'-l)\theta_0}
  \;\exp(-\frac{1}{4} (l'-l)^2 \Delta^2)
$$
and finally:
$$
<\cpl{l}^{\xi}>
 \; \propto \;
 \somme{l'=0}{\infty}\;(2l'+1)\;\cl{l'}
  \frac{\theta_0}{\sqrt{ll'}}
  \;\frac{\sin((l'-l)\theta_0)}{(l'-l)\theta_0}
  \;\exp(-\frac{1}{4} (l'-l)^2 \Delta^2)
$$
The value of a $<\cpl{l}^{\xi}>$ is given by a kind of convolution:
\begin{itemize}
\item The width of the $sinc$ goes like $\frac{1}{\theta_0}$,
      so as $\theta_0$ goes to zero, the width will become large.
\item The width of the gaussian goes like $\frac{1}{\Delta}$
      ($\sigma = \frac{\sqrt{2}}{\Delta}$),
      so the larger the width of the smoothing function (large $\Delta$),
      the smaller the width of the gaussian in the last formula.
\end{itemize}
The oscillations described in the previous section (see figure~\ref{clfrksi1})
are due to the contribution of the low multipoles at high $l$:
if there is no apodization, as there is a lot of power in low $l$ multipoles,
the high $l$ multipoles will get a lot of power transfered from the low ones
and the oscillations will be large: \\
- If the apodization is too sharp, the width of the gaussian will be
very large and the ``convolution'' will be dominated
by the $sinc$. The ``sinc'' decrease slowly ($1/l$),
the transfert of power from the low multipole to the high ones will be important
and so will be the oscillations.
We need a smooth apodization function. \\
- If $\theta_0$ is large
the width of the $sinc$ will be small
and the oscillation of the $sinc$ will remain imprinted
on the spectrum. We must take $\theta_0$ not near $\theta_{lim}$. \\

\ \newline
The figure~\ref{clksith1} shows the effect of various apodization
parameters on the predicted $\cl{l}$ (formula~\ref{clestksiap}).
The apodizations are applied to the ``theoretical angular correlation function''
with $\theta_{lim}\sim 19\;deg$ corresponding to $17\x17\;deg^2$ spherical cap
maps.
\begin{itemize}
\item for $\theta_0=15\;deg, \Delta=0.5\;deg$: the spectrum oscillates at low $l$
due to the sharpness of the filter. At high $l$ it is undistinguishable
from the theorical one.
\item for $\theta_0=10\;deg, \Delta=5\;deg$: the angular correlation spectrum
is not apodized enough near $\theta_{lim}$ and, as explained above,
the low multipoles $\cl{l}$ values correlate with the high multipole ones
leading to oscillations at high $l$ in the reconstructed spectrum.
\item $\theta_0=1\;deg, \Delta=2\;deg$ shows that one cannot lower too much
$\theta_0$ without loosing information.
The reconstructed spectrum becomes highly biaised and smoothed.
\item $\theta_0=10\;deg, \Delta=2\;deg$ shows an example of
good apodization. The reconstructed spectrum differs only at very low
multipole values where the size of the map becomes too small.
\end{itemize}
Using~\ref{covclksi} with~\ref{covclksimlltap},
the top plot of figure~\ref{clksith2} shows the variance $l(l+1)\sigma_{\cpl{l}^{\xi}}/2\pi$
for ($\theta_0=10\;deg, \Delta=2\;deg$) and ($\theta_0=10\;deg, \Delta=5\;deg$):
because of the poor apodization of the second filter, the variance increases
a lot at high multipoles. \\
The bottom plot of figure~\ref{clksith2} shows the correlation function for $l=1500$
for ($\theta_0=10\;deg, \Delta=2\;deg$) and ($\theta_0=5\;deg, \Delta=2\;deg$):
the width of the correlation peak depends on the value of $\theta_0$
and is greater for the second filter as the accessible separation angle
in the angular correlation spectrum is lower in that case.

\subsection{Practical $\cl{l}$ reconstruction from the angular \\ correlation spectrum.}\label{clfrksiprat}

\begin{figure}
   \centering
   \includegraphics[width=150mm]{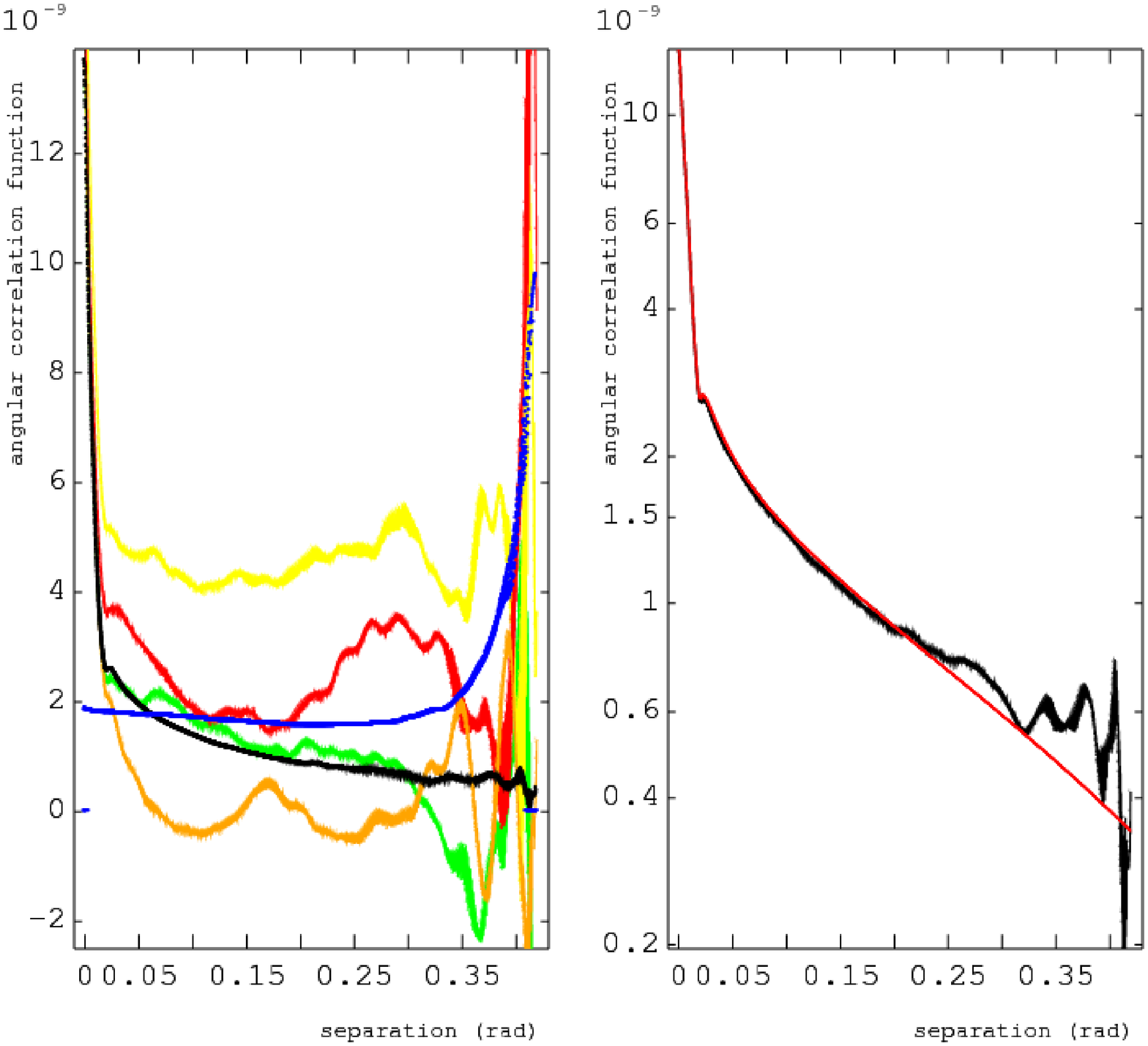}
   \caption[Reconstructed angular correlation function from simulation]
         {
Reconstructed angular correlation function from $173$ simulations of
$17\;deg \;\x17\;deg$ squared maps: \\
$\bullet$ {\it Left figure} \\
- the yellow, orange, green and red curves are reconstructed
angular function on $4$ individual generations. \\
- the black curve is the mean angular function computed
from the $173$ generations. \\
- the blue curve is the dispersion of the angular functions computed
from the $173$ generations. \\
$\bullet$ {\it Right figure} \\
- the black curve is the mean angular function computed
from the $173$ generations. \\
- the red curve is the expected angular function computed
directly from the input $\cl{l}$ spectrum
        }
   \label{simangksi}
\end{figure}

 The above analytical results were applied to 
a set of $173$ simulated $17\x17\;deg^2$ square maps to
check the quality of the $\cl{l}$ reconstruction.  
Computing the correlation spectrum is time consuming,
therefore, for practical purposes, the temperature field
has been calculated on a rectangular grid with constant $\theta$
and $\phi$ intervals of $1 \;arcmin$, across the equator. 
The temperature field is obtained using SYNFAST (see \cite{Gors05})
with $NSIDE=4096$ and attributing to the grid nodes the temperature
value of the nearest HEALPIX map cell center.
No telescope lobe nor measurement noise was introduced. 
The separation angle is computed only from the $\theta$ and $\phi$
differences which allows fast computation of the
angular correlation histogram.
We have checked that, for the small maps under consideration,
this procedure does not introduce
any bias by simulating maps with the same area but
largely elongated along the equator.
Reconstructed angular correlation spectra are shown
in figure~\ref{simangksi} as functions 
of $\gamma$, not $cos(\gamma)$, because the high angular momentum information is 
at small values of $\gamma$. The spectra are very different,
one from the other, due to the large low $l$ value variance of the $\cl{l}$.

Reconstructing the $\cl{l}$ spectrum from these correlation histograms requires
some care.
First, at high $l$ values, the Legendre polynomials oscillate rapidly and the 
calculation of the integral~(\ref{clestksiap}) must be performed with
a very small angular step. 
For that purpose, the correlation spectrum histograms have been
oversampled by a factor $10$
and the intermediate correlation function values obtained
from a third order spline calculation.
Secondly, the angular step of the correlation histogram must be smaller than
the map step, otherwise some information is lost. For instance, at small 
separation angle, it is necessary to distinguish the angular separation angle
of adjacent cells from the one of cells which are neighbours on a diagonal.
We have used a sampling four times finer than the interval between adjacent map
cells ($\frac{1}{4}\;arcmin$).

\begin{figure}
   \centering
   \includegraphics[width=150mm]{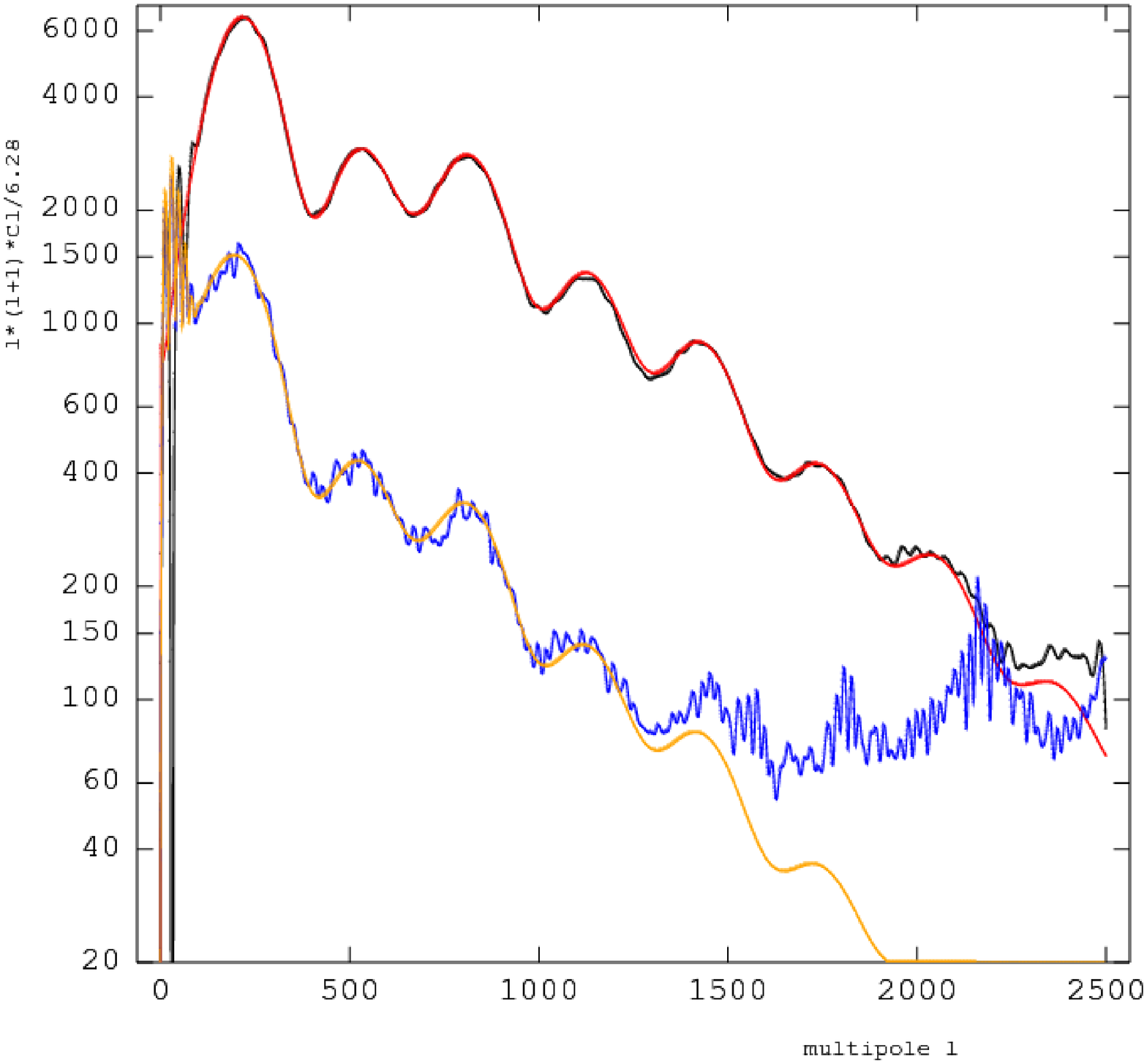}
   \caption[Reconstructed $<\cpl{l}^{\xi}>$ for integration with apodization]
         {
Reconstructed $l(l+1)<\cpl{l}^{\xi}>/2\pi$ from $\fcorp(\gamma)$
integration with apodization
on $173$ simulations of $17\;deg \;\x17\;deg$ squared maps: \\
- the black curve shows the mean reconstructed $l(l+1)<\cpl{l}^{\xi}>/2\pi$
  for  apodization $(\theta_0= 10\;deg,\;\Delta = 2\;deg)$ \\
- the red curve shows the $l(l+1)\cl{l}/2\pi$ input spectrum \\
- the blue curve shows the dispersion $\sigma_{\cpl{l}^{\xi}}$
  of the reconstructed $\cpl{l}^{\xi}$ on the generations
  for the same apodization \\
- the orange curve shows the dispersion $\sigma_{\cpl{l}^{\xi}}$
  predicted by~(\ref{covclksi}) and~(\ref{covclksimlltap})
  for a $17\x17\;deg^2$ spherical cap
        }
   \label{simclr1}
\end{figure}

\begin{figure}
   \centering
   \includegraphics[width=150mm]{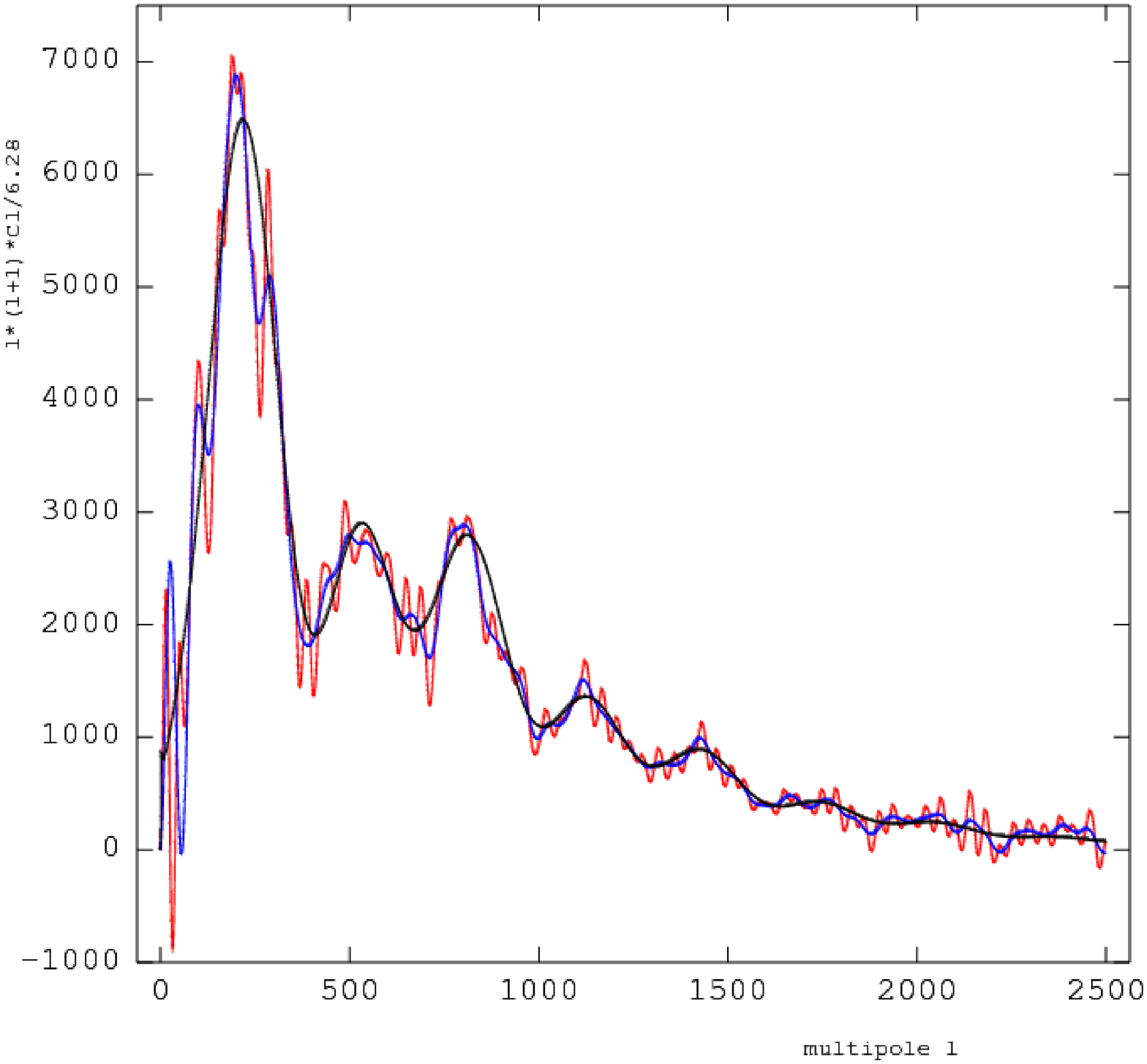}
   \caption[Reconstructed $\cpl{l}^{\xi}$ for integration with apodization]
         {
Reconstructed $l(l+1)\cpl{l}^{\xi}/2\pi$ for integration with apodization
for one individual generation of a $17\;deg \;\x17\;deg$ squared map: \\
- the red curve is for apodization $(\theta_0= 10\;deg,\;\Delta = 2\;deg)$ \\
- the blue curve is for apodization $(\theta_0= 5\;deg,\;\Delta = 2\;deg)$ \\
- the black curve shows the $l(l+1)\cl{l}/2\pi$ input spectrum
        }
   \label{simclr2}
\end{figure}

\begin{figure}
   \centering
   \includegraphics[width=150mm]{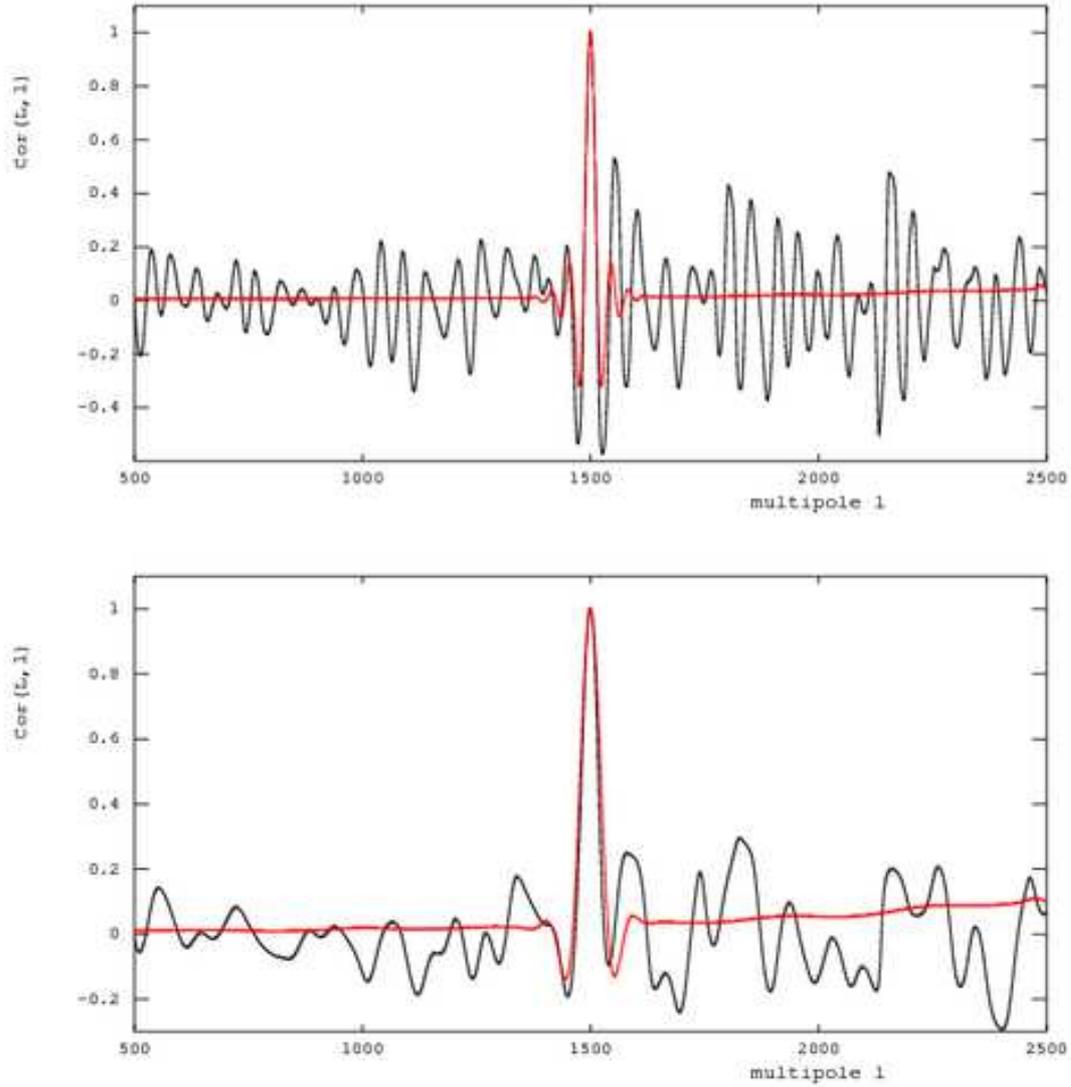}
   \caption[Reconstructed correlation of $\cpl{l}^{\xi}$ for integration with apodization]
         {
Reconstructed correlation of $\cpl{l}^{\xi}$ for integration with apodization for $L=1500$: \\
- {\it Top figure} is for apodization $(\theta_0= 10\;deg,\;\Delta = 2\;deg)$ \\
- {\it Bottom figure} is for apodization $(\theta_0= 5\;deg,\;\Delta = 2\;deg)$ \\
- blacks curves are the reconstructed $Cor(l,L)$ on the $173$ generations of
  $17\;deg \;\x17\;deg$ squared maps \\
- red curves are the predicted $Cor(l,L)$ for a
  $17\;\x17\;deg^2$ spherical cap
        }
   \label{simclr3}
\end{figure}

With these cautions, reconstructed $\cl{l}$ are shown in figure~\ref{simclr1}
and~\ref{simclr2}.
The filter shapes have been chosen according to the discussion of the preceding
section. Figure~\ref{simclr1} shows the average of the reconstructed $\cl{l}$ spectra 
over the $173$ simulated maps, while figure~\ref{simclr2} shows one particular
simulation.
Figure~\ref{simclr1} compares also the calculated variance of the $\cl{l}$
(formula~\ref{covclksi})
with the variance deduced from the $173$ simulations.
They agree well for $l \lesssim 1500$.
The discrepancy at larger $l$ values may be due to the HEALPIX map resolution,
and to the histogram binwidth definition.
This is reflected by the systematic bias of the average 
reconstructed $\cl{l}$ spectrum for $l > 2200$.
Figure~\ref{simclr2} compares the reconstructed $\cl{l}$ spectra obtained with two
different apodisation cut-off parameters. The one corresponding to 
$\theta_0=5\;deg$ looks better than the one obtained with $\theta_0=10\;deg$,
but this reflects only the fact that the $\cl{l}$ values are correlated on a
larger scale in the first case, washing out some fluctuations,
as shown in figure~\ref{simclr3}.

\clearpage

\clearpage
\section{Conclusion.}

In these notes we have computed the biases and covariance of the $\cl{l}$
reconstructed from small sky maps.
This was done for two methods, the widely used one based on FFT analysis
of the temperature field, and the one which uses the angular correlation
spectrum. These two methods are complementary. 
Estimators of the $\cl{l}$ have been defined and we have shown how they can be 
computed numerically, as well as their covariance matrix and correlations.
These calculations are simpler to perform in the case of sperical cap maps,
due to the high degree of symetry. We have shown, using simulated sky maps,
that most of the results do not depend on the map shape at first order.
Most of the biases introduced by the use of small maps can be
studied numerically without having to simulate large amounts of sky maps.
We have also studied the complicated dependency of the 
reconstructed $\cl{l}$ variance over the map size.
In the case of the correlation spectrum we have shown how its apodisation
works.

\clearpage
\appendix
\addcontentsline{toc}{chapter}{APPENDICES:}
\chapter*{APPENDICES}
\renewcommand{\thesection}{\Alph{section}}
\renewcommand{\theequation}{\thesection.\arabic{equation}}
\setcounter{equation}{0}

\section{Computation of the angular correlation integral.} \label{appyyd}

Let's compute the internal integral $\mathcal{I}$ of the angular correlation
function of section~\ref{basicksi}:
$$
\mathcal{I}
 \eg
\integ{S^2\x S^2}{}
\;\conj{\ylm{l_1}{m_1}}(\Omega_1)\ylm{l_2}{m_2}(\Omega_2)\;d\Omega_1 d\Omega_2
\;\delta(\V{\Omega_1}.\V{\Omega_2}-\cos(\gamma))
$$
We rotate the original coordinate axis $Oxyz$ to  $OXYZ$
such that $\V{\Omega_1}=(\theta_1,\phi_1)$ be the unit
vector along $OZ$.
Such a rotation is performed by first rotating by an angle
$\phi_1$ around $Oz$.
The $Oy$ axis moves to  $Oy_1$.
Then a rotation with angle $\theta_1$ around $Oy_1$ is performed.
The rotation which transforms $Oxyz$ to $OXYZ$ is the rotation
of Euler angles $R(\alpha=\phi_1,\beta=\theta_1,0)$.
Let's $(\theta_2,\phi_2)$ be the coordinates of $\V{\Omega_2}$
in $Oxyz$ et $(\Theta_2,\Phi_2)$ in $OXYZ$. \\
The transformation law of the spherical harmonics is (see \cite{Mess64}):
$$
\ylm{l_2}{m_2}(\Theta_2,\Phi_2)
\eg
\somme{m'_2}{}\;\ylm{l_2}{m'_2}(\theta_2,\phi_2)
\;\drot{l_2}{m'_2m_2}(R(\phi_1,\theta_1))
$$
The inverse rotation gives:
$$
\ylm{l_2}{m_2}(\theta_2,\phi_2)
\eg
\somme{m'_2}{}\;\ylm{l_2}{m'_2}(\Theta_2,\Phi_2)
\;\drot{l_2}{m'_2m_2}(R^{-1}(\phi_1,\theta_1))
$$
As $R$ is a unitary transform
$\;
\drot{l}{m'm}(R^{-1}(\phi,\theta))
\eg
\conj{\drot{l}{mm'}}(R(\phi,\theta))
\;$
so we obtain:
$$
\ylm{l_2}{m_2}(\theta_2,\phi_2)
\eg
\somme{m'_2}{}\;\ylm{l_2}{m'_2}(\Theta_2,\Phi_2)
\;\conj{\drot{l_2}{m_2m'_2}}(R(\phi_1,\theta_1))
$$
Let's go back to $\mathcal{I}$.
As the transform is a rotation, the jacobian is equal to $1$ and:
$$
d\Omega_1 d\Omega_2
\eg
d\cos(\theta_1)d\phi_1\;d\cos(\theta_2)d\phi_2\;
\eg
d\cos(\theta_1)d\phi_1\;d\cos(\Theta_2)d\Phi_2\;
$$
The scalar product is invariant under rotation, so the argument
of the delta distribution remains unchanged:
$\;\V{\Omega_1}.\V{\Omega_2}=\cos(\Theta_2)$.
$$ \begin{array}{rcl}
\mathcal{I}
&=&
\integ{S^2\x S^2}{}
\;\conj{\ylm{l_1}{m_1}}(\theta_1,\phi_1)
\somme{m'_2}{}\;\ylm{l_2}{m'_2}(\Theta_2,\Phi_2)
\;\conj{\drot{l_2}{m_2m'_2}}(R(\phi_1,\theta_1))
\;d\cos(\theta_1)d\phi_1\;d\cos(\Theta_2)d\Phi_2\; \\
&&
  \qquad\delta(\cos(\Theta_2)-\cos(\gamma)) \\
&=&
\integ{S^2\x S^2}{}
\;\conj{\ylm{l_1}{m_1}}(\theta_1,\phi_1)
\somme{m'_2}{}\;\ylm{l_2}{m'_2}(\gamma,\Phi_2)
\;\conj{\drot{l_2}{m_2m'_2}}(R(\phi_1,\theta_1))
\;d\cos(\theta_1)d\phi_1\;d\Phi_2\;
\end{array} $$
Since
$\;
\ylm{l_2}{m'_2}(\gamma,\Phi_2)
\eg
\sqrt{\frac{2l+1}{4\pi}\frac{(l_2-m'_2)!}{(l_2+m'_2)!}}
\;\plm{l_2}{m'_2}(\cos(\gamma))\;e^{im'_2\Phi_2}
$, \\
for the integral over $\Phi_2$ to be non zero, we must have $m'_2=0$.
$$
\mathcal{I}
\eg
\integ{S^2}{}
\;\conj{\ylm{l_1}{m_1}}(\theta_1,\phi_1)
\;\sqrt{\frac{2l+1}{4\pi}}\;\plm{l_2}{0}(\cos(\gamma))
\;\conj{\drot{l_2}{m_20}}(R(\phi_1,\theta_1))
\;2\pi\;d\cos(\theta_1)d\phi_1
$$
with
$$
\drot{l_2}{m_20}(R(\phi_1,\theta_1))
\eg
\sqrt{\frac{4\pi}{2l+1}}
\;\conj{\ylm{l_2}{m_2}}(\theta_1,\phi_1)
$$
and $\plm{l_2}{0}(\cos(\gamma))\eg \pl{l_2}(\cos(\gamma))$, we obtain:
$$ \begin{array}{rcl}
\mathcal{I}
&=&
\integ{S^2}{}
\;\conj{\ylm{l_1}{m_1}}(\theta_1,\phi_1)
\;\sqrt{\frac{4\pi}{2l+1}}\;\ylm{l_2}{m_2}(\theta_1,\phi_1)
\;\sqrt{\frac{2l+1}{4\pi}}\;\pl{l_2}(\cos(\gamma))
\;2\pi\;d\cos(\theta_1)d\phi_1 \\
&=&
2\pi\;\delta_{l_1l_2}\;\delta_{m_1m_2}\;\pl{l_2}(\cos(\gamma))
\end{array} $$
\encadre{
\begin{equation}
\mathcal{I}
  \eg
2\pi\;\delta_{l_1l_2}\;\delta_{m_1m_2}\;\pl{l_2}(\cos(\gamma))
\end{equation}
}

\clearpage
\section{Computation of the $\cpl{l}$ covariance for a portion \\ of sphere.} \label{appclblmlm}

We have seen (cf section~\ref{clpartbias}) that we can write:
$$
\aplm{l}{m} \eg \somme{l'm'}{}\;B_{lm;l'm'}\;\alm{l'}{m'}
$$
with either
$$
B_{lm;l'm'} \eg
  (-1)^m \somme{l'',m''}{}\;\blm{l''}{m''}
  \sqrt{\frac{(2l+1)(2l'+1)(2l''+1)}{4\pi}}
    \wigj{l'}{l}{l''}{m'}{-m}{m''}\wigj{l'}{l}{l''}{0}{0}{0}
$$
or
$$
B_{lm;l'm'}
\eg
\integ{A}{}\;\ylm{l'}{m'}\conj{\ylm{l}{m}}\;d\Omega
$$
The relation $\conj{\ylm{l}{-m}}=(-1)^m\ylm{l}{m}$ leads to:
$$
\conj{B_{lm;l'm'}} \eg B_{l'm';lm} \eg (-1)^{m+m'}\;B_{l-m;l'-m'} \;\in\; \EnsC
$$
\Main Let's compute the ensemble average $<\cpl{l}>$:
$$\begin{array}{rcl}
\cpl{l}
  &=&
  \frac{1}{2l+1}\;\somme{m=-l}{+l}\;\conj{\aplm{l}{m}}\;\aplm{l}{m} \\
  &=&
  \frac{1}{2l+1}\;\somme{m}{}\;\somme{l'm'}{}\;\somme{l''m''}{}
    \conj{B_{lm;l'm'}}\;B_{lm;l''m''}\;\conj{\alm{l'}{m'}}\;\alm{l''}{m''} \\
<\cpl{l}>
  &=&
  \frac{1}{2l+1}\;\somme{m}{}\;\somme{l'm'}{}\;\somme{l''m''}{}
    \conj{B_{lm;l'm'}}\;B_{lm;l''m''}\;<\conj{\alm{l'}{m'}}\;\alm{l''}{m''}> \\
  &=&
  \frac{1}{2l+1}\;\somme{m}{}\;\somme{l'm'}{}\;\somme{l''m''}{}
    \conj{B_{lm;l'm'}}\;B_{lm;l''m''}\;\delta_{l'l''}\delta{m'm''}\;\cl{l}
\end{array}$$
\encadre{$$
<\cpl{l}>
  \eg
  \frac{1}{2l+1}\;\somme{m=-l}{+l}\;\somme{l'm'}{}\;\abs{B_{lm;l'm'}}^2\;\cl{l}
$$}
\Main Now let's compute the $\cpl{l}$  covariance:
$$\begin{array}{rcl}
\cpl{l}\;\conj{\cpl{L}}
 &=&
  \frac{1}{(2l+1)(2L+1)}
  \;\;\somme{m}{}\;\somme{l'm'}{}\;\somme{l''m''}{}
  \;\;\somme{M}{}\;\somme{L'M'}{}\;\somme{L''M''}{} \\
 && \qquad\qquad
  \;B_{lm;l'm'}\conj{B_{lm;l''m''}}
  \;B_{LM;L'M'}\conj{B_{LM;L''M''}}
  \;\alm{l'}{m'}\conj{\alm{l''}{m''}}
  \;\alm{L'}{M'}\conj{\alm{L''}{M''}}
\end{array}$$
Using the ensemble average of $4$ $\alm{l}{m}$ products
for a {\it real gaussian} temperature field:
$$ \begin{array}{rcl}
<\alm{l_1}{m_1}\conj{\alm{l_2}{m_2}}\alm{l_3}{m_3}\conj{\alm{l_4}{m_4}}>
  &=&
\cl{l_1}\cl{l_3}\;
\left(
\delta_{l_1l_2}\delta_{m_1m_2}\delta_{l_3l_4}\delta_{m_3m_4}
\;+\;
\delta_{l_1l_4}\delta_{m_1m_4}\delta_{l_2l_3}\delta_{m_2m_3}
\right) \\
  &&
\;+\;
(-1)^{m_1+m_2}\;\cl{l_1}\cl{l_2}\;
\delta_{l_1l_3}\delta_{m_1-m_3}\delta_{l_2l_4}\delta_{m_2-m_4}
\end{array} $$
We obtain:
$$\begin{array}{rcl}
<\cpl{l}\;\conj{\cpl{L}}>
 &=&
  \frac{1}{(2l+1)(2L+1)}
  \;\somme{}{}\cdots\somme{}{}
  \;B_{lm;l'm'}\conj{B_{lm;l''m''}}
  \;B_{LM;L'M'}\conj{B_{LM;L''M''}} \\
 && \qquad\qquad
  \left\{
  \cl{l'}\cl{L'}\;
  (
  \delta_{l'l''}\delta_{m'm''}\delta_{L'L''}\delta_{M'M''}
  +\delta_{l'L''}\delta_{m'M''}\delta_{l''L'}\delta_{m''M'}
  ) 
  \right.\\
 && \qquad\qquad
  \left.
  + (-1)^{m'+m''}\;\cl{l'}\cl{l''}\;
  \delta_{l'L'}\delta_{m'-M'}\delta_{l''L''}\delta_{m''-M''}
  \right\} \\
\end{array}$$
$$\begin{array}{rcl}
<\cpl{l}\;\conj{\cpl{L}}>
 &=&
  \frac{1}{(2l+1)(2L+1)}\;\somme{m}{}\;\somme{M}{}\;\{ \\
 &&
 \somme{l'm'L'M'}{}\;
   \cl{l'}\cl{L'}\;
  \;B_{lm;l'm'}\conj{B_{lm;l'm'}}\;B_{LM;L'M'}\conj{B_{LM;L'M'}} \\
 &&
 +\;\somme{l'm'L'M'}{}\;
   \cl{l'}\cl{L'}\;
  \;B_{lm;l'm'}\conj{B_{lm;L'M'}}\;B_{LM;L'M'}\conj{B_{LM;l'm'}} \\
 &&
 +\;(-1)^{m'+m''}\;\somme{l'm'l''m''}{}\;
   \cl{l'}\cl{l''}\;
  \;B_{lm;l'm'}\conj{B_{lm;l''m''}}\;B_{LM;l'-m'}\conj{B_{LM;l''-m''}} \\
 &&
 \}
\end{array}$$
The first term corresponds to $\;<\cpl{l}><\cpl{L}>$. \\
One could exchange $l'',m''$ to $L',M'$ in the third term sum:
$$\begin{array}{rcl}
V(l,L)
 &=&
  <\cpl{l}\;\conj{\cpl{L}}>\;-\;<\cpl{l}><\cpl{L}> \\
 &=&
  \frac{1}{(2l+1)(2L+1)}\;\somme{m}{}\;\somme{M}{}
  \;\somme{l'm'}{}\;\somme{L'M'}{}\;\cl{l'}\cl{L'}\;\{ \\
 &&
  \;B_{lm;l'm'}\conj{B_{lm;L'M'}}\;B_{LM;L'M'}\conj{B_{LM;l'm'}} \\
 &&
  +\;(-1)^{m'+M'}
  \;B_{lm;l'm'}\conj{B_{lm;L'M'}}\;B_{LM;l'-m'}\conj{B_{LM;L'-M'}} \\
 &&
 \}
\end{array}$$
Then exchanging $M$ to $-M$ in the second term sum
and using the symetry relation for the $B_{lm;l'm'}$:
$$\begin{array}{rcl}
V(l,L)
 &=&
  \frac{1}{(2l+1)(2L+1)}\;\somme{m}{}\;\somme{M}{}\
  \;\somme{l'm'}{}\;\somme{L'M'}{}\;\cl{l'}\cl{L'}\;\{ \\
 &&
  \;B_{lm;l'm'}\conj{B_{lm;L'M'}}\;B_{LM;L'M'}\conj{B_{LM;l'm'}} \\
 &&
  +\;(-1)^{m'+M'}
  \;B_{lm;l'm'}\conj{B_{lm;L'M'}}\;B_{L-M;l'-m'}\conj{B_{L-M;L'-M'}} \\
 &&
 \} \\
 &=&
  \frac{1}{(2l+1)(2L+1)}\;\somme{m}{}\;\somme{M}{}
  \;\somme{l'm'}{}\;\somme{L'M'}{}\;\cl{l'}\cl{L'}\;\{ \\
 &&
  \;B_{lm;l'm'}\conj{B_{lm;L'M'}}\;B_{LM;L'M'}\conj{B_{LM;l'm'}} \\
 &&
  +
  \;B_{lm;l'm'}\conj{B_{lm;L'M'}}\;\conj{B_{LM;l'm'}}B_{LM;L'M'} \\
 &&
 \} \\
 &=&
  \frac{2}{(2l+1)(2L+1)}\;\somme{m}{}\;\somme{M}{}
  \;\somme{l'm'}{}\;\somme{L'M'}{}\;\cl{l'}\cl{L'}\;\{ \\
 &&
  \;B_{lm;l'm'}\conj{B_{lm;L'M'}}\;B_{LM;L'M'}\conj{B_{LM;l'm'}} \\
 &&
 \} \\
 &=&
  \frac{2}{(2l+1)(2L+1)}\;\somme{m}{}\;\somme{M}{}
  \; \left( \somme{l'm'}{}\;\cl{l'}\;B_{lm;l'm'}\conj{B_{LM;l'm'}} \right)
  \; \left( \somme{L'M'}{}\;\cl{L'}\;\conj{B_{lm;L'M'}}\;B_{LM;L'M'} \right) \\
 &=&
  \frac{2}{(2l+1)(2L+1)}\;\somme{m}{}\;\somme{M}{}
  \; \left( \somme{l'm'}{}\;\cl{l'}\;B_{lm;l'm'}\conj{B_{LM;l'm'}} \right)
  \; \conj{\left( \somme{L'M'}{}\;\cl{L'}\;B_{lm;L'M'}\;\conj{B_{LM;L'M'}} \right)} \\

\end{array}$$
We finaly obtain:
$$\begin{array}{rcl}
V(l,L)
 &=&
  <\cpl{l}\;\conj{\cpl{L}}>\;-\;<\cpl{l}><\cpl{L}> \\
 &=&
  \frac{2}{(2l+1)(2L+1)}\;\somme{m=-l}{+l}\;\somme{M=-L}{+L}\;
  \;\abs{
  \somme{l'm'}\;\cl{l'}\;B_{lm;l'm'}\;\conj{B_{LM;l'm'}}
  }^2
\end{array}$$
If we replace the $B_{lm;l'm'}$ by their values, we obtain:
$$\begin{array}{rcl}
V(l,L)
 &=&
  \frac{2}{(2l+1)(2L+1)}\;\somme{m=-l}{+l}\;\somme{M=-L}{+L}
  \;\left| \;\somme{l'm'}{}\;\cl{l'} \right. \\
 &&
  \x\left[
  \somme{b\beta}{}\;\blm{b}{\beta}\;(-1)^m
  \;\sqrt{\frac{(2l+1)(2l'+1)(2b+1)}{4\pi}}
  \;\wigj{l'}{l}{b}{m'}{-m}{\beta}\;\wigj{l'}{l}{b}{0}{0}{0}
  \right] \\
 &&
  \left.
  \x\left[
  \somme{d\delta}{}\;\conj{\blm{d}{\delta}}\;(-1)^M
  \;\sqrt{\frac{(2L+1)(2l'+1)(2d+1)}{4\pi}}
  \;\wigj{l'}{L}{d}{m'}{-M}{\delta}\;\wigj{l'}{L}{d}{0}{0}{0}
  \right]
  \right| ^2 \\
 &=&
  \frac{2}{(4\pi)^2}\;\somme{m=-l}{+l}\;\somme{M=-L}{+L}\;
  \;\left|
  \;\somme{l'm'}{}\;(2l'+1)\;\cl{l'} 
  \x\somme{b\beta}{}\somme{d\delta}{}\;\sqrt{(2b+1)(2d+1)}
  \;\blm{b}{\beta}\conj{\blm{d}{\delta}}
  \right. \\
 &&
  \left.
  \qquad\qquad\x
  \wigj{l'}{l}{b}{m'}{-m}{\beta}
  \wigj{l'}{L}{d}{m'}{-M}{\delta}
  \wigj{l'}{l}{b}{0}{0}{0}
  \wigj{l'}{L}{d}{0}{0}{0}
  \right| ^2
\end{array}$$
That formula is not very usefull because it involves the computation
of a general $3j$ symbol which is time consuming.
This is why we have to add another constraint:
{\it the spherical symetry of the portion of sphere under study}. \\
Note that we would have obtain exactly the same result by doing the
computation with the other definition for the $B_{lm;l'm'}$.
In this case,
because it would involve integral of product of $\ylm{l}{m}$
other a portion of sphere of general shape,
the computation time would be enormous .

\clearpage
\section{Computation of the angular correlation distribution \\ for a portion of sphere.} \label{appksipart}

\begin{figure}[pht]
   \begin{center}
   \psfig{file=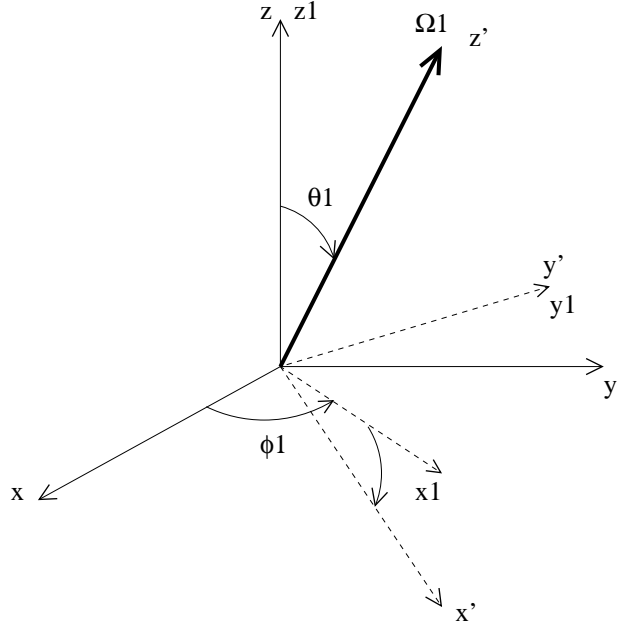,width=8cm}
   \caption[Euler rotation]
         {Euler rotation.}
   \label{coresteuler}
   \end{center}
\end{figure}

Let's compute
$$ \begin{array}{rcl}
\mathcal{N}^A(\gamma)\;\fcorp(\gamma)
 &=&
\somme{\cdots}{}
\;\conj{\alm{l_1}{m_1}}\conj{\blm{L_1}{M_1}}
\;\alm{l_2}{m_2}\blm{L_2}{M_2} \\
 &&
\quad\x
\integ{S^2\x S^2}{}\;d\Omega_1 d\Omega_2
\;\delta(\V{\Omega_1}.\V{\Omega_2}-\cos(\gamma))
\;\conj{\ylm{l_1}{m_1}}(\Omega_1)\conj{\ylm{L_1}{M_1}}(\Omega_1)
\;\ylm{l_2}{m_2}(\Omega_2)\ylm{L_2}{M_2}(\Omega_2)
\end{array} $$
where the sum $\somme{}{}$
runs on $\{l_1,m_1,L_1,M_1,l_2,m_2,L_2,M_2\}$. \\
We perform the integration on $\Omega_2$ by rotating
the frame $(Oxyz)$ into $(Ox'y'z')$ such that
the axis $Oz'$ be on $\V{\Omega_1}$
(see figure~\ref{coresteuler})

We perform the Euler rotation $R(\alpha,\beta,\gamma)$
of the frame\footnote{
Here $\gamma$ is the third angle of the Euler rotation
not the separation angle of the angular correlation function.
}:
\begin{itemize}
\item Rotation of $\alpha=\phi_1$ around $Oz$:
      $(O,x,y,z)\;\rightarrow\;(O,x_1,y_1,z_1=z)$
\item Rotation of $\beta=\theta_1$ around $Oy_1$:
      $(O,x_1,y_1,z_1=z)\;\rightarrow\;(O,x',y'=y_1,z')$
\item Rotation of  $\gamma=0$ around $Oz_2$ (i.e. the identity).
\end{itemize}
So the rotation is:
$R(\alpha,\beta,\gamma)\eg R(\phi_1,\theta_1,0)$. \\
We define $(\theta,\phi)$ to be the polar coordinates of the frame $(O,x,y,z)$
and $(\theta',\phi')$ the polar coordinates in the new frame $(O,x',y',z')$.
$$\begin{array}{rcl}
\ylm{l}{m}(\theta',\phi')
  &=&
  \somme{m'}{}\;\drot{l}{m'm}(\phi_1,\theta_1,0)\;\ylm{l}{m'}(\theta,\phi) \\
\ylm{l}{m}(\theta,\phi)
  &=&
  \somme{m'}{}\;\conj{\drot{l}{mm'}}(\phi_1,\theta_1,0)\;\ylm{l}{m'}(\theta',\phi')
\end{array}$$
Performing the rotation we get:
$$ \begin{array}{rcl}
\mathcal{N}^A(\gamma)\;\fcorp(\gamma)
 &=&
\somme{\cdots}{}
\;\conj{\alm{l_1}{m_1}}\conj{\blm{L_1}{M_1}}
\;\alm{l_2}{m_2}\blm{L_2}{M_2}
\;\integ{S^2}{}\;d\Omega_1 
\;\conj{\ylm{l_1}{m_1}}(\Omega_1)\conj{\ylm{L_1}{M_1}}(\Omega_1) \\
 &&
\qquad\x
\integ{S^2}{}\;d\Omega_2'
\;\delta(\V{\Omega_1'}.\V{\Omega_2'}-\cos(\gamma))
\;\conj{\drot{l_2}{m_2m_2'}}\ylm{l_2}{m_2'}(\Omega_2')
\conj{\drot{L_2}{M_2M_2'}}\ylm{L_2}{M_2'}(\Omega_2')
\end{array} $$
where the sum runs on
$\{l_1,m_1,L_1,M_1,l_2,m_2,L_2,M_2,m_2',M_2'\}$ \\
and where we simplify the notation
$\;\drot{l}{m'm}\equiv\drot{l}{m'm}(\phi_1,\theta_1,0)$. \\
We have $d\Omega_2'=d\cos(\theta_2')d\phi_2'$.
The integral on $d\cos(\theta_2')$ and the function $\delta(\cdots)$
leads to the replacement of $\theta_2'$ by $\gamma$.
Writing $\ylm{l}{m}(\theta,\phi)=\hplm{l}{m}(\theta)e^{im\phi}$, we have
for the integral on $\phi_2'$:
$$ \begin{array}{rcl}
\integ{0}{\pi}\integ{0}{2\pi}\;d\cos(\theta_2')d\phi_2'
  \;\ylm{l_2}{m_2'}(\Omega_2')\ylm{L_2}{M_2'}(\Omega_2')
 &=&
\integ{0}{2\pi}\;d\phi_2'
  \;\hplm{l_2}{m_2'}(\gamma)\hplm{L_2}{M_2'}(\gamma)
  \;e^{i(m_2'+M_2')\phi_2'} \\
 &=&
 2\pi\delta_{m_2',-M_2'}
 \;\hplm{l_2}{m_2'}(\gamma)\hplm{L_2}{M_2'}(\gamma)
\end{array} $$
Thus we obtain:
$$ \begin{array}{rcl}
\mathcal{N}^A(\gamma)\;\fcorp(\gamma)
 &=&
2\pi\;\somme{\cdots}{}
\;\conj{\alm{l_1}{m_1}}\alm{l_2}{m_2}
\;\conj{\blm{L_1}{M_1}}\blm{L_2}{M_2}
\;\hplm{l_2}{m_2'}(\gamma)\hplm{L_2}{-m_2'}(\gamma) \\
 &&
\qquad\x
\integ{S^2}{}\;d\Omega_1 
\;\conj{\ylm{l_1}{m_1}}(\Omega_1)\conj{\ylm{L_1}{M_1}}(\Omega_1)
\;\conj{\drot{l_2}{m_2m_2'}}\;\conj{\drot{L_2}{M_2-m_2'}}
\end{array} $$
where the sum runs on
$\{l_1,m_1,L_1,M_1,l_2,m_2,L_2,M_2,m_2'\}$. \\
Now we perform the ensemble average, and as
$\;
<\conj{\alm{l_1}{m_1}}\alm{l_2}{m_2}>
 =
\cl{l_1}\delta_{l_1l_2}\delta_{m_1m_2}
$,
we get:
$$ \begin{array}{rcl}
<\mathcal{N}^A(\gamma)\;\fcorp(\gamma)>
 &=&
2\pi\;\somme{\cdots}{}
\;\conj{\blm{L_1}{M_1}}\blm{L_2}{M_2}
\;\cl{l_1}
\;\hplm{l_1}{m_2'}(\gamma)\hplm{L_2}{-m_2'}(\gamma) \\
 &&
\qquad\x
\integ{S^2}{}\;d\Omega_1 
\;\conj{\drot{l_1}{m_1m_2'}}\conj{\ylm{l_1}{m_1}}(\Omega_1)
\;\conj{\drot{L_2}{M_2-m_2'}}\conj{\ylm{L_1}{M_1}}(\Omega_1)
\end{array} $$
where the sum runs on
$\{l_1,m_1,L_1,M_1,L_2,M_2,m_2'\}$. We have:
$$
\somme{m_1}{}\;\conj{\drot{l_1}{m_1m_2'}}\conj{\ylm{l_1}{m_1}}(\Omega_1)
  \eg
\conj{\left( \somme{m_1}{}\;\drot{l_1}{m_1m_2'}\ylm{l_1}{m_1}(\Omega_1) \right)}
  \eg
\conj{\ylm{l_1}{m_2'}}(\Omega_1')
$$
and by definition, $\V{\Omega_1'}$ is on $Oz'$,
($\theta_1'=0$), so
$\;\conj{\ylm{l_1}{m_2'}}(\Omega_1')=\hplm{l_1}{m_2'}(0)e^{-im_2'\phi_1'}$. \\
As $\hplm{l_1}{m_2'}(0)=\hplm{l_1}{0}(0)\delta_{m_2'0}$, we obtain:
$$
<\mathcal{N}^A(\gamma)\;\fcorp(\gamma)>
 \eg
2\pi\;\somme{\cdots}{}
\;\conj{\blm{L_1}{M_1}}\blm{L_2}{M_2}
\;\cl{l_1}
\;\hplm{l_1}{0}(\gamma)\hplm{L_2}{0}(\gamma)\hplm{l_1}{0}(0)
\integ{S^2}{}d\Omega_1
\conj{\drot{L_2}{M_20}}(\phi_1,\theta_1,0)\conj{\ylm{L_1}{M_1}}(\Omega_1)
$$
where the sum runs on
$\{l_1,L_1,M_1,L_2,M_2\}$. \\
But we have:
$
\drot{L_2}{M_20}(\phi_1,\theta_1,0)
 \eg
\sqrt{\frac{4\pi}{2L_2+1}}\;\conj{\ylm{L_2}{M_2}}(\theta_1,\phi_1)
$
$$ \begin{array}{rcl}
<\mathcal{N}^A(\gamma)\;\fcorp(\gamma)>
 &=&
2\pi\;\somme{\cdots}{}\;\sqrt{\frac{4\pi}{2L_2+1}}
\;\conj{\blm{L_1}{M_1}}\blm{L_2}{M_2}
\;\cl{l_1}
\;\hplm{l_1}{0}(\gamma)\hplm{L_2}{0}(\gamma)\hplm{l_1}{0}(0) \\
 &&
\qquad\qquad\x
\integ{S^2}{}\;d\Omega_1 
\;\ylm{L_2}{M_2}(\theta_1,\phi_1)\conj{\ylm{L_1}{M_1}}(\Omega_1) \\
 &=&
2\pi\;\somme{\cdots}{}\;\sqrt{\frac{4\pi}{2L_2+1}}
\;\conj{\blm{L_1}{M_1}}\blm{L_2}{M_2}
\;\cl{l_1}
\;\hplm{l_1}{0}(\gamma)\hplm{L_2}{0}(\gamma)\hplm{l_1}{0}(0)
\;\delta_{L_2L_1}\delta_{M_2M_1} \\
 &=&
2\pi\;\somme{\cdots}{}\;\sqrt{\frac{4\pi}{2L_1+1}}
\;\conj{\blm{L_1}{M_1}}\blm{L_1}{M_1}
\;\cl{l_1}
\;\hplm{l_1}{0}(\gamma)\hplm{L_1}{0}(\gamma)\hplm{l_1}{0}(0)
\end{array} $$
where the final sum runs on $\{l_1,L_1,M_1\}$. \\
Using the definition of $\bl{l}$ (see~\ref{blmask})
as well as
$\hplm{l}{m}=\sqrt{\frac{2l+1}{4\pi}}\sqrt{\frac{(l-m)!}{(l+m)!}}\plm{l}{m}$,
$\plm{l}{0}=\pl{l}$ and $\pl{l}(0)=1$, we obtain:
$$ \begin{array}{rcl}
<\mathcal{N}^A(\gamma)\;\fcorp(\gamma)>
 &=&
\left(
\somme{l_1}{}\;\frac{2l_1+1}{4\pi}\;\cl{l_1}\;\pl{l_1}(\gamma)
\right)
\x
\left(
2\pi
\;\somme{L_1}{}\;(2L_1+1)\;\bl{L_1}\;\pl{L_1}(\gamma)
\right)
\end{array} $$
To compute
$\;
\mathcal{N}^A(\gamma)\eg \integ{S^2\x S^2}{}\;d\Omega_1 d\Omega_2
\;\conj{W^A(\Omega_1)}\;W^A(\Omega_2)
\;\delta(\V{\Omega_1}.\V{\Omega_2}-\cos(\gamma))
\;$
we do the same computation as before with
$\;
T(\Omega)=1=\sqrt{4\pi}\ylm{0}{0}(\Omega)
\;$
so
$\;\alm{l_1}{m_1}=\sqrt{4\pi}\delta_{l_10}\delta_{m_10}\;$
and
$\;\cl{l_1}=4\pi\delta_{l_10}\;$. \\
We thus obtain ($\pl{0}(\theta)=1$):
$$
\mathcal{N}^A(\gamma)
  \eg
2\pi\;\somme{L}{}\;(2L+1)\;\bl{L}^A\;\pl{L}(\gamma)
$$
and finally:
\encadre{$$
<\fcorp(\gamma)>
  \eg
\left( \somme{l}{}\;\frac{2l+1}{4\pi}\;\cl{l}\;\pl{l}(\gamma) \right)
\frac{\somme{L}{}\;(2L+1)\;\bl{L}\;\pl{L}(\gamma)}
     {\somme{L}{}\;(2L+1)\;\bl{L}^A\;\pl{L}(\gamma)}
$$}

\clearpage
\section{Computation of the angular correlation integral \\ for a portion of sphere.} \label{appyydpart}

Let's compute the integral:
$$ \begin{array}{rcl}
I_{l'm'}^{lm}(c)
  &=&
\integ{A\x A}{} d\Omega d\Omega'\;\delta(\V{\Omega}.\V{\Omega'}-c)
\;\ylm{l'}{m'}(\Omega')\conj{\ylm{l}{m}}(\Omega) \\
  &=&
\integ{S^2\x S^2}{} d\Omega d\Omega'\;\delta(\V{\Omega}.\V{\Omega'}-c)
\;W^A(\Omega')W^A(\Omega)
\;\ylm{l'}{m'}(\Omega')\conj{\ylm{l}{m}}(\Omega) \\
  &=&
\somme{LML'M'}{}\;\conj{\blm{L}{M}}\blm{L'}{M'}
\integ{S^2\x S^2}{} d\Omega d\Omega'\;\delta(\V{\Omega}.\V{\Omega'}-c)
\;\ylm{l'}{m'}(\Omega')\ylm{L'}{M'}(\Omega')
\conj{\ylm{l}{m}}(\Omega)\conj{\ylm{L}{M}}(\Omega) \\
\end{array} $$
Using the formula of the product of two $\ylm{l}{m}$
relative to the Clebsch-Gordan coefficients
and remembering that the coefficients are real:
$$ \begin{array}{rcl}
I_{l'm'}^{lm}(c)
  &=&
\somme{LML'M'}{}\;\conj{\blm{L}{M}}\blm{L'}{M'}
\;\somme{\lambda\mu\lambda'\mu'}{}
\;\sqrt{\frac{(2l+1)(2L+1)}{4\pi(2\lambda+1)}}
\;\sqrt{\frac{(2l'+1)(2L'+1)}{4\pi(2\lambda'+1)}} \\
 && \qquad\x
\;\clg{lLmM}{\lambda\mu}\clg{lM00}{\lambda 0}
\;\clg{l'L'm'M'}{\lambda'\mu'}\clg{l'L'00}{\lambda' 0} \\
 && \qquad\qquad\x
\;\integ{S^2}{}\;d\Omega d\Omega'\;\delta(\V{\Omega}.\V{\Omega'}-c)
\;\ylm{\lambda'}{\mu'}(\Omega')\conj{\ylm{\lambda}{\mu}}(\Omega)
\end{array} $$
The remaining integral has been computed in appendix~\ref{appyyd}
$$
\mathcal{I}
  \eg
2\pi\;\delta_{\lambda\lambda'}\;\delta_{\mu\mu'}\;\pl{\lambda}(c)
$$
Thus
$$ \begin{array}{rcl}
I_{l'm'}^{lm}(c)
  &=&
2\pi\;\somme{LML'M'}{}\;\conj{\blm{L}{M}}\blm{L'}{M'}
\;\somme{\lambda\mu}{}
\;\pl{\lambda}(c) \\
 && \qquad\x
\;\sqrt{\frac{(2l+1)(2L+1)}{4\pi(2\lambda+1)}}
\;\clg{lLmM}{\lambda\mu}\clg{lM00}{\lambda 0} \\
 && \qquad\qquad\x
\;\sqrt{\frac{(2l'+1)(2L'+1)}{4\pi(2\lambda+1)}}
\;\clg{l'L'm'M'}{\lambda\mu}\clg{l'L'00}{\lambda 0}
\end{array} $$
It is real and can be rewritten as integrals of three $\ylm{l}{m}$.
$$ \begin{array}{rcl}
I_{l'm'}^{lm}(c)
  &=&
2\pi\somme{LML'M'}{}\conj{\blm{L}{M}}\blm{L'}{M'}
\somme{\lambda\mu}{}
\pl{\lambda}(c)
\integ{S^2}{}d\Omega
\conj{\ylm{l}{m}}(\Omega)\conj{\ylm{L}{M}}(\Omega)\ylm{\lambda}{\mu}(\Omega)
\integ{S^2}{}d\Omega'
\ylm{l'}{m'}(\Omega')\ylm{L'}{M'}(\Omega')\conj{\ylm{\lambda}{\mu}}(\Omega') \\
  &=&
2\pi\somme{\lambda\mu}{}\pl{\lambda}(c)
\somme{LML'M'}{}
\integ{S^2}{}d\Omega
\conj{\ylm{l}{m}}(\Omega)\conj{\blm{L}{M}}\conj{\ylm{L}{M}}(\Omega)\ylm{\lambda}{\mu}(\Omega)
\integ{S^2}{}d\Omega'
\ylm{l'}{m'}(\Omega')\blm{L'}{M'}\ylm{L'}{M'}(\Omega')\conj{\ylm{\lambda}{\mu}}(\Omega') \\
  &=&
2\pi\somme{\lambda\mu}{}\;\pl{\lambda}(c)
\;\integ{S^2}{}\;d\Omega
\;\conj{W^A(\Omega)}\conj{\ylm{L}{M}}(\Omega)\ylm{\lambda}{\mu}(\Omega)
\integ{S^2}{}\;d\Omega'
\;W^A(\Omega')\ylm{L'}{M'}(\Omega')\conj{\ylm{\lambda}{\mu}}(\Omega') \\
  &=&
2\pi\somme{\lambda\mu}{}\;\pl{\lambda}(c)
\;\integ{A\x A}{}\;d\Omega
\;\conj{\ylm{l}{m}}(\Omega)\ylm{\lambda}{\mu}(\Omega)
\integ{A\x A}{}\;d\Omega'
\;\ylm{l'}{m'}(\Omega')\conj{\ylm{\lambda}{\mu}}(\Omega') \\
\end{array} $$
Remembering the $B_{lml'm'}$ definitions (see section~\ref{clpartbias}):
$$
I_{l'm'}^{lm}(c)
  \eg
2\pi\;\somme{LM}{}\;\pl{L}(c)\;B_{LMlm}\;\conj{B_{LMl'm'}}
$$






\clearpage

\addcontentsline{toc}{chapter}{Bibliography.}
\begin{thebibliography}{}

\bibitem[Benoit et al, 2003]{Ben03} Benoit,A. et al (The Archeops Collaboration), A\&A,399, p.L19-L23 (2003)

\bibitem[Brink et al, 1962]{Brink62} Brink,D.M.; Satchler,G.R., Angular Momentum (Clarendon Press 1962)

\bibitem[De Bernardis et al, 2000]{Bern00} De Bernardis, P., Ade, P. A. R., Bock, J. J., et al. 2000, Nature, 404, 955

\bibitem[G{\`o}rski et al, 2005]{Gors05} G{\`o}rski, K. M., Hivon, E., Banday, A. J., et al. 2005, ApJ , 622, 759  (see also http://healpix.jpl.nasa.gov)

\bibitem[Gradshtein et al, 1980]{Grad80} Gradshtein I.S.; Ryzhik,I.M.; Jeffrey,A., Table of integrals, series, and products (Academic Press 1980)

\bibitem[Hanany et al, 2000]{Han00} Hanany, S., Ade, P., Balbi, A., et al. 2000, ApJ , 545, L5

\bibitem[Hivon et al, 2002]{Hiv02} Hivon et al, The Astrophysical Journal, Volume 567, Issue 1, pp. 2-17 (2002).

\bibitem[Masi et al, 2006]{Mass06} S.Masi, et al. 17th ESA Symposium on European Rocket and Balloon Programmes and Related Research, 30 May-2 June 2005, Sandefjord, Norway. ESA Publications  Division, ISBN 92-9092-901-4, 2005, p  581-586.

\bibitem[Messiah, 1964]{Mess64}Messiah,A. m\'ecanique quantique Tome 2 (Editions DUNOD, paris 1964)

\bibitem[Natoli et al, 1997]{Muc97} P. Natoli, P.F. Muciaccia, and N. Vittorio, ApJ Letter 488, L63 (1997).

\bibitem[Seljak et al, 1996]{Sel96} Seljak, U., Zaldarriaga, M., 1996 ApJ, 469, 437 (see also http://www.cmbfast.org)

\bibitem[Tegmark, 1997]{Tegm97} Tegmark, M. 1997, PrD , 56, 4514

\bibitem[Wandelt et al, 2001]{Wan01} Wandelt et al, Physical Review D (Particles, Fields, Gravitation, and Cosmology), Volume 64, Issue 8, 15 October 2001

\bibitem[Zaldarriaga et al, 2000]{Zal00} Zaldarriaga, M., Seljak, U., 2000 ApJS, 129, pp 431-434

\end{thebibliography}
\end{document}